\documentclass[useAMS,usenatbib]{mn2e}
\usepackage{times}
\usepackage{graphicx}
\usepackage{amssymb}
\usepackage{longtable}

\title[Star formation \& quenching in MaNGA]{Are galactic star formation and quenching governed by local, global or environmental phenomena?}
\author[Asa F. L. Bluck et al.]{Asa F. L. Bluck$^{1,2,3,*}$, Roberto Maiolino$^{1,2}$, Sebastian F. S\'anchez$^4$, Sara L. Ellison$^5$, \newauthor Mallory D. Thorp$^5$, Joanna M. Piotrowska$^{1,2}$, Hossen Teimoorinia$^{6,5}$, Kevin A. Bundy$^{7}$
\\$^1$ Kavli Institute for Cosmology, University of Cambridge, Madingley Road, Cambridge, CB3 0HA, UK
\\$^2$ Cavendish Laboratory - Astrophysics Group, University of Cambridge, 19 JJ Thomson Avenue, Cambridge, CB3 0HE, UK
\\$^3$ Hughes Hall College, University of Cambridge, Wollaston Road, CB1 2EW, UK
\\$^4$ Instituto de Astronomia, Universidad Nacional Autonoma de Mexico, A. P. 70-264, C.P. 04510, Mexico, D.F., Mexico
\\$^5$ Department of Physics \& Astronomy, University of Victoria, Finnerty Road, Victoria, BC V8P 1A1, Canada
\\$^6$ NRC Herzberg Astronomy and Astrophysics, 5071 West Saanich Road, Victoria, BC, V9E 2E7, Canada
\\$^7$ UCO/Lick Observatory, University of California, Santa Cruz, 1156 High St., Santa Cruz, CA 95064, USA
\\$*$ Email: asa.bluck@mrao.cam.ac.uk }

\begin{document}

\maketitle

\begin{abstract}

We present an analysis of star formation and quenching in the SDSS-IV MaNGA-DR15, utilising over 5 million spaxels from $\sim$3500 local galaxies. We estimate star formation rate surface densities ($\Sigma_{\rm SFR}$) via dust corrected $H\alpha$ flux where possible, and via an empirical relationship between specific star formation rate (sSFR) and the strength of the 4000 \AA \hspace{0.05cm} break (D4000) in all other cases. We train a multi-layered artificial neural network (ANN) and a random forest (RF) to classify spaxels into `star forming' and `quenched' categories given various individual (and groups of) parameters. We find that global parameters (pertaining to the galaxy as a whole) perform collectively the best at predicting when spaxels will be quenched, and are substantially superior to local/ spatially resolved and environmental parameters. Central velocity dispersion is the best single parameter for predicting quenching in central galaxies. We interpret this observational fact as a probable consequence of the total integrated energy from AGN feedback being traced by the mass of the black hole, which is well known to correlate strongly with central velocity dispersion. Additionally, we train both an ANN and RF to estimate $\Sigma_{\rm SFR}$ values directly via regression in star forming regions. Local/ spatially resolved parameters are collectively the most predictive at estimating $\Sigma_{\rm SFR}$ in these analyses, with stellar mass surface density at the spaxel location ($\Sigma_*$) being by far the best single parameter. \it{Thus, quenching is fundamentally a global process but star formation is governed locally by processes within each spaxel.}

\end{abstract}
\begin{keywords}
Galaxies: formation, evolution, environment, structures, bulge, disk; star formation; observational cosmology
\end{keywords}

%
%   INTRO
%

\section{Introduction}

Understanding the formation of stars within galaxies, and the subsequent end of star formation via the  `quenching' process, remain two of the most important unresolved problems in extragalactic astrophysics. Only $\sim$10\% of baryonic matter is currently in stars (Fukugita \& Peebles 2004, Shull et al. 2012), and hence the vast majority of baryons are in principle available as an abundant resource for future star formation. Moreover, the vast majority of baryons in high mass galaxies reside in a hot gas halo surrounding the galaxy (e.g. Forman et al. 1985, Paolillo et al. 2002, Xia et al. 2002, Fabian et al. 2006), from which gas cooling and subsequent accretion onto the galaxy is naturally expected. 

Since the virial temperature of high mass haloes is $>>10^6$K (e.g. Sarazin 1986, Tozzi \& Norman 2001, Muanwong et al. 2002, Majerowicz et al. 2002), the baryons in the hot gas halo must be in a predominantly ionised form, and hence the dominant cooling mechanisms will be through collisionally de-excited metal line emission (at $T \lessapprox 2 \times10^{7}$ K) and thermal bremsstrahlung (at $T \gtrapprox 2 \times10^{7}$ K). The rate of cooling from both of these processes is proportional to the square of the density (or, more precisely, $\Gamma_{\rm cool} \propto n_{e} n_{I} \sim \rho^{2}$, e.g. Fabian 1994, Voigt et al. 2002, McNamara \& Nulsen 2007). Additionally, the density of plasma must increase with increasing mass of dark matter halo. Thus, naively one would expect higher cooling rates and hence higher rates of gas accretion and star formation in more massive galaxies (Fabian 1994, 1999, 2012). This is not observed, and in fact the opposite is found to be true: high mass galaxies are more frequently quenched (non star forming) than lower mass galaxies (e.g. Baldry et al. 2006, Peng et al. 2010, 2012). Furthermore, typical cooling times for high mass clusters are predicted to be $\sim$1 Gyr, which is far less than that observed (e.g. McNamara \& Nulsen 2007). These serious inconsistencies between fundamental theoretical predictions and observations have been dubbed the `cooling problem', or more flamboyantly the `cooling catastrophe' (e.g. Ruszkowski et al. 2002).

The obvious solution to the cooling problem is through heating (although see Martig et al. 2009 for an alternative quenching mechanism). That is, the increased cooling rate of higher mass haloes must be offset by an (even greater) increased source of heat in those systems (e.g. Springel et al. 2005, Croton et al. 2006, Bower et al. 2006, 2008). Much of the modern history of the field of galactic star formation and quenching has been focused on answering precisely which heating mechanism(s) are feasible, given modern observations of galaxy properties (e.g. Bell et al. 2012, Wake et al. 2012, Cheung et al. 2012, Fang et al. 2013 Woo et al. 2013, Bluck et al. 2014, Lang et al. 2014, Omand et al. 2014, Bluck et al. 2016, Teimoorinia et al. 2016, Bluck et al. 2019) and sophisticated simulations of the physics of galaxy evolution (e.g. Vogelsberger et al. 2014a,b, Somerville \& Dave 2015, Schaye et al. 2015, Brennan et al. 2017, Henriques et al. 2015, 2017, 2019). 

Initially it was speculated that the star formation process might be self-quenching in the sense that star formation itself provides the heat needed to offset cooling in massive systems, most probably through supernova explosions. However, this idea has been widely rejected for two reasons. First, the total energy available from supernovae in massive galaxies is far too low to offset the expected cooling rate (e.g. Croton et al. 2006, Bower et al. 2006, 2008), and furthermore individual supernova explosions have far too little power to expel gas fully from high mass galaxy haloes (e.g. Henriques et al. 2019). Second, since star formation rates substantially decrease in high mass galaxies, it has never been clear how an offset heating from supernova Type-II feedback can persist after quenching. If supernova Type-II regulate star formation, eventually giant ellipticals ought to accrete vast amounts of gas and rejuvenate in a spectacular manner, which is, of course, not observed. However, it remains possible that supernova Type-Ia can impact the haloes of quenched galaxies (e.g. Matteucci et al. 1986, 2006), although simple energetics arguments suggest this cannot be the sole mechanism (e.g. Croton et al. 2006, Bower et al. 2008).

Dekel \& Birnboim (2006) proposed an alternative mechanism for late time heating in massive galaxies, via the shock heating of cold gas flows into massive haloes above a critical mass of $M_{\rm Halo} \gtrapprox 10^{12} M_{\odot}$. Observational support for this theoretical process was found in Woo et al. (2013), whereby the quenched fraction of central galaxies was found to scale more tightly with halo mass than stellar mass. However, both Bluck et al. (2014) and Woo et al. (2015) agree that measurements of the central density of galaxies is far superior to halo mass in parameterizing central galaxy quenching. Furthermore, in Bluck et al. (2016) we find that varying halo mass at a fixed central velocity dispersion (by up to three orders of magnitude) engenders no significant variation in the quenched fraction at all, even around the expected threshold of halo mass quenching. These studies provide strong observational reasons to explore alternatives to halo mass quenching. Additionally, cosmological hydrodynamical simulations find that shock heating by itself is insufficient to quench massive galaxies early enough, in order to obtain agreement with the observed stellar mass function (e.g., Vogelsberger et al. 2014a,b, Schaye et al. 2015).

Finally, heating from active galactic nuclei (AGN) has been proposed as a possible solution to the problem of why massive galaxies stop forming stars (e.g. Croton et al. 2006, Bower et al. 2006, 2008, Sijacki et al. 2007, Henriques et al. 2015, 2019). The distinct advantage of AGN is that they provide more than enough energy over the lifetime of a galaxy to fully offset cooling in even the most massive haloes by $\sim$50-100 times over (as determined in Silk \& Rees 1998 via theoretical considerations, and in Bluck et al. 2011 via a statistical analysis of deep X-ray data). Additionally, early studies have identified an over-abundance of AGN in galaxies with intermediate levels of star formation, indicating a possible link between presence of AGN and a decline in star formation (e.g. Kauffmann et al. 2003, S\'anchez et al. 2004, Nandra et al. 2007). However, understanding the impact of AGN lifetime and duty cycle on these results has proved problematic to their interpretation (e.g. Hickox et al. 2009, 2014). Ultimately, the main issue with AGN feedback is precisely {\it how} will the energy couple to the hot gas halo, i.e. by what mechanism(s) will AGN convert some fraction of their vast energy into heating the hot gas halo, leading to an offset of cooling and eventual quenching of the galaxy?  

Two modes of AGN feedback have been proposed, a high luminosity/ high Eddington ratio `quasar-mode' (e.g. Nesvadba et al. 2008, Hopkins et al. 2008, 2010, Feruglio et al. 2010, Maiolino et al. 2012) and a low luminosity/ low Eddington ratio `radio-mode' (e.g. Croton et al. 2006, Bower et al. 2006, 2008, Fabian 2012). The former arises from powerful winds driven by the most X-ray luminous AGN in the Universe, which can potentially blow out large quantities of gas from the galaxy (e.g. Tombesi et al. 2010, Feruglio et al. 2010, Cicone et al. 2012, 2014, 2015, Heckman et al. 2014). The latter arise from late-time low luminosity radio heating, most probably via jets (e.g. Di Matteo et al. 2000, Fabian et al. 2006, McNamara \& Nulsen 2007, Fabian 2012). Modern observations of quasar-driven outflows find a substantial mass of gas is expelled from the galaxy, but in general it does not achieve high enough kinetic energies to escape the dark matter halo potential (e.g. Nesvadba et al. 2008, Veilleux et al. 2013, Cicone et al. 2014, 2015). Nonetheless, even if bound, the circulation of gas through the circum-galactic medium (CGM) can still input a significant amount of energy in the form of heat (e.g., Fluetsch et al. 2019). However, although the quasar-mode is powerful (high luminosity) it is also a rare event, most probably a one-time process in the lifetime of a galaxy (and certainly does not extend into the quenched phase of galaxy evolution). Thus, it is hard to envisage how AGN feedback through the quasar-mode can quench star formation in galaxies in the long term, since cooling and accretion from the hot gas halo can resume again after the quasar episode. 

Radio-mode AGN feedback, on the other hand, provides episodic heating throughout the lifetime of a high mass galaxy, and hence can potentially explain a lack of late time cooling. Observational support of this picture is quite sparse, possibly due to radio mode feedback being a low luminosity process (and hence intrinsically hard to observe). Nevertheless, a significantly higher fraction of low luminosity radio emitters is found in high mass galaxies (e.g. Di Matteo et al. 2000, 2003, Hickox et al. 2009, 2014), and there is substantial observational evidence for radio bubbles and X-ray cavities in high mass gaseous haloes (e.g. McNamara et al. 2000, Fabian et al. 2006, McNamara \& Nulsen 2007, Fabian 2012, Hlavacek-Larrondo et al. 2012, 2015, 2018). Despite the relative lack of direct observational confirmation (especially in intermediate mass haloes), radio-mode AGN heating has become a ubiquitous ingredient in modern cosmological hydrodynamical simulations (e.g., Vogelsberger et al. 2014a,b, Schaye et al. 2015, Nelson et al. 2018) and in modern semi-analytic models (e.g. Somerville \& Dave 2014, Henriques et al. 2015, 2019). Without AGN feedback in the radio-mode, the number of high stellar mass galaxies is severely over-predicted in models, compared to observations from wide-field galaxy surveys.

To combat the issue of radio mode AGN feedback being a low luminosity process with a very long duty cycle (i.e. not amenable to being `caught in the act'), we developed an approach to investigate the total integrated impact of AGN feedback on galaxy quenching (see Bluck et al. 2014, 2016). Specifically, we utilise the mass of the supermassive black hole as a fossil record of the total integrated energy released from the black hole (following highly general theoretical arguments from Soltan 1982 and Silk \& Rees 1998). Since dynamical measurements of supermassive black hole masses are difficult to acquire, and are only measured in $\sim$100 systems  (e.g. Saglia et al. 2016), we utilise the empirical $M_{BH} - \sigma$ relationship (Ferrarese \& Merritt 2000, McConnell \& Ma 2013, Saglia et al. 2016) to estimate black hole masses in large samples of galaxies. The most important result from these studies is that black hole mass is the tightest correlator to quenching in central galaxies, and is substantially superior to both stellar and halo mass, and to morphology and environment. In Teimoorinia, Bluck \& Ellison (2016) we develop a machine learning approach to rank parameters in terms of their importance to quenching. Again, we find central velocity dispersion (and hence estimated black hole mass) to be the most successful parameter for predicting quenching in central galaxies. Using a much smaller sample of dynamically measured black hole masses, Terrazas et al. (2016, 2017) find that black hole mass traces quenching more closely than stellar mass, in line with our findings. 

In addition to the fraction of quenched galaxies depending on mass (so called `mass quenching', Peng et al. 2010), environment has also been shown to be a significant correlator to quenching, particularly for satellite galaxies (e.g. Peng et al. 2010, 2012, Woo et al. 2013, 2015, 2017, Bluck et al. 2014, 2016). Satellites can experience a wide variety of physical processes which may in principle remove gas and engender quenching. These include, ram pressure stripping of the hot gas halo, ram pressure stripping of cold gas within the galaxy, galaxy-galaxy and host halo tidal stripping of gas, and the removal of satellite galaxies from nodes in the cosmic web (e.g. Balogh et al. 2004, Cortese et al. 2006, van den Bosch et al. 2007, 2008, Tasca et al. 2009, Wetzel et al. 2013, Henriques et al. 2015). These environmental processes are expected to act in addition to the intrinsic processes discussed at length throughout this introduction. Thus, satellites may be both mass and environment quenched (in the parlance of Peng et al. 2010), but central galaxies are primarily affected by mass quenching.

Currently, a revolution is underway in extragalactic astronomy, whereby integral field units (IFUs) are being utilised to yield spatially resolved spectroscopy of local and high redshift galaxies in substantial numbers for the first time (e.g. Cappellari et al. 2011, S\'anchez et al. 2012, Bryant et al. 2015, Bundy et al. 2015). By far the largest of the new generation of wide-field IFU surveys is the Mapping Nearby Galaxies at Apache Point Observatory survey (MaNGA, Bundy et al. 2015), which is the primary data source of this paper. IFU spectroscopy allows a spatially resolved view of galaxies, including (importantly for this work) star formation and quenching. Potentially, resolved spectroscopy of star forming and quenched galaxies can lead to powerful new constraints on quenching models, as well as new insights into the processes of star formation on kpc-scales. Indeed, these goals were amongst the primary justifications for these surveys to be carried out (e.g. Bundy et al. 2015, Bryant et al. 2015).

Recent studies utilising resolved spectroscopy have shown that massive star forming and green valley galaxies have increasing sSFR (=SFR/$M_{*}$) and/or $\Delta$SFR radial profiles, which has been widely interpreted as a possible signature of inside-out quenching (e.g. Tacchella et al. 2015, Gonz\'alez Delgado et al. 2016, Belfiore et al. 2017, 2018, Ellison et al. 2018, Spindler et al. 2018, S\'anchez et al. 2018, Woo \& Ellison 2019). However, S\'anchez et al. (2018) and Wang et al. (2019) have pointed out that in most cases galaxies with rising sSFR profiles still have declining SFR profiles, and hence the reason for the increasing sSFR profiles is not due to quenching per se but rather due to steep (above exponential) declines in mass density within galaxies. That is, rising sSFR profiles may be largely attributable to the presence of a bulge structure (although the control sample of Ellison et al. 2018 is largely immune to this issue). 

The vast majority of resolved star formation studies from IFU surveys restrict their analysis primarily to star forming regions within galaxies, and hence truly quenched systems, and quenched regions within galaxies, are largely ignored (although see Woo \& Ellison 2019 for a notable exception). The reason behind this omission is to restrict the analysis to reliable measurements of star formation. However, an unintended result of this approach is to have highly biased samples, which are systematically incomplete in high mass and spheroidal galaxies. In this paper we take a different approach, classifying spaxels into star forming and quenched categories by their measured star formation rates from emission lines (where possible), or else inferred indirectly via the strength of the 4000 \AA \hspace{0.05cm} break (for line-less regions and regions with AGN contamination). Consequently, we are able to investigate quenching for a {\it complete} sample of galaxies, and spaxels within galaxies. We carefully validate our SFR measurements against various alternatives, including via multiple single stellar population (SSP) model fitting. 

The primary motivation for this paper is to expand our prior work on ranking galaxy parameters as predictors of star formation and quenching (Bluck et al. 2014, 2016, 2019 \& Teimoorinia et al. 2016) to a spatially resolved view of galaxies. More specifically, we include both spatially resolved parameters and spatially resolved measurements of star formation and quenching within galaxies. With this approach we answer two fundamental questions: 1) is {\it star formation} governed by local, global or environmental processes?; and 2) is {\it quenching} governed by local, global, or or environmental processes? Of course, the answers to these two questions may be very different. Beyond this, we also determine which specific observables on all scales (from local to environmental) impact star formation and quenching, and asses the similarities and differences between central and satellite galaxies. To answer these questions we employ a sophisticated machine learning approach utilising both a multi-layered artificial neural network and a random forest to analyse over 5 million spaxels from the latest data release of the MaNGA survey. Our approach provides robust constraints to theoretical models of star formation and quenching, and we end this paper with a substantial discussion of which theoretical mechanisms remain viable in light of our analysis.

The paper is structured as follows. In Section 2 we discuss our data sources. In Section 3 we present our method to assign star formation rates to all galaxy spaxels. In Section 4 we present our results on star formation and quenching, including our machine learning analysis. We present extensive testing of the stability of our main results in Appendix A. In Section 5 we investigate quenching across the stellar - halo - black hole mass parameter space, and interpret our results with analytic theory (derived in Appendix B). Finally, in Section 6 we provide a brief summary of the major contributions of this paper. Throughout the paper, we adopt a spatially flat $\Lambda$CDM cosmology with the following parameters: $\Omega_M$ = 0.3, $\Omega_\Lambda$ = 0.7, $H_0$ = 70 km/s/Mpc.

%
%   DATA
%

\section{Data}

\subsection{MaNGA DR15 \& P{\small IPE}3D}

\begin{figure*}
\includegraphics[width=1.0\textwidth]{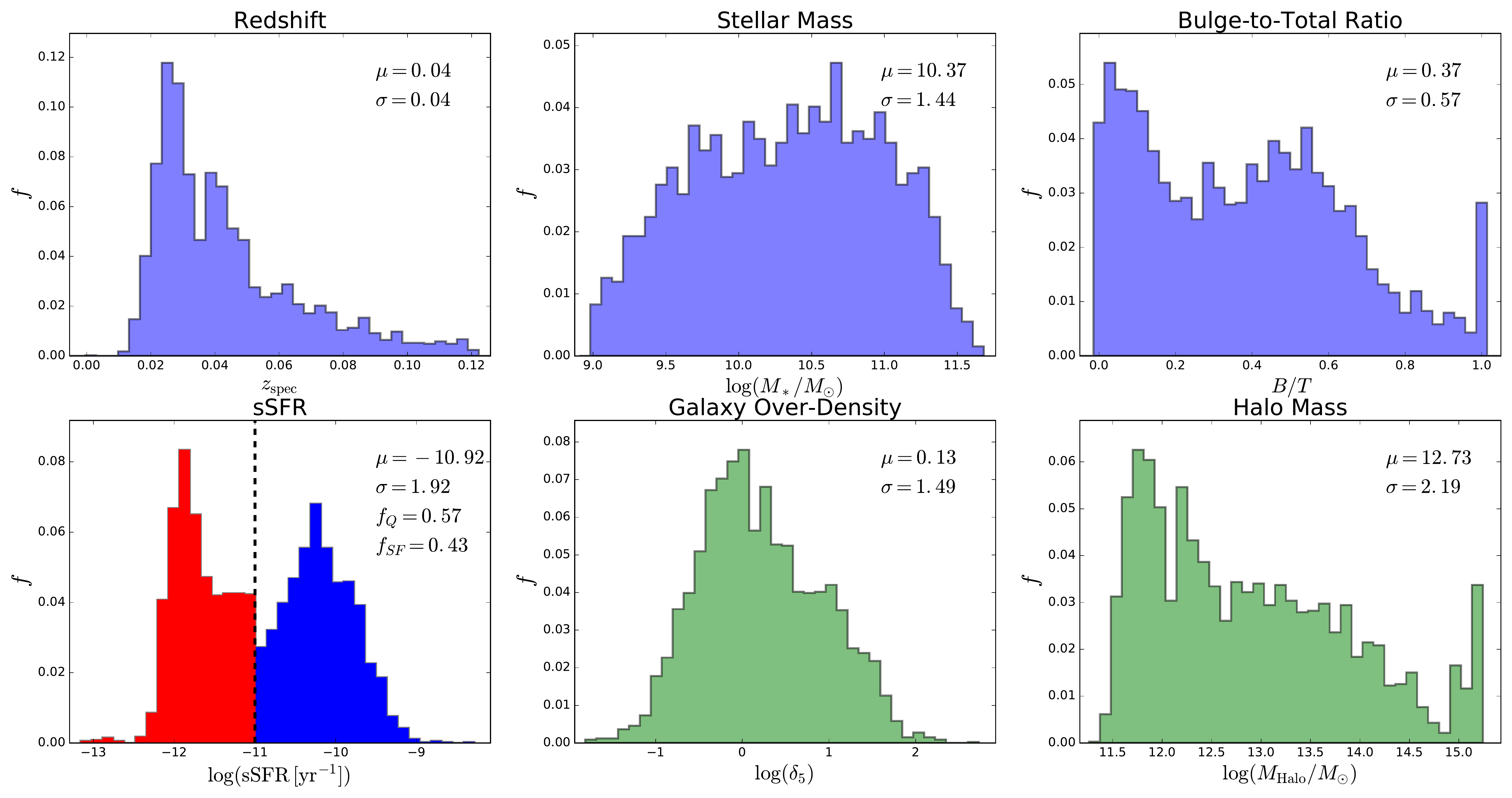}
\caption{Distributions of galaxy (shown in light blue) and environmental (shown in light green) properties, from top left to bottom right: spectroscopic redshift ($z_{\rm spec}$), stellar mass ($M_{*}$), bulge-to-total stellar mass ratio ($B/T$), sSFR (= SFR / $M_{*}$), galaxy over-density evaluated at the 5th nearest neighbour ($\delta_5$), and group halo mass ($M_{\rm Halo}$). Additionally, the median and interquartile range of each parameter is displayed on each panel as $\mu$ and $\sigma$, respectively. For sSFR (bottom left panel), we also show a crude separation into star forming and quenched systems based on a simple cut at $\log({\rm sSFR}) = -11$. Our final sample consists of 57\% quenched (shown in red) and 43\% star forming (shown in blue) systems.}
\end{figure*}

We use as our primary data source the publicly available Sloan Digital Sky Survey data release 15  (SDSS DR15, Aguado et al. 2019). Specifically, we utilise the Mapping Nearby Galaxies at Apache Point Observatory survey (MaNGA, Bundy et al. 2015)\footnote{Website: www.sdss.org/dr15/manga/}. MaNGA is an ongoing SDSS-IV project utilising the Apache Point Observatory (APO) to observe 10 000 galaxies with integral field units, drawn from the SDSS legacy galaxy parent sample. The advantage of selecting MaNGA galaxies from the SDSS is that we additionally have access to a wide array of ancillary data, including stellar masses, morphologies, and environmental measurements (see Section 2.2). The observing strategy and survey design of MaNGA is explained in detail in Law et al. (2015). In this sub-section we give only a very brief overview. In the DR15, resolved spectroscopy and intermediate data products are provided for $\sim$4700 galaxies (almost a factor of two increase in number over the previous data release). Consequently, the latest MaNGA release provides the largest sample of spatially resolved spectroscopic measurements in local ($z \sim 0.1$) galaxies to date, which affords a distinct advantage to the statistical methods employed throughout this paper. 

MaNGA makes use of the twin Baryon Oscillation Spectroscopic Survey (BOSS) spectrographs, bundling together the two arcsec BOSS fibres into hexagonal IFUs used to feed the spectrographs. As such, MaNGA provides spatially resolved spectroscopy across each of the observed galaxies, which can be used to construct spatially resolved maps of kinematics, emission and absorption lines as well as derived parameters such as stellar mass surface density and star formation rate surface density (see below). The IFUs vary in diameter from 12 arcsecs to 32 arcsecs, depending on the angular size of the target galaxy (which involves a variation in the number of fibres used from 19 to 127 per IFU). MaNGA galaxies are selected with a flat mass distribution with $\log_{10}(M_{*}/M_{\odot}) > 9$, and the variable IFU size is designed to cover 1.5 effective radii for 2/3 of the target sample. For the remaining 1/3 of the target galaxies, this criterion is expanded to 2.5 effective radii. As such, a significant fraction of all targeted galaxies are visible within the IFU, and furthermore a representative sample of local (z $\sim$ 0.1) galaxies is obtained, with a broad range in mass, morphology, star formation and environment (see Bundy et al. 2015 for full details).

We analyse the MaNGA datacubes using the {\sc Pipe3D} pipeline (S\'anchez et al. 2016b), which is designed to fit the stellar continuum with multiple single stellar population (SSP) models, and to measure the nebular emission lines of galaxies in contemporary integral field spectroscopy (IFS) data. These data are all publicly available as part of the SDSS DR15 MaNGA release\footnote{Website: www.sdss.org/dr15/manga/manga-data/manga-pipe3d-value-added-catalog/}. This pipeline is based on the {\sc FIT3D } fitting package (S\'anchez et al. 2016a). The current implementation of {\sc Pipe3D} adopts the GSD156 library of simple stellar populations (Cid-Fernandes et al. 2013) that comprises 156 templates covering 39 stellar ages (from 1Myr to 14.1Gyr), and four metallicities (Z/Z$\odot$=0.2, 0.4, 1, and 1.5). These templates have been extensively used within the CALIFA collaboration (e.g. Perez et al. 2013, Marino et al. 2013), and in other surveys (e.g. Haines et al. 2015, Ibarra et al. 2016). Details of the fitting procedure, dust attenuation curve, and uncertainties on the processing of the stellar populations are given in S\'anchez et al. (2016a,b). {\sc Pipe3D} has been successfully used in the analysis of IFS from many different surveys, including CALIFA (e.g. S\'anchez-Menguiano et al. 2016), MaNGA (Ibarra-Medel et al. 2016; Barrera-Ballesteros et al. 2018), SAMI (S\'anchez et al. 2019a), and AMUSING (S\'anchez-Menguiano et al. 2018), and it has been extensively tested against other analysis tools for both the derivation of the stellar population properties (S\'anchez et al. 2016a, S\'anchez et al. 2019b) and emission line parameters (e.g. Belfiore et al. 2019). Consequently, in this section we give only a brief overview of the most important details.

Prior to any analysis a spatial binning is performed in order to increase the S/N without altering substantially the original shape of the galaxy. Two criteria are adopted to guide the binning process: (i) a desired S/N for the binned spectra, and (ii) a maximum difference in the flux intensity between adjacent spaxels. The first criterion selects a S/N per \AA\ of 50, that corresponds to the limit above which the recovery of the stellar population properties have uncertainties of $\sim$10-15\% (S\'anchez et al. 2016a). The second criterion selects a maximum difference in the flux intensity of 15\%. This corresponds to the typical flux variation along an exponential disk of the average size of our galaxies in a range of 1-2 kpc, and shorter scale-lengths for more early-type galaxies.

As a result of the binning process, the original spaxels (with a size of 0.5$\arcsec$$\times$0.5$\arcsec$) are aggregated into tessellas of variable size. The typical size of the tessellas range between 2-5 spaxels in most of the cases, with a few larger ones in the outer regions of the galaxies (e.g. Ibarra-Medel et al. 2016). Contrary to other binning schemes, the original shape of the galaxy is better preserved by our adopted procedure, not mixing adjacent regions corresponding to clearly different structures (e.g., arm/ inter-arms). The disadvantage is that it does not provide an homogeneous S/N distribution across the entire FoV and the S/N limit is not reached in all the final bins/ voxels. The S/N limit of 50 was selected based on the extensive simulations described in S\'anchez et al. (2016a) in order to recover reliably the star formation histories (SFHs) and stellar properties in general.  For lower S/N those properties are recovered in a less precise but still accurate way. The tessellas with lower S/N are found mostly in the outer regions, where there are still a large number of individual bins. Therefore, averaging the stellar properties (including the SFHs) either radially or integrated across the entirety of the field-of-view (FoV) provides uncertainties similar to the ones from individual but larger S/N bins. This has been demonstrated explicitly in Ibarra-Medel et al. (2016). Ultimately, our adopted procedure provides a more accurate SFH than what would be derived from co-adding all the spectra within the FoV into a single voxel and analysing it, according to recent results from Ibarra-Medel et al. (2019).

P{\small IPE}3D adopts a two stage fitting procedure whereby first the kinematic parameters are derived (stellar velocity and velocity dispersion) along with the dust attenuation ($A_{V,*}$), and second multi-SSP models are fit to the stellar continuum. A Monte Carlo iteration of the second step with varying input spectra (based on their measured errors) is used to find the optimal coefficients for the linear fit, and their respective errors. The stellar population parameters (e.g., stellar metallicity and stellar mass density) are taken as the linearly weighted sum of the best fit SSP models, as normal. For example, the stellar mass of a given spatial binning of spaxels is given by:

\begin{equation}
M_{*} = \sum\limits_{i}\big({c_{i} (M/L)_{i}}\big) \times L = \langle M/L \rangle L
\end{equation}

\noindent where the summation is applied over the full sample of SSP models ($i$) with the weighted contribution of each model given by $c_{i}$. $(M/L)_i$ indicates the mass-to-light ratio of each model, and L indicates the total optical luminosity in the spatial binning of spaxels. Similar procedures are implemented for the metallicity and age parameters, which are given in a luminosity weighted and mass weighted form, whereby the latter is additionally weighted by the $M/L$ ratio of each stellar population (as above).

In addition to the P{\small IPE}3D data products for MaNGA DR15 (which we adopt as our primary resource of spatially resolved information within MaNGA galaxies), we also consider the MaNGA Data Analysis Pipeline (DAP) parameters in order to validate both methods. Generally speaking, there is excellent agreement between P{\small IPE}3D and the DAP for parameters which are measured in both pipelines. Full details on the DAP measurements are provided in Law et al. (2016), Yan et al. (2016) and Westfall et al. (2019). The DAP methodology is qualitatively similar to P{\small IPE}3D, but differs in the detail of the implementation. Ultimately, the primary reason we adopt P{\small IPE}3D as opposed to the DAP in this study is because we have need of stellar mass surface densities in our analysis, which are reliably provided by P{\small IPE}3D but are not available in the DAP. Additionally, we have a preference to use the same data source where possible (primarily for the sake of internal consistency), hence our choice to focus exclusively on the P{\small IPE}3D outputs for this paper. However, we have confirmed that all of our results and conclusions would be identical if we had instead opted to use the DAP measurements for emission line fluxes and spectral indices.

\subsection{SDSS Ancillary Data}

Our goal in this paper is to assess which parameters impact galactic star formation and quenching, and hence we must look at a wide variety of parameters, not restricted solely to spatially resolved measurements (discussed above). As such, we match the MaNGA DR15 galaxy sample to the full SDSS DR7 spectroscopic sample, yielding $\sim$4200 secure matches (with galaxy centre separation $<$ 2 arcsec on sky). We additionally require that each galaxy is present in the MPA SDSS value added catalogue (Brinchmann et al. 2004), the Yang et al. (2007, 2008, 2009) group catalogues, the Mendel et al. (2014) stellar mass and structural catalogues, the Simard et al. (2011) morphological catalogues, and the MaNGA Data Reduction Pipeline (DRP) file (Law et al. 2016). Finally, we require that there is a `good' (i.e. non-Null, non-NaN, and not flagged with a warning) measurement of each of the parameters we require from these catalogues. Application of all of these cuts yields a final sample of 3523 galaxies (2550 centrals and 973 satellites).

All global and environmental parameters are taken from the SDSS value added catalogues (as referenced above). Full details on the parameters used in this study are provided in many other publications, most recently in Section 2 of Bluck et al. (2019). As such, we will not present a detailed explanation of each parameter here. Instead, we briefly mention the source of each parameter below. 

Global stellar masses and bulge-to-total stellar mass ratios are taken from the structural catalogues of Mendel et al. (2014), which are derived via spectral energy distribution (SED) fitting to multi-wavelength photometric ($u,g,r,i,z$) bulge-disc decompositions performed with the {\small GIM2D} package (Simard et al. 2002). Note that these are mass weighted structures (as opposed to light weighted morphologies, which are more common in the literature). Central - satellite classification, halo masses, and distances from the central galaxy for satellites are taken from the SDSS group catalogues of Yang et al. (2007, 2008, 2009). Halo masses are inferred from an abundance matching technique applied to the total stellar mass of the group, and groups are identified via an iterative friends-of-friends algorithm. Nearest neighbour local density measurements are derived in Mendel et al. (2013) following the standard procedure of Baldry et al. (2006). We adopt values set at 3rd, 5th and 10th nearest neighbour thresholds. Geometric parameters of galaxies (e.g. axis ratios, position angles, and effective half-light radii) are taken from the MaNGA DRP file and from the morphological catalogues of Simard et al. (2011), derived with {\small GIM2D} fitting of single S\'ersic (1963) profiles and bulge-disc decompositions, where appropriate. Global star formation rates are taken from Brinchmann et al. (2004), which are computed via emission lines where possible, and via the empirical sSFR - D4000 relation for the remainder of the sample.

We display the distributions of a number of key galactic and environmental parameters in Fig. 1. We emphasise that we have a very broad range in galaxy properties, with a diverse population of galaxies spanning from pure discs to spheroids, with a fairly flat stellar mass distribution from $M_* = 10^{9.5} - 10^{11.5} M_{\odot}$, and an approximately even sampling of star forming and quiescent systems. This high level of diversity in a sample with resolved spectroscopy is truly unprecedented, and hence represents an ideal test-bed for investigating the processes associated with galaxy transformations (including quenching). Nonetheless, we stress that our results are necessarily sample dependent, and thus significant changes to the host galaxy population may reasonably yield different results. Due to the range in redshifts, the physical size of the region of a galaxy observed within each spaxel (with fixed angular size of 0.5 arcsec) is 0.46 $\pm$ 0.26 kpc (mean $\pm$ standard deviation).

%
%   METHOD
%

\section{SFR Method}

Star formation rate surface densities ($\Sigma_{\rm SFR}$) are not provided in either the MaNGA DAP or P{\small IPE}3D data products for DR15. As such, we must compute them before analysing star formation and quenching within MaNGA galaxies. In this section we explain how we compute $\Sigma_{\rm SFR}$ for all galaxy spaxels in our sample. We adopt a two stage approach: 1) $\Sigma_{\rm SFR}$ is derived from emission lines, where possible (see Section 3.1); and 2) $ \Sigma_{\rm SFR}$ is estimated from the strength of the 4000 \AA \hspace{0.05cm} break and the stellar mass surface density ($\Sigma_*$) in the spaxel in all other cases (see Section 3.2). We explain in detail our prescription for each method below. It is crucial for our quenching analysis that every galaxy spaxel in MaNGA DR15 has a $\Sigma_{\rm SFR}$ value, and we achieve this by trading off accuracy for completeness. For our star formation analysis, we restrict the sample to robust measurements in star forming regions, where we trade off completeness for accuracy (note the inversion). Ultimately, the specific scientific goal of a given analysis will determine which sample is appropriate to use, as explained further throughout this section. We validate our SFR method against two alternative approaches in Appendix A, and demonstrate that none of our key results or conclusions are dependent on the precise implementation of the SFR method.

\subsection{$\Sigma_{\rm SFR}$ from $H\alpha$ for high-S/N Star Forming Spaxels}

We take the line fluxes for $H\alpha$, $H\beta$, [OIII]$\lambda$5007 and [NII]$\lambda$6584 from the public P{\small IPE}3D DR15 data cubes. Line fluxes are corrected for Galactic extinction and the underlying stellar absorption through simultaneous stellar continuum fitting and subtraction. The P{\small IPE}3D line fluxes are in excellent agreement with the equivalent measurements from the MaNGA DAP, with mean differences of less than 0.02 dex and standard deviations between the datasets of less than 0.05 dex, for each line. First, we compute the S/N ratio of each line, given the errors included by the pipeline. We then select all spaxels with S/N $>$ 3, in each of the above mentioned emission lines. As expected, the S/N ratio in $H\beta$ is the limiting case for most spaxels. However, the $H\beta$ line is essential to measure accurately in our analysis for two reasons: 1) we use it in our dust correction prescription (via the Balmer decrement); and 2) we use it in our classification of spaxels into `star forming',  `composite' and  `AGN'. We compare to an alternative method not utilising H$\beta$ flux in the spaxel in Appendix A.  

A total of 1.07 million spaxels meet our signal-to-noise cuts, out of a total of 5.34 million `galaxy' spaxels (a fraction of 20\%). We define a spaxel to belong to a galaxy if it has a non-zero segmentation map index (i.e. Pipe3D deems the spaxel to belong to a galaxy) and additionally the spaxel has a mass surface density $\Sigma_* > 10^6 M_{\odot} / {\rm kpc}^2$ (chosen to be approximately consistent with the expected mass surface densities at $\sim$2.5 effective radii in low mass galaxies). Even if we were to take a S/N threshold of one (i.e. pushing the data to the absolute limit), only 2.09 million spaxels would be recovered, leading to a completeness of less than 40\%, and extremely high errors in the low-S/N cases. We return in Section 3.2 to the issue of estimating $\Sigma_{\rm SFR}$ values for the majority of galaxy spaxels, without sufficiently high S/N ratios to estimate reliably through emission lines. 

For the 20\% of spaxels with high-S/N ($>$3) emission line fluxes, we dust correct each line, assuming an intrinsic Balmer decrement of $H\alpha$/$H\beta$ = 2.86. We compute the extinction at wavelength $X$ from the measured Balmer decrement using the following procedure (see also Barrera-Ballesteros et al. 2018):

\begin{equation}
A(X) = -2.5 \times \log_{10}\big(\frac{f_{H\alpha}/f_{H\beta}}{2.86}\big) \times \frac{K_{X}}{K_{H\alpha} - K_{H\beta}}
\end{equation}

\noindent where,

\begin{equation}
K_{X} \equiv A(X) / A(V)
\end{equation}

\noindent and the dust reddening $A(X)$ is defined such that the observed apparent magnitude is given by

\begin{equation}
m_{X,{\rm obs}} = m_{X, {\rm int}} + A(X)
\end{equation}

\noindent where $m_{X, {\rm int}}$ is the intrinsic apparent magnitude. Thus, the dust corrected flux is given simply by

\begin{equation}
f_{X,{\rm corr}} = f_{X,{\rm obs}} \times 10^{A(X)/2.5}
\end{equation}

\noindent In the above expressions, X represents the wavelength of each of the emission lines in turn, i.e $H\alpha$, $H\beta$, [OIII]$\lambda$5007 and [NII]$\lambda$6584. The V-band normalised extinction at each wavelength ($K_X$) is provided by the extinction curve of Cardelli et al. (1989), with $R_{V} = 3.1$. We have checked that our choice of extinction curve does not significantly impact our results, and confirm that, e.g., the use of a Calzetti et al. (2000) attenuation curve yields almost identical results for the vast majority of spaxels, when we adopt the same value for $R_{V}$ (as also found in S\'anchez et al. 2016a).

\begin{figure}
\includegraphics[width=0.54\textwidth]{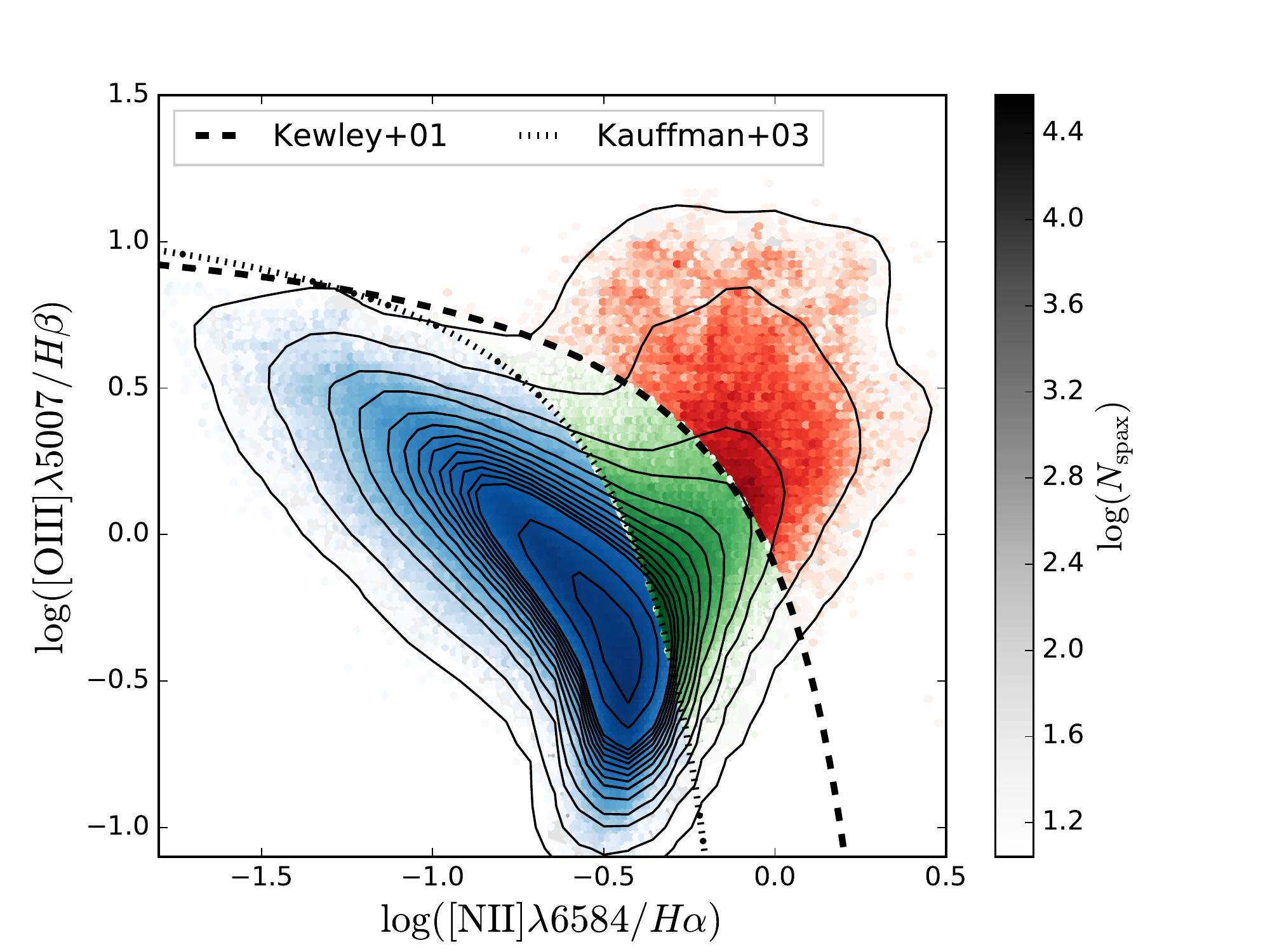}
\caption{The classification of spaxels into `Star Forming' (shown in blue), `Composite' (shown in green) and `AGN' (shown in red), from their location on the Baldwin, Phillips \& Terlevich (1981) emission line diagnostic diagram. The y-axis shows the ratio of the dust corrected line fluxes of [OIII]$\lambda$5007 to H$\beta$, and the x-axis shows the dust corrected line ratio of the fluxes of [NII]$\lambda$6584 to H$\alpha$. Classification of the spaxels is made using the theoretical lines of Kewley et al. (2001) and Kauffmann et al. (2003), as shown by the dashed and dotted lines, respectively. Only spaxels with S/N $>$ 3 in all BPT lines are shown in this figure. }
\end{figure}

We next classify our sample of high-S/N emission line spaxels into regions where ionising radiation is dominated by star formation, AGN, or a mix of both. To achieve this, we utilise the well studied Baldwin, Phillips \& Terlevich (1981, BPT) emission line diagnostic diagram, which plots the [OIII]$\lambda$5007/$H\beta$ flux ratio against the [NII]$\lambda$6584/$H\alpha$ flux ratio. In Fig. 2 we show the location of our emission line spaxels on the BPT diagram. Additionally we show the theoretical lines from Kewley et al. (2001) and Kauffmann et al. (2003), used to classify spaxels by their ionisation state. More precisely, we class spaxels which lie above the Kewley et al. (2001) line to be  `AGN' (shown in red), and those which lie below the Kauffmann et al. (2003) line to be  `star forming' (shown in blue). Spaxels which lie between the two lines are defined to be  `composite' (shown in green), which contain a contribution to the line strengths from both AGN and star formation. In Appendix A we compare the $\Sigma_{\rm SFR}$ values derived from this method to an alternate method for classifying spaxels via EW(H$\alpha$), and the results are essentially identical.

Emission lines may only be used to infer star formation rates reliably for spaxels without any AGN contamination (i.e. those classed as `star forming'). There are 0.93 million spaxels with S/N $>$ 3 which are classed as star forming in the BPT diagnostic diagram, representing just 17\% of galaxy spaxels from the MaNGA DR15. Note that even if we used an alternative approach to infer dust content and AGN contamination (see Appendix A), a lack of S/N in H$\alpha$ would still leave $\sim$half of the spaxels without an SFR surface density estimate, and the remainder would have less accurate measurements. 

For the high-S/N BPT `Star Forming' spaxels (without significant AGN contamination to their line fluxes), we compute star formation rates utilising the relationship of Kennicutt (1998), assuming a Salpeter IMF, via the following calibration:

\begin{equation}
{\rm SFR}(H\alpha) [M_{\odot}/{\rm yr}] = 7.9 \times 10^{-42} L_{H\alpha} [{\rm ergs \hspace{0.05cm} s}^{-1}]
\end{equation}

\noindent where the $H\alpha$ luminosity is given from the dust corrected $H\alpha$ flux and spectroscopic redshift (evaluating the luminosity distance for a spatially flat $\Lambda$CDM cosmology). Finally, we convert the SFR within each spaxel to $\Sigma_{\rm SFR}$ as:

\begin{equation}
\Sigma_{\rm SFR}(H\alpha) = \frac{{\rm SFR}(H\alpha)} {(0.5 \times D_A)^2}
\end{equation}

\noindent where the MaNGA pixel scale is 0.5"/pix, and $D_A$ is the angular diameter distance (derived from the spectroscopic redshift, assuming our adopted cosmology). We apply an inclination correction to this parameter by adding $\log(b/a)$ to each value, where $b/a$ is the axis ratio of the galaxy as determined by a S\'ersic fit to r-band photometric data (taken from the DRP file). We have checked that none of our results or conclusions are highly sensitive to the inclination correction, and indeed we recover essentially identical results with a face-on restricted sample.

We estimate the error on the emission line  $\Sigma_{\rm SFR}$ values to be $\sim$0.1-0.2 dex, via a standard propagation of uncertainty from the emission line fluxes, redshifts, and the inherent uncertainty on the Kennicutt (1998) calibration. Additional systematic uncertainty is engendered from the choice of the IMF. However, we mitigate this issue in the current work by using a Salpeter IMF consistently throughout all measurements of parameters.

\begin{figure*}
\includegraphics[width=0.49\textwidth]{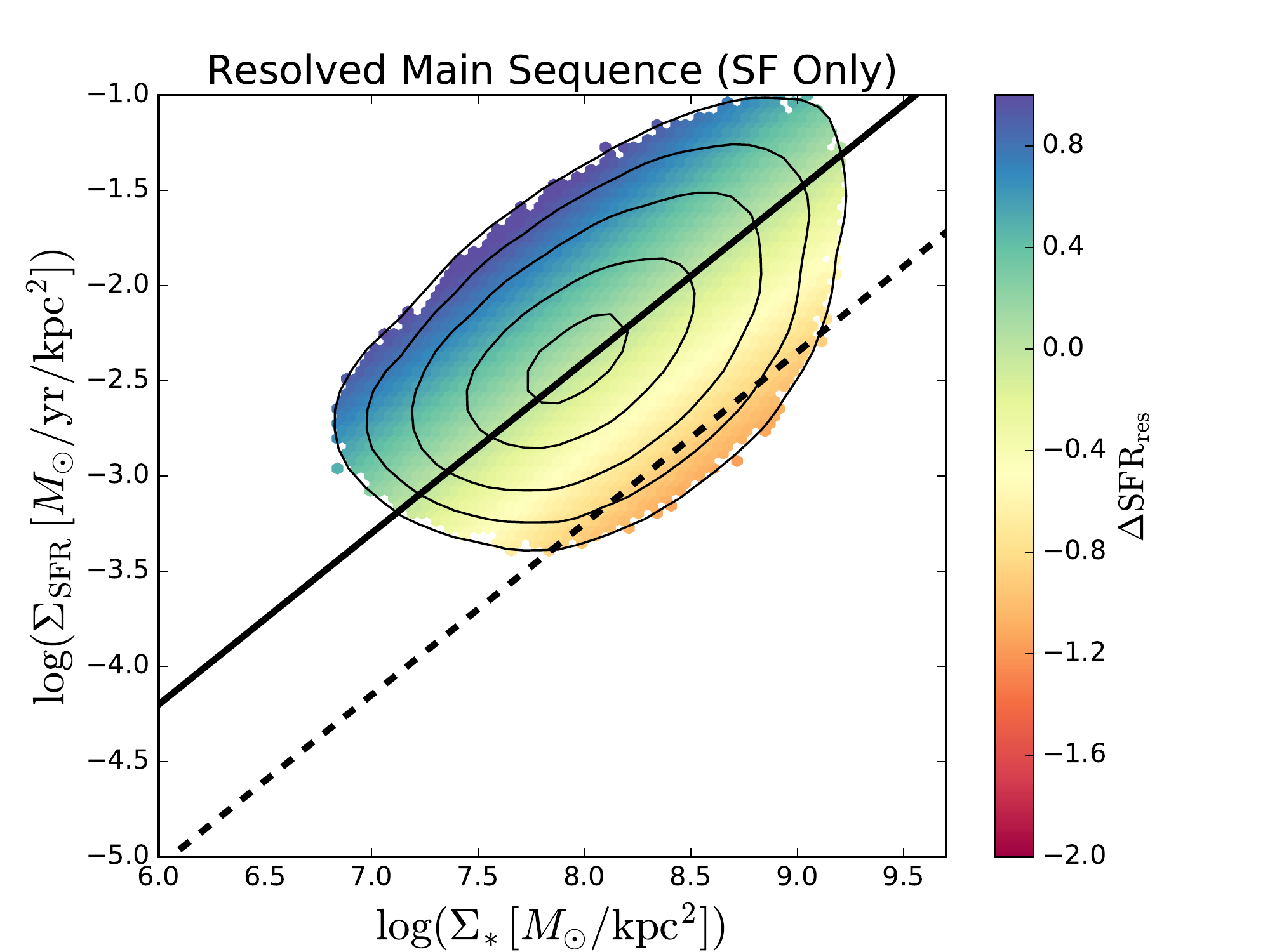}
\includegraphics[width=0.49\textwidth]{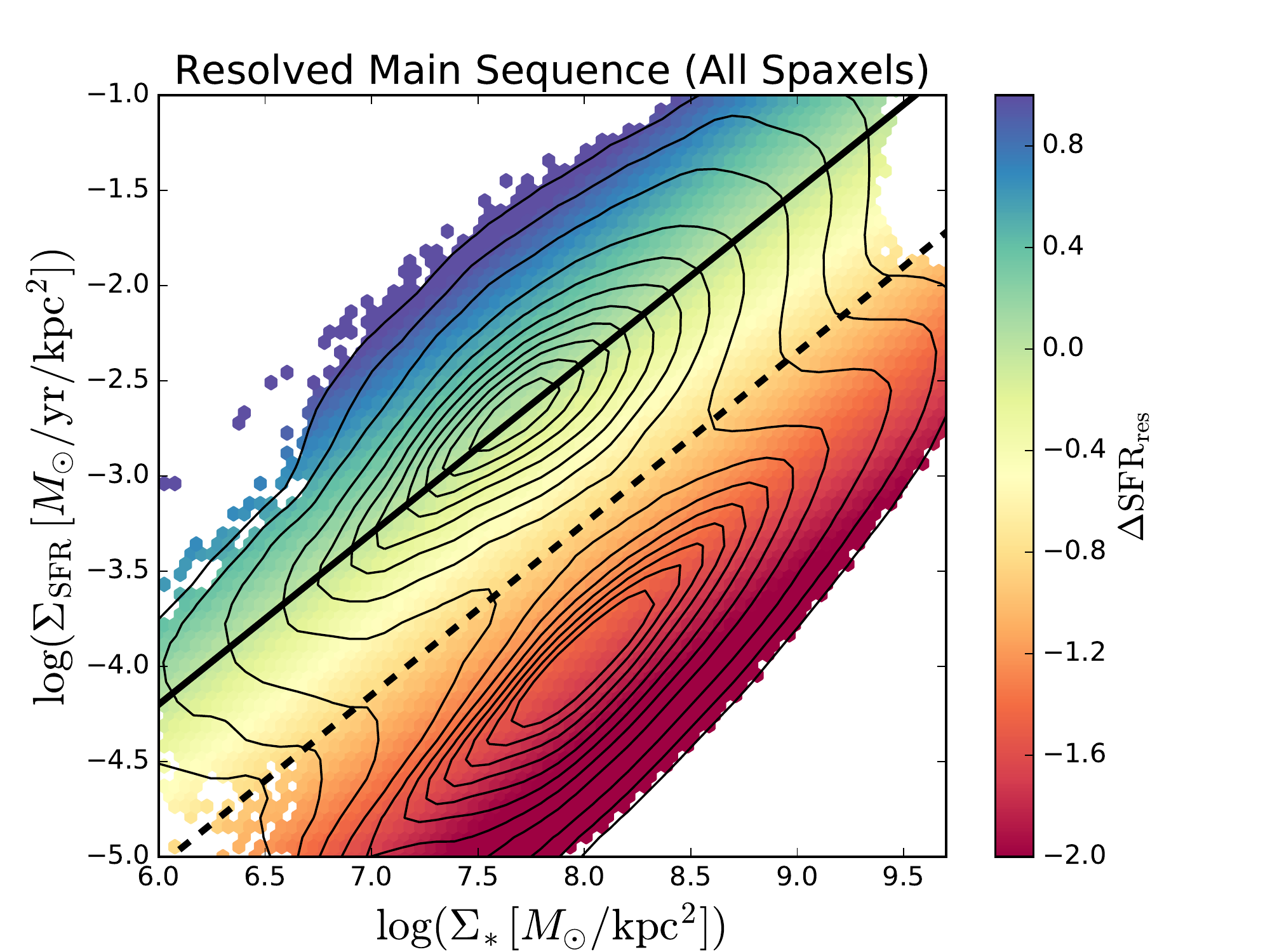}
\caption{{\it Left panel:} The resolved star forming main sequence ($\Sigma_{\rm SFR} - \Sigma_{*}$ relation) for BPT classified star forming spaxels with S/N $>$ 3 (17\% of the total galaxy spaxels). For this sample, star formation rates are determined from emission lines (see Section 3.1). {\it Right panel:} The resolved star forming main sequence for all galaxy spaxels ($>$99\%). Star formation rate surface densities are estimated for low S/N and non-star forming spaxels via the empirical sSFR - D4000 relation (see Section 3.2 \& Fig. 4). Both panels are colour coded by the logarithmic distance each spaxel resides at from the resolved main sequence ($\Delta$SFR$_{\rm res}$), and density contours are overlaid. A least squares linear fit to the star forming spaxels is shown as a solid black line, and the threshold of quenched spaxels (located at the minimum of the density contours) is shown as a dashed black line.}
\end{figure*}

In the left-hand panel of Fig. 3, we show the resolved main sequence ($\Sigma_{\rm SFR} - \Sigma_{*}$ relation) for the subset of MaNGA spaxels with S/N $>$ 3 emission lines, which are furthermore identified as star forming on the BPT emission line diagnostic diagram (shown in Fig. 2). There is a clear positive trend between $\Sigma_{\rm SFR}$ and  $\Sigma_{*}$ for the star forming sample (as has been seen elsewhere, e.g. Ellison et al. 2018 and references therein). We perform a least squares linear regression fit of the resolved star forming main sequence, finding the following functional form of the relationship:

\begin{equation}
\log_{10}(\Sigma_{\rm SFR, MS}) = 0.90(\pm 0.22) \times \log_{10}(\Sigma_{*})  -  9.57(\pm 1.93)
\end{equation}

\noindent This best fit line is shown as a thick solid black line in Fig. 3, and approximately indicates the locus of the resolved main sequence relation. The measured gradient and offset are in good general agreement with several values from the literature (e.g. S\'anchez et al. 2013, Cano Diaz et al. 2016). The mean logarithmic distance from the main sequence relation ($\Delta$SFR$_{\rm res}$) is indicated by the colour of each hexagonal bin in the $\Sigma_{\rm SFR} - \Sigma_{*}$ plane, labelled by the colour bar. The dashed black line indicates the threshold of quenched spaxels, which is motivated and defined in the following sub-section.

It is important to note that although the $\Sigma_{\rm SFR}$ values shown in the left-panel of Fig. 3 are robust and reliable they are also highly incomplete, representing just 17\% of the full galaxy spaxel dataset. Furthermore, we do not recover a random subset of spaxels with this approach, instead we are biased to high star formation rates. Consequently, in Fig. 3 (left panel) we are systematically missing spaxels which are forming stars at rates significantly lower than the resolved main sequence, i.e. regions which are `quenched'. Additionally, we systematically miss spaxels at the low $\Sigma_*$ end of the resolved star forming main sequence. Both of these issues strongly motivate the need to estimate $\Sigma_{\rm SFR}$ for spaxels without strong emission lines, and for regions with AGN contamination.

\subsection{$\Sigma_{\rm SFR}$ from $D4000$ for all other Galaxy Spaxels}

In this section we construct a simple, empirically motivated method to estimate $\Sigma_{\rm SFR}$ values in star forming regions where $H\alpha$ flux cannot be used (e.g. due to a low S/N or AGN contamination), and to classify spaxels broadly into star forming and quenched categories. We validate this approach against photometric SFR surface densities derived from SSP model fitting in Appendix A. All of the main results and conclusions of this work are identical whichever method is used. However, for the main body of the text we prefer to use the simple empirical method described here because a) it is possible to apply direct to the spectra and hence is much easier to reproduce in future studies; and b) it is much less model dependent than derived photometric SFRs from SSP fitting. Additionally, we validate our classification prescription against detections vs. non-detections in  $H\alpha$ and luminosity weighted stellar age (from {\sc Pipe3D}). All tests lead to essentially identical conclusions in the following results sections, and hence our chosen method is not a source of significant bias or error in the analyses of this paper. 

In order to estimate $\Sigma_{\rm SFR}$ in spaxels without strong emission lines, and for emission line regions with contamination from AGN, we must look for a non-emission line parameter which satisfies three criteria. Specifically, our sought parameter must be:

\begin{enumerate}
\item correlated with $\Sigma_{\rm SFR}$ in star forming regions
\item effective at identifying quenched regions
\item measured reliably in all galaxy spaxels
\end{enumerate}

\noindent The above might seem to be a tall order; however, fortunately there exists at least one available parameter which achieves all of these requirements. Inspired by Brinchmann et al. (2004) using single aperture spectroscopy in the SDSS, and following more recent spatially resolved analyses with earlier MaNGA data releases (e.g., Spindler et al. 2018, Wang et al. 2019), we adopt the strength of the 4000 \AA \hspace{0.05cm} break (D4000) as our desired alternative to emission lines for estimating $\Sigma_{\rm SFR}$. We define the D4000 index to be:

\begin{equation}
D4000 \equiv \int\limits_{4050(1+z)}^{4250(1+z)} \hspace{-0.1cm} f_{\lambda} \hspace{0.05cm} d\lambda \hspace{0.2cm} \big/ \hspace{0.1cm} \int\limits_{3750(1+z)}^{3950(1+z)} \hspace{-0.1cm} f_{\lambda} \hspace{0.05cm}d\lambda
\end{equation}

\noindent where $f_{\lambda}$ is the flux density at each wavelength $\lambda$, and the limits on the integrals are chosen to span a small range of wavelengths either side of the (rest frame) 4000\AA \hspace{0.05cm} break. Note that we take the $f_{\lambda}$ definition of flux density (as opposed to the $f_{\nu}$ definition), which is arbitrary since it only affects the values by a constant scaling factor. We also adopt the broad definition of the 4000\AA \hspace{0.05cm} break (D4000), instead of the narrow definition (D$_{n}$4000). The reason for this is twofold. First, the broad definition is less sensitive to issues related to velocity dispersion smearing than the narrow definition (see, e.g., S\'anchez et al. 2016b), and, second, it is the sole 4000 \AA \hspace{0.05cm} break measurement provided in P{\small IPE}3D precisely because of the first issue. The second issue is mentioned simply because we have a nominal desire to restrict our current analysis to the P{\small IPE}3D data products for the sake of self-consistency. Of course, we also have access to the MaNGA DAP measurements and the raw spectra as well. Using these, we have tested that the $\Sigma_{\rm SFR}$ values computed from the broad and the narrow 4000 \AA \hspace{0.05cm} break indices (as well as the $f_{\lambda}$ and $f_{\nu}$ definitions) are consistent, and indeed there is no significant difference between them. More specifically, we recover $\Sigma_{\rm SFR}$ values which are bias free (b $<$ 0.01) and with a scatter comparable to the error on the measurement ($\sigma <$ 0.5 dex). Thus, our results and conclusions would be identical regardless of the precise definition of the index chosen. As such, we only show results for the broad $f_{\lambda}$ definition in this paper. For completeness, we note that this particular index is sometimes labelled B4000 to avoid confusion with other D4000 definitions (see Gorgas et al 1999).

\begin{figure*}
\includegraphics[width=0.49\textwidth]{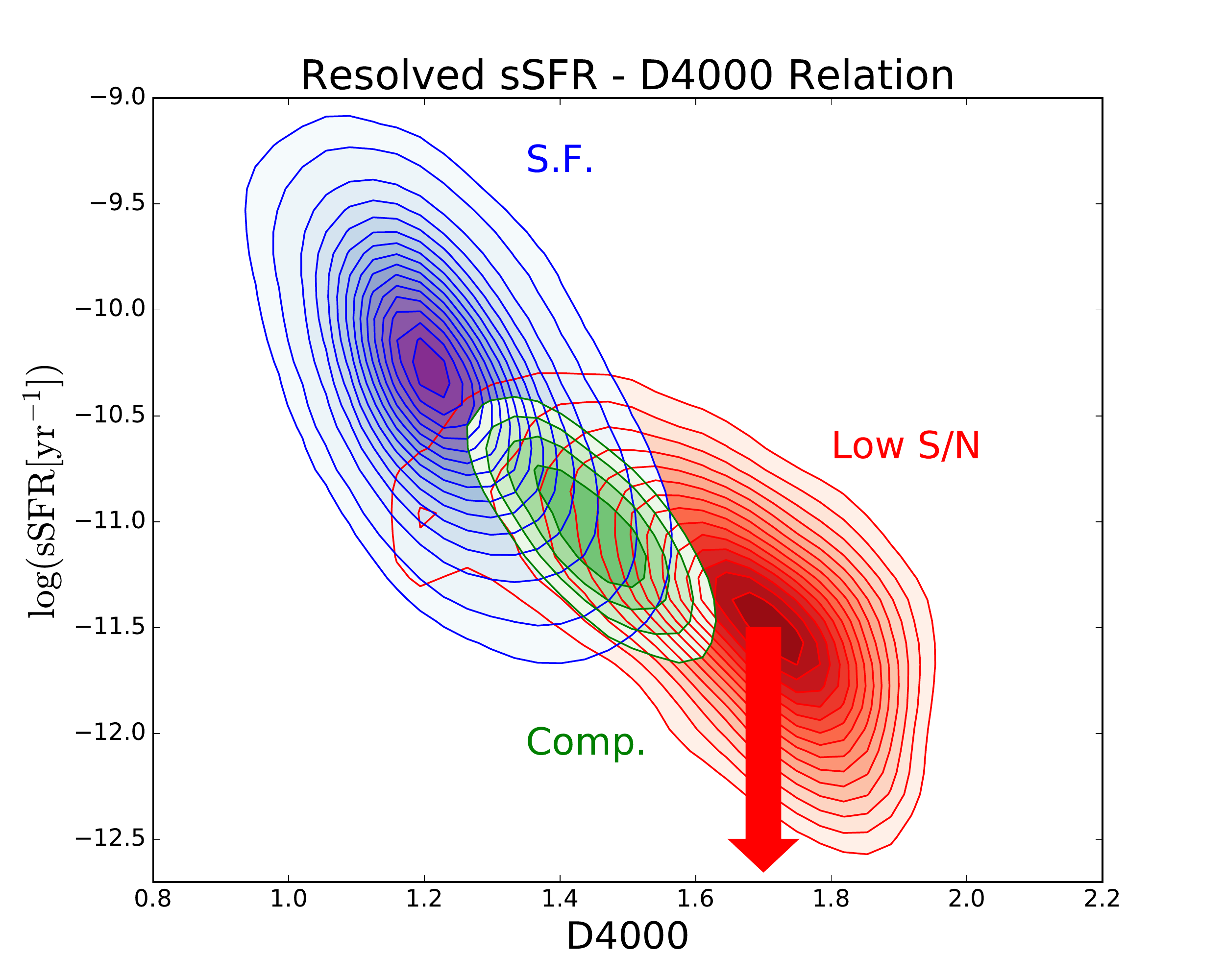}
\includegraphics[width=0.49\textwidth]{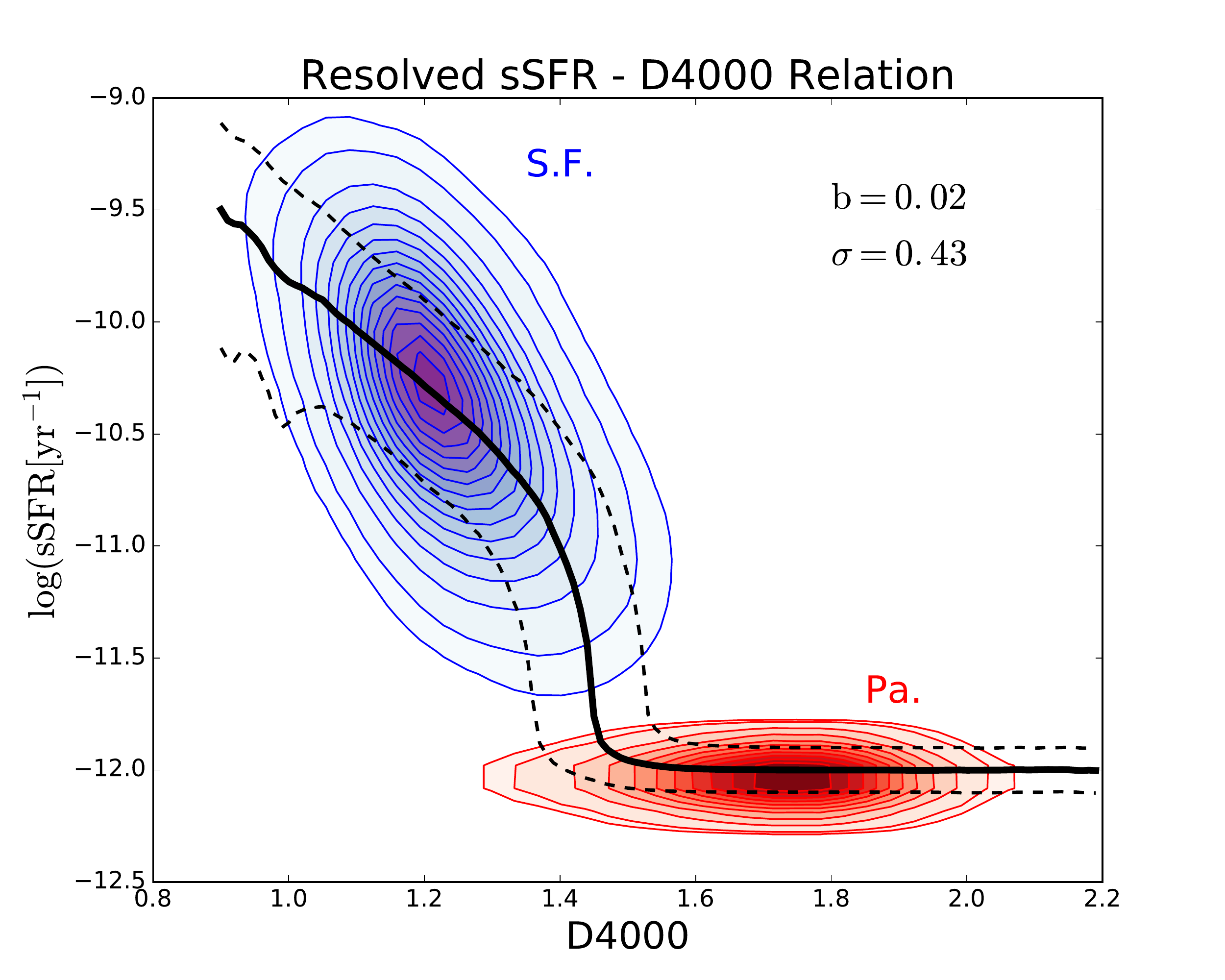}
\caption{{\it Left panel:} The empirical relationship between sSFR (= $\Sigma_{\rm SFR}(H\alpha) / \Sigma_{*}$) and the D4000 index, for all spaxels with BPT emission line S/N $>$ 3 which are furthermore classified as star forming (shown as blue contours, labelled as `S.F.'). Additionally, the upper limits in sSFR at each D4000 value are shown for low-S/N and lineless spaxels (red contours, labelled  `Low S/N').  Clearly, the star forming and low-S/N spaxels are well separated in the D4000 index, as expected since the latter is a good estimator of the age of the stellar population. Additionally, the sSFR - D4000 relation is shown for composite regions (as green contours), for comparison. {\it Right panel:} Genuinely passive spaxels (defined to have $H\alpha$ S/N $<$ 1 and old stellar populations) are here artificially relocated to a fixed value of sSFR = $10^{-12} {\rm yr}^{-1}$ (shown in red, and labelled `Pa.'). The median relation for the full (star forming + passive) sample is shown as a thick black line, with the dashed lines indicating the 1$\sigma$ dispersion. At D4000 $<$ 1.45 spaxels are invariably star forming, with sSFR given reasonably accurately by the relation with D4000; whereas at D4000 $>$ 1.45 spaxels are invariably passive with unknown exact sSFRs, except that they must be (substantially) lower than the star forming population.}
\end{figure*}

To see how the D4000 index satisfies our three conditions, in the left-hand panel of Fig. 4 we plot resolved sSFR (= $\Sigma_{\rm SFR}(H\alpha) / \Sigma_{*}$) as a function of D4000. For the high-S/N star forming sub-sample (the same sample as plot in the left panel of Fig. 3) we find a strong relationship between sSFR and D4000 (shown as blue contours and labelled S.F.), whereby higher D4000 values yield lower sSFR values. Quantitatively, we find a moderately strong anti-correlation between sSFR and D4000 of $r_{\rm Pearson}$ = -0.54 (with a scatter of $\sigma$ $\sim$ 0.4 dex). Thus, our first condition is met: D4000 provides a means to estimate sSFR (and hence $\Sigma_{\rm SFR}$) for star forming regions due to the empirical correlation between these parameters.

Additionally, in the left-panel of Fig. 4 we show the upper limits in sSFR (from the $\Sigma_{\rm SFR}(H\alpha)$ detection limits) for all of the low-S/N spaxels. This subset will contain genuinely quenched/ passive regions as well a regions with very high dust obstruction and/ or low $\Sigma_*$. The mean D4000 of the low-S/N spaxels is 1.68$\pm$0.01 and the mean D4000 for the high-S/N star forming spaxels is 1.23$\pm$0.01 (where the errors are given by the standard error on the mean). Thus, star forming and quiescent spaxels are generally well separated in D4000, which aids the identification of quenched regions, satisfying our second criterion (although see below for a more robust test). Finally, D4000 is measured in $>$ 99\% of spaxel binnings (with a S/N $>$ 5), hence our third and final criterion is also satisfied by this approach. As a result of the above success with our three conditions, we may use some (as yet to be quantified) relationship between sSFR and D4000 to estimate sSFR and hence $\Sigma_{\rm SFR}$ in all galaxy spaxels, as desired. 

As a sanity check, in the left-panel of Fig. 4 we also show the location of high-S/N BPT `composite' spaxels. These lie at intermediate D4000 values to star forming and quiescent spaxels, but lie slightly higher in sSFR than expected for their D4000 values, compared to the star forming sample. This is as expected, since the sSFR values computed for spaxels with some AGN contamination must be interpreted as effective upper-limits, given the unconstrained contribution to their line fluxes from AGN ionising photons. We do not use spaxels with AGN contamination in our sSFR - D4000 calibration, unlike in some other approaches (e.g., Brinchmann et al. 2004). However, we do apply the sSFR-D4000 relation to AGN contaminated (as well as line-less) regions for completeness. The excellent agreement in $\Sigma_{\rm SFR}$ values between SSP model fitting and our D4000 estimates lends further confidence to this approach (see Appendix A). 

To avoid issues of some quiescent (low-S/N) spaxels falling below our detection threshold as a result of heavy dust obscuration or as a result of having low $\Sigma_*$ (e.g., in the outskirts of low mass galaxies), we add two more constraints on the low-S/N data to isolate genuinely quenched regions within galaxies. Specifically, we require that  $H\alpha$ is non-detected at the level of S/N $<$ 1 (i.e. there is no measurable $H\alpha$ emission), and additionally require that the luminosity weighted age of the stellar population is `old'. The first constraint ensures that the lack of emission line flux is specifically associated with a lack of observed star formation. The second criterion ensures that the lack of observed star formation is most probably a result of an old stellar population, with little-to-no ongoing star formation, as opposed to a lack of emission lines due to heavy dust extinction or a spaxel residing in the outskirts of a galaxy. We take our threshold for `old' to be Age$_{\rm lum}$ = 3 $\times 10^9$ yr. This value is motivated because it is the minimum of the bimodal distribution in age output by the  P{\small IPE}3D pipeline. Note that the specific numeric value is highly model dependent, but that the relative measurement of ages between spaxels is much more robust. Thus, the separation into old and young regions is a relative method, designed here to avoid including genuinely star forming regions (which must have relatively young stellar populations as measured in light) into our passive sample. For truly passive spaxels, the overlap in D4000 with the star forming sample is very small, with $<$ 10\% of spaxels residing in the D4000 region corresponding to ambiguous levels of sSFR, which may naturally be interpreted as a resolved green valley. In most analyses, we remove this small population in any case.

In the right-panel of Fig. 4 we show again the density contours for the high-S/N star forming sample (shown in blue), but now include only the genuinely passive spaxels (shown in red). The sSFR values of passive spaxels, which were shown as upper limits in the left-panel of Fig. 4, are here collapsed to a fixed value of $\log_{10}({\rm sSFR})$ = -12 (as in Brinchmann et al. 2004 \& Spindler et al. 2018). This value is arbitrary (although it corresponds approximately to the mean upper limit in sSFR of the passive sample). Varying the exact value of the passive sSFR limit has no effect on our classification of spaxels into quenched and star forming categories, and only a minor impact on the recovered $\Sigma_{\rm SFR}$ values (restricted entirely to the spatially resolved green valley region, which we usually remove from our analyses). We introduce a small amount of random scatter ($\sigma$ = 0.1 dex) into the passive spaxel sSFR values, to aid in visualising these regions here. Nothing quantitative can be inferred about the sSFRs of the passive population, except that they must be substantially lower than the star forming population. Thus, it is highly robust to conclude that these regions within galaxies are forming stars substantially lower than the resolved main sequence (i.e. that they are quenched), but it is not possible to conclude precisely how little star formation they experience. This fact must be embraced by the scientific methodology when utilising these values.

In the right-panel of Fig. 4 we show the running median (solid black line) and $\pm$ 1$\sigma$ range (dashed black lines) in the sSFR - D4000 relationship, for a concatenated sample of high-S/N star forming spaxels and (genuinely) passive spaxels (with fixed nominal low sSFR values). This relationship shows clearly the D4000 threshold at which quenched spaxels dominate the sample, which occurs at D4000 = 1.45. That is, regions within galaxies with D4000 $<$ 1.45 are typically star forming (and can be modelled reasonably accurately through the sSFR - D4000 correlation); whereas regions within galaxies with D4000 $>$ 1.45 are typically quenched. For the quenched population we may reliably infer that star formation is low relative to the resolved main sequence, but the exact values of star formation (likely well below the detection threshold) is, of course, unknown. We now have all we need to estimate $\Sigma_{\rm SFR}$ from D4000. We assign $\Sigma_{\rm SFR}$ values from D4000 as follows:

\begin{equation}
\Sigma_{\rm SFR}({\rm D4000}) = {\rm sSFR(D4000)} \times \Sigma_{*}
\end{equation}

\noindent where sSFR(D4000) is taken as the median value of sSFR at D4000 from the running median (solid black line in Fig. 4), in bins of $\delta_{\rm D4000}$ = 0.01. We additionally include scatter at the level of $\sigma$ = 0.33 dex. This is largely an aesthetic choice, as it prevents all quenched spaxels from occupying the same exact ridge line on the $\Sigma_{\rm SFR} - \Sigma_*$ plane, which aids in visualisation later on in the paper. Moreover, this value of scatter is chosen to be lower than the error on the recovery of $\Sigma_{\rm SFR}$ for star forming regions, i.e. it does not worsen the accuracy of the fit. 

We test the recovery of $\Sigma_{\rm SFR}$ from D4000 with the measured values from $H\alpha$, for high-S/N star forming regions. The D4000 method recovers $\Sigma_{\rm SFR}$ with essentially no bias (b = 0.02 dex) and a standard deviation of $\sigma$ = 0.43 dex. Thus, the error on $\Sigma_{\rm SFR}$(D4000) is approximately a factor of two higher than the error on $\Sigma_{\rm SFR}$($H\alpha$) (which is a comparable result as found for the SDSS in Rosario et al. 2016, and in IFS studies in Gonz\'alez Delgado et al. 2016 \& S\'anchez et al. 2018). Consequently, our D4000 method to measure star formation rate densities in all spaxels yields completeness ($\sim$100\% vs. 17\%), but comes at a price in terms of accuracy (0.4 dex vs. 0.2 dex). Hence, the sample one chooses to work with must be motivated by the specific science goals of the study. For example, can one accept a reduced accuracy for the sake of completeness, or would one be better to use a highly reliable but highly biased sub-sample? In the following results sections of this paper we will take both approaches, where appropriate.

Finally, we construct a master $\Sigma_{\rm SFR}$ parameter, which is taken from the emission line method of Section 3.1 if all BPT emission lines are found to have S/N $>$ 3 and additionally if the spaxel lies in the star forming region of the BPT diagram. For all other cases (low S/N regions, and regions with any AGN contamination) we estimate $\Sigma_{\rm SFR}$ from the D4000 method, explained in this section. We now have a complete sample of $\Sigma_{\rm SFR}$ with which to work with. To illustrate the power of this sample, we show the complete $\Sigma_{\rm SFR} - \Sigma_{*}$ relationship for all spaxels in Fig. 3 (right panel), placed next to the same figure for the high-S/N star forming spaxels only (left panel of Fig. 3). It is striking how different the two figures look. Specifically, the complete sample fills in both quenched regions (lying below the resolved main sequence) and the low $\Sigma_*$ end of the resolved main sequence. As before, this figure is colour coded by the logarithmic distance each spaxel resides at from the star forming main sequence ridge-line (eq. 8, shown as a thick black line). Additionally, we show as a dashed black line the minimum of the density contours, which indicates the threshold of quenched spaxels within galaxies. It can now be clearly seen that virtually no quenched regions within galaxies are included in the high-S/N star forming region (as one might have reasonably expected). Thus, if our goal is to probe quenching, we must focus on the complete sample.

Ultimately, the sSFR - D4000 calibration works because the D4000 index is a sensitive tracer of the presence (or absence) of young stars in the spectrum of a given region of a galaxy. There are, of course, weak degeneracies with metallicity and dust extinction inherent in the method (and indeed as used throughout the vast number of SDSS publications, from Brinchmann et al. 2004 onwards). However, we emphasize here that an approach utilising photometric $\Sigma_{\rm SFR}$ values from SSP model fitting (which explicitly fits for metallicity and extinction as well as age) leads to essentially identical results and conclusions to our relatively simple H$\alpha$ - D4000 hybrid approach (see Appendix A for full details). In the final analysis, the metallicity and dust degeneracies are both contained in the 0.43 dex scatter, which is more than tight enough to be highly useful in this study.

\subsection{$\Delta$SFR \& Spaxel Classification}

Due to the fact that there is a positive relationship between $\Sigma_{\rm SFR}$ and $\Sigma_*$ for star forming regions within galaxies, and furthermore, given that the exponent on this relationship is not exactly equal to one, it is not possible to determine whether a region is forming stars at a rate consistent with the main sequence or not from $\Sigma_{\rm SFR}$, or even sSFR, alone (i.e. the distributions of  $\Sigma_{\rm SFR}$ and sSFR have $\Sigma_*$ dependent minima). To combat this subtle issue, we adopt the approach of Bluck et al. (2014, 2016) but here applied to spatially resolved data. Specifically, we construct a new statistic which measures the logarithmic distance each spaxel resides at from the resolved star forming main sequence ridge line (which is qualitatively similar to the approach of Ellison et al. 2018). This parameter is defined for each spaxel `$i$' as:

\begin{equation}
\Delta{\rm SFR}_{\rm res, i} = \log_{10}\big(\Sigma_{\rm SFR, i}\big) - \log_{10}\big(\Sigma_{\rm SFR, MS} (\Sigma_{*, i})\big)
\end{equation} 

\noindent where $ \Sigma_{\rm SFR, i}$ is the star formation rate surface density of each spaxel in turn, evaluated via emission lines or through the sSFR - D4000 relationship (as appropriate). $\Sigma_{\rm SFR, MS}(\Sigma_{*, i})$ indicates the expectation value for $\Sigma_{\rm SFR}$ (given the $\Sigma_*$ value of each spaxel), which is quantified by the main sequence ridge line (defined in eq. 8).

The distribution of $\Delta{\rm SFR}_{\rm res}$ is shown in Fig. 5 for the full sample (thick black line) and for the high-S/N star forming sample (light magenta line). Regions within galaxies which are forming stars exactly on the resolved main sequence will have $\Delta{\rm SFR}_{\rm res}$ = 0 by definition. As has been seen for the full galaxy distribution (e.g. Bluck et al. 2016) the distribution in this statistic is highly bimodal for the full sample. However, given the fixed upper limits of sSFR used in our definition of $\Sigma_{\rm SFR}$ for quiescent regions, the lower $\Delta{\rm SFR}_{\rm res}$ peak should really just be considered as a place-holder for relatively low values. Given an arbitrarily high level of accuracy in probing star formation, one would expect the quenched peak to spread out along the x-axis of Fig. 5, extending in principle all the way to -$\infty$ in log-space. This issue must be dealt with, either by considering the low $\Delta{\rm SFR}_{\rm res}$ values as essentially indistinguishable (as in the analyses of this paper), or else by attempting to break the degeneracy by some other means (e.g., via the age of the stellar population).

The distribution in $\Delta{\rm SFR}_{\rm res}$ shown in Fig. 5 clearly separates into two regions: a star forming peak around $\Delta{\rm SFR}_{\rm res}$ = 0 and a quenched peak around $\Delta{\rm SFR}_{\rm res}$ = -1.7. In this work we utilise a simple empirically motivated method to separate star forming and quenched regions within galaxies by the minimum of the $\Delta{\rm SFR}_{\rm res}$ distribution (as in Bluck et al. 2014, 2016 for the galaxy-wide SDSS sample). This minimum occurs at $\Delta{\rm SFR}_{\rm res}$=-0.85. Thus, values higher than this threshold are defined as star forming (note: in a different sense to the BPT diagram), and values lower than this threshold are defined to be quenched. However, values very close to the threshold are somewhat ambiguous. Hence, we additionally define a buffer-zone of $\sim$ 10\% of the range in $\Delta{\rm SFR}_{\rm res}$ centred on the star forming - quenched threshold. Thus, the full sample of spaxels is categorised into three classes:

\begin{enumerate}
\item Star Forming: $\Delta{\rm SFR}_{\rm res} > -0.6$
\item Green Valley: $-1.1 < \Delta{\rm SFR}_{\rm res} < -0.6$
\item Quenched: $\Delta{\rm SFR}_{\rm res} < -1.1$
\end{enumerate}

\noindent Each of these three regions are coloured in Fig. 5 (blue, green, and red, respectively). It is clear that for the high-S/N BPT star forming sample (magenta line), there are essentially no truly quenched spaxels at all, and only a very minor population of green valley regions. Thus, it is imperative to utilise the full/ complete sample of spaxels in order to probe quenching. We use these classes for training and validation in the machine learning classification analysis (presented in Section 4.2).

\begin{figure}
\includegraphics[width=0.49\textwidth]{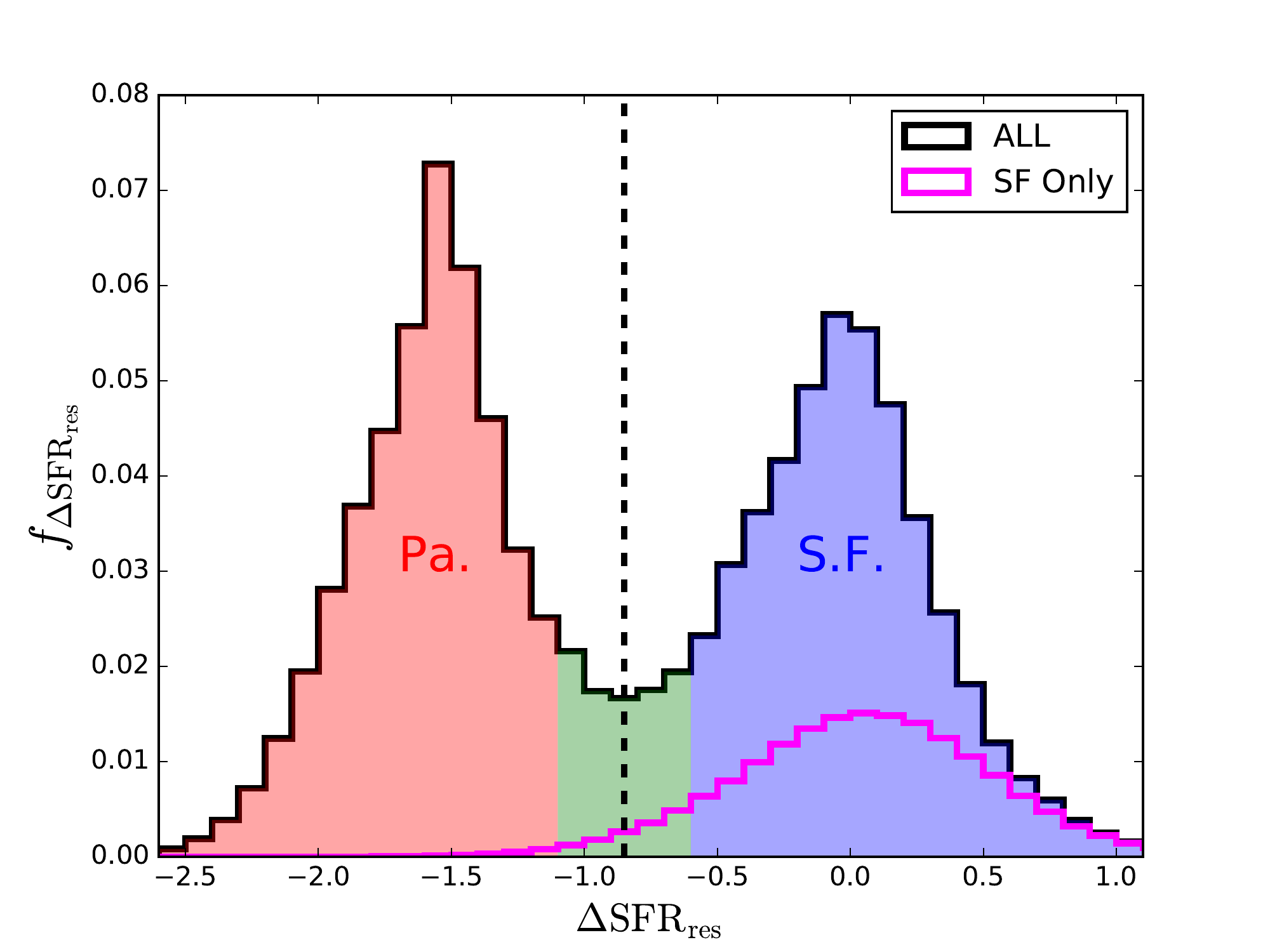}
\caption{The distribution of the resolved $\Delta$SFR parameter for all spaxels (shown as a thick black line) and for the high S/N star forming sub-population (shown as a magenta line). The regions where galaxies are forming stars on the main sequence, significantly below the main sequence, and at an intermediate level are shown in blue, red and green, respectively. The dashed line at $\Delta$SFR = -0.85 indicates the minimum of the distribution, and we adopt this value for our classification threshold.}
\end{figure}

%
%   RESULTS
%

\section{Results}

\subsection{General Trends in Resolved Star Formation \& Quenching}

\begin{figure*}
\includegraphics[width=1\textwidth]{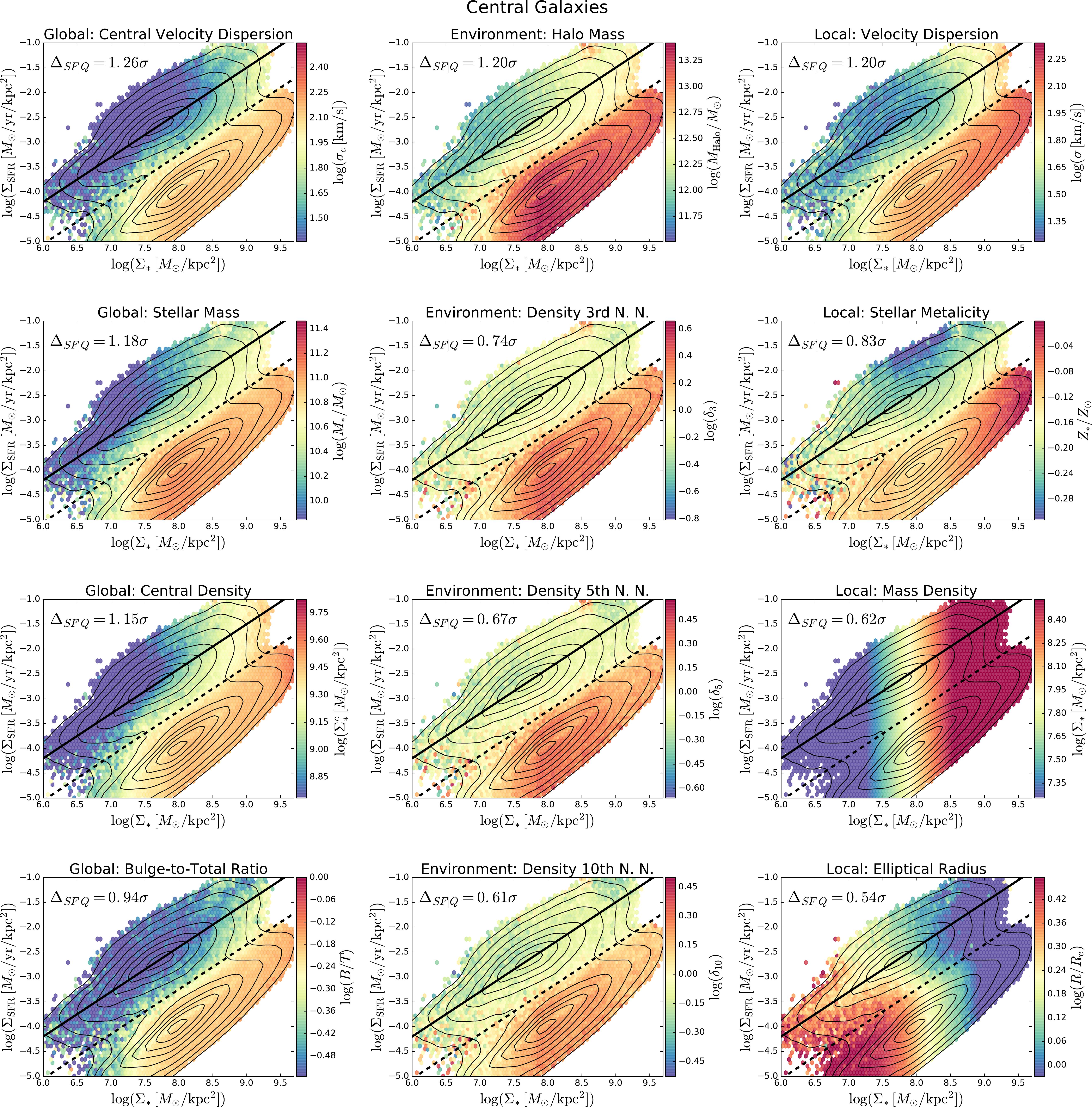}
\caption{The resolved main sequence relationship for {\it central galaxies}, shown for all spaxels. Each panel is split into small hexagonal bins, coloured by the mean value of a variety of global, environmental and local (spatially resolved) parameters, as indicated by the individual panel titles and the colour bar labels. The figure is organised as follows: the left column shows global parameters, the middle column shows environmental parameters, and the right column shown local parameters. Each row is ordered from largest to smallest difference between star forming and quenched spaxels. To quantify the difference in each parameter between star forming and quenched spaxels, we present the $\Delta_{SF|Q}$ statistic on each panel (see eq. 12). High values of this statistic indicate a large difference between star forming and quenched spaxels in the parameter under investigation, and low values represent a small or negligible difference. For fairness of presentation, each colour bar is centred on the median value and spans a [-1$\sigma$, +1$\sigma$] range in each parameter. As in Fig. 3, the solid black lines indicates the main sequence relation, and the dashed black lines indicates the division between star forming and quenched spaxels.}
\end{figure*}

\begin{figure*}
\includegraphics[width=1\textwidth]{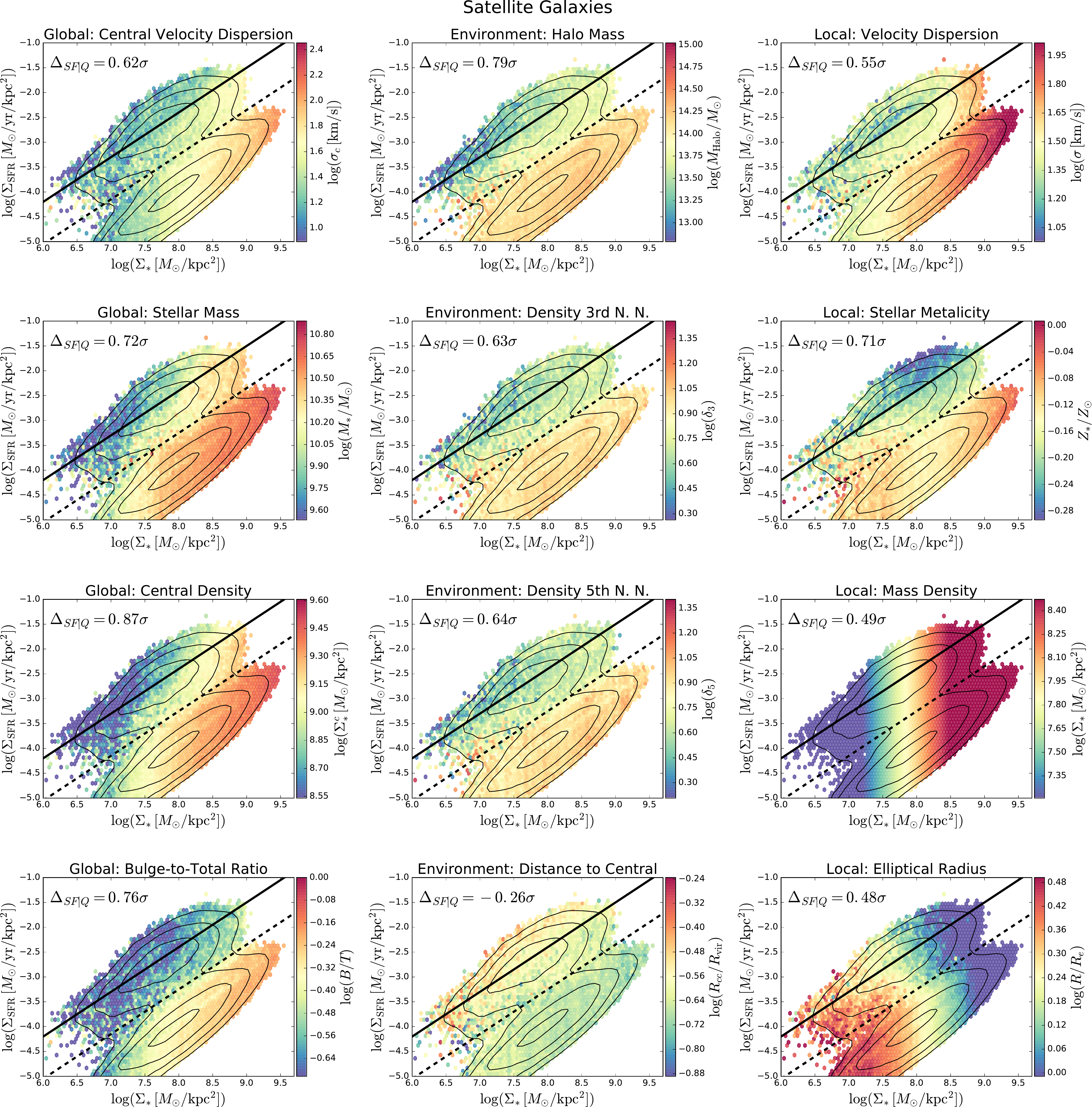}
\caption{This figure is almost identical to Fig. 6, but now showing the results for {\it satellite galaxies}. Note that this figure is ordered as for centrals (to aid in comparison), and hence there is no longer a clear decline in $\Delta_{SF|Q}$ down each column. Additionally, we add the projected distance to the central galaxy as an additional environmental parameter for satellites, which replaces local density evaluated at the 10th nearest neighbour in this figure.}
\end{figure*}

In this section we explore in a general way the dependence of resolved star formation quenching on a variety of variables, separated into the following sets: local (spatially resolved), global (one value per galaxy, pertaining to the galaxy as a whole), and environmental (one value per galaxy, pertaining to the environment in which the galaxy resides). More specifically, we consider the following parameters:

\begin{enumerate}
\item {\bf Global Parameters:} central velocity dispersion evaluated within 1kpc ($\sigma_c$)\footnote{We also consider central parameters defined through the effective radius (i.e. evaluating the average $\Sigma_*$ and $\sigma$ within $R_e/4$ and $R_e/2$). These alternatives lead to identical rankings, and hence we do not consider them further here.}; total stellar mass ($M_*$); central mass density within 1kpc ($\Sigma_*^c$); and bulge-to-total stellar mass ratio ($B/T$)
\item {\bf Environmental Parameters:} group halo mass, evaluated from an abundance matching technique ($M_{H}$); local galaxy density evaluated at the 3rd, 5th and 10th nearest neighbour ($\delta_3$, $\delta_5$ and $\delta_{10}$); and distance to the central galaxy for satellites ($D_{c}$)
\item  {\bf Local Parameters:} velocity dispersion of the spaxel ($\sigma$); stellar mass surface density of the spaxel ($\Sigma_{*}$); stellar metallicity of the spaxel ($Z_*$); and location of the spaxel within the galaxy (semi-major axis $R/R_e$\footnote{The semi-major axis of the unique ellipse with given axis ratio (b/a) and position angle ($\theta_{p.a.}$) which passes through the centre of each spaxel is determined via a standard de-projection technique. Note that the ranking of the Euclidean distance ($r_E$) and the semi-major axis are very similar in all of our analyses, and differ only slightly in regression (where the semi-major axis performs better than the Euclidean distance from the centre).}). 
\end{enumerate}

All of these parameters have been chosen not to be trivially connected to star formation in any manner, and additionally to be reflective of interesting physical processes within galaxies. To this end, we have avoided line fluxes and equivalent widths, colours and magnitudes, and parameterisations of models (e.g. S\'ersic indices). Additionally, we have exclusively chosen parameters which can be measured accurately in {\it both} star forming and quenched spaxels, in order to determine how each parameter impacts the probability of a region within a galaxy being quenched. As a result of this, we exclude gas-phase properties from our main analysis (although see Section 4.2.3 for a brief analysis of gas-phase parameters in star forming, emission-line regions). Additionally, for some tests in the machine learning section (Section 4.2), we also consider the total star formation rate of the galaxy (SFR) and the distance each galaxy resides at from the star forming main sequence ($\Delta$MS).

In Fig. 6 we present the resolved star forming main sequence relationship for all (star forming and quenched) spaxels from central galaxies. The $\Sigma_{\rm SFR} - \Sigma_*$ plane is divided into small hexagonal bins colour coded by the mean of each parameter under investigation. The colour bar for each panel runs from -1$\sigma$ to +1$\sigma$ for each parameter. This bespoke structuring of the colour bars ensures a fair visual comparison between parameters, even though their values are wildly different in magnitude and range. As in Fig. 3, the solid black line indicates a least squares fit to the star forming main sequence (eq. 8), the dashed black line indicates the location of the minimum of the density contours (which sets the threshold for quenched spaxels), and density contours are shown as faint black lines. 

Fig. 6 is structured as follows. The first column groups all of the global parameters considered, the middle column groups all of the environmental parameters considered, and the right column groups all of the local (spatially resolved) parameters considered. Within each column, parameters are ordered from most to least different between the star forming and quenched populations. We quantify this difference as:

\begin{equation}
\Delta_{SF|Q}  = \frac{ {\rm med}({\rm Var})_Q - {\rm med}({\rm Var})_{SF}}{\sigma_{\rm Var}}  
\end{equation}

\noindent where ${\rm med}({\rm Var})_{SF}$ indicates the median value of variable `Var', evaluated for the star forming spaxels only (i.e. those residing above the dashed black line); ${\rm med}({\rm Var})_{Q}$ indicates the median value of variable `Var' evaluated for quenched spaxels only (i.e. those residing below the black dashed line); and  `Var' indicates each of the twelve variables under consideration in this work, in turn. Crucially, this difference is normalised by the dispersion of the variable across the full set of all spaxels (i.e. star forming and quenched), which is indicated by $\sigma_{\rm Var}$. This normalisation accounts for different magnitudes of values and range within each parameter, allowing a fair comparison between each parameter. Hence, the $\Delta_{SF|Q}$ statistic is given in units of the dispersion of each variable. Additionally, we take logarithmic values of each parameter here, and throughout this work, to avoid issues of comparing logarithmic and linear units.

On each panel in Fig. 6 we present the $\Delta_{SF|Q}$ statistic, which quantifies what can be readily intuited by eye: variation in the colour-bar parameter between star forming and quenched spaxels reduces significantly for each column, moving from top to bottom in the figure. For central galaxies the parameter which is most different between star forming and quenched spaxels is central velocity dispersion. This fact is reminiscent of the key result from Teimoorinia, Bluck \& Ellison (2016) where central velocity dispersion is found to be the most constraining single variable for predicting whether central galaxies will be quenched or star forming. However, this prior study did not have access to spatially resolved information and hence we significantly extend and expand upon that early work here. Generally speaking, global parameters tend to perform better at distinguishing between star forming and quenched spaxels than environmental parameters or local parameters. Nonetheless, there are environmental and local parameters which are effective at discriminating between star forming and quenched spaxels. The most distinguishing environmental parameter is group halo mass, and the most distinguishing local parameter is velocity dispersion at the spaxel location. These parameters are known to correlate very strongly with global parameters, and hence interpreting their success is a subtle problem (see Section 4.4 for further discussion on this point).

It is particularly interesting to note that central mass surface density (evaluated as the mean value within 1 kpc) is significantly more discriminating of star forming and quenched spaxels than the stellar mass density at the location of the spaxel. It is also interesting that central velocity dispersion (evaluated within 1kpc) is more discriminating than velocity dispersion of the spaxel. These results suggest that, although there are far more local parameters than global parameters (one per spaxel vs. one per galaxy), it is global parameters rather than local parameters which impact the spatially resolved quenching of centrals more than the conditions at the spaxel location. All of the environmental parameters are not particularly discriminating of quenching for centrals, with the sole exception of halo mass, which of course for centrals is highly correlated with stellar mass and other global parameters. Thus, in general, global parameters appear to be more discerning of spatially resolved quenching than local or environmental parameters for centrals (from our set of 12 parameters, excluding gas-phase properties).

In Fig. 7 we reproduce Fig. 6 for satellite galaxies, which are defined as any group members which are less massive than the central. This figure is structured in an identical manner to Fig. 6, to aid in comparison. The most obvious difference between central and satellite galaxies in these figures is that in general all parameters vary less between star forming and quenched spaxels for satellites than for centrals. This is interesting because it suggests that no single parameter is particularly effective at constraining whether spaxels will be forming stars or not for satellites. More specifically, central velocity dispersion is much less different between star forming and quenched spaxels for satellites than it is for centrals. However, global parameters still appear to be more discriminating of star forming and quenched spaxels than local or environmental parameters, although the relative difference between global and environmental parameters is reduced.

The preceding analysis is instructive, but it lacks quantitative rigour. In the next section we turn our focus to a machine learning approach designed to answer two pertinent questions: 1) which individual and groups of parameters are most effective at predicting whether spaxels will be star forming or quenched?; and 2) which individual and groups of parameters are most effective at estimating $\Sigma_{\rm SFR}$ in star forming regions? For the first question we must consider all spaxels (i.e. star forming and quenched) in order to make any progress at all. For the second question we are able to proceed by restricting our analysis to the resolved main sequence of star forming spaxels only, where we have more reliable measurements of $\Sigma_{SFR}$. Note that these questions are (very) different. The physical processes responsible for star formation and quenching are likely different, and are certainly not required to be the same.

\subsection{Machine Learning Analysis \& Rankings}

We develop a sophisticated machine learning technique utilising both artificial neural networks (ANNs) and a random forest (RF) to classify MaNGA spaxels into star forming and quenched classes, based on a variety of global, environmental and local parameters, treated individually in the ANN case and as a group in the RF case. We also employ our ANN and RF to predict actual $\Sigma_{\rm SFR}$ values for star forming spaxels through a regression analysis. In this section we present our detailed results from these two analyses, after describing our methodology.

Our decision to run separate analyses for star formation and quenching is motivated by two separate concerns. First, the $\Sigma_{\rm SFR}$ values in star forming regions are reasonably well constrained and hence regression is possible (and desirable). However, in quiescent regions, we can only reliably determine whether a given spaxel is star forming or quenched. Thus, analysing star formation is naturally approached through regression, but quenching can only be studied via classification in our data (and indeed in essentially all extant observational data). Second, even given arbitrarily accurate $\Sigma_{\rm SFR}$ values in all regions of MaNGA galaxies, philosophically it would still be a mistake to combine the star formation and quenching analyses. The reason for this is that almost any value of $\Sigma_{\rm SFR}$ might be deemed star forming or quenched, as a result of varying $\Sigma_*$ in the region (e.g., see Fig. 3). Ultimately, this is closely analogous to the difference between SFR and sSFR in global studies of galaxies: the latter can readily indicate quenching, but the former cannot by itself. Finally, in retrospect, the parameters which are best at predicting quenching and ongoing levels of star formation turn out to be very different, which would be completely missed in a combined analysis. 

In this section we concentrate on central galaxies. There are three reasons for this. First, many studies have found profound differences between centrals and satellites in terms of star formation and quenching (e.g. Peng et al. 2012, Bluck et al. 2014, 2016, Woo et al. 2015), and hence separating these populations is essential. Second, centrals outnumber satellites by $\sim$ 3:1 in our MaNGA sample and hence we have considerably greater statistical power for training networks with centrals than satellites. Third, a preliminary machine learning analysis of satellite galaxies yielded ambiguous results (most likely as a consequence of the relatively small sample size) and hence there is little novel to show in that regard yet. We plan to revisit the following machine learning analyses with satellite galaxies in the coming years, once a higher number of these systems are observed with MaNGA. We do note here that in terms of star formation, satellites performed similarly to centrals in our preliminary analysis, but in terms of quenching the two populations exhibit major differences in the parameters which are most effective for classifying spaxels.

\subsubsection{Machine Learning Methodology}

We utilise an approach very similar to Bluck et al. (2019) combining ANN classification and ANN regression to analyse a large astrophysical dataset. Our goal is to infer how strongly different parameters are connected to resolved star formation and quenching in a fully self-consistent and model independent manner. ANNs have the distinct advantage over other statistical techniques in that they do not assume anything about the underlying structure, correlation, monotonicity, or connections of the input data. Moreover, sophisticated multi-layered networks (utilising so called `deep learning') act as a universal function generator (e.g. Wichchukit \& O'Mahony 2010) making them ideal to analyse multiple parameters acting in concert to give rise to a given physical effect, in this case star formation and quenching. However, the main limitation of ANN is that they do not reveal the relative importance of parameters when used in a group. To combat this, we additionally analyse star formation and quenching via an RF approach. The combination of both yields greater insight than either alone.

We perform our machine learning analysis in P{\small YTHON} using the powerful S{\small CIKIT}-L{\small EARN} package. For ANN we incorporate MLPC{\small LASSIFIER} as our primary classification tool and MLPR{\small EGRESSOR} as our primary regression tool. For both classification and regression we design a network with two hidden (deep) layers, structured with neurons sequenced as 12:6. This represents 18 nodes (109 weights for single variables) for the analysis of several million independent data values. These specifications are chosen via direct experimentation with the data, and represent the optimal performance in the testing sample, without over-fitting. We note that for individual parameters, the performance is not strongly dependent on the network complexity and over-fitting is extremely rare. Following our experience with applying neural networks to astrophysical data, we adopt a {\it relu}-activation function and adopt the {\small ADAM} numerical solver (which is recommended in the documentation, but see Bluck et al. 2019 for further justification). 

Additionally, we test the network performance for all variables used simultaneously to explore the maximum potential of the data with varying levels of network sophistication. For the run with all variables we find that, unlike for the individual runs, over-fitting is a much more serious issue. To combat this we reduce the network complexity to a single hidden layer with just 5 nodes (representing 81 weights to set for the fully connected network). We find that increasing the network complexity up to these values leads to continuously increasing performance of the network in the test sample with these data, yet increasing beyond this leads to no significant further improvement in the test sample, and increased over-fitting as evidenced by discrepancies between the training and test sample performances. Thus, we determine that our ANN is optimised (converged and efficient) for the MaNGA dataset.

For both the classification and regression analyses we first perform feature scaling on all of the input data by subtracting off the median value and normalising by the interquartile range, which is more robust to outliers than standard mean subtraction and standard deviation normalisation. Thus, all data is converted to a unitless and scale free format. We then randomly select 50\% of the available data (from the full spaxel set for classification, and the star forming sub-sample for regression) to train the network. The remaining 50\% of spaxels are used to test the performance of the network on data which the network never interacts with during training. Crucially, we select spaxels from different galaxies for training and testing, which is important for preventing over-fitting on global or environmental parameters (although is not strictly necessary for local/ spatially resolved parameters). We run the full training and validation process ten times over, for ten randomly selected training samples, yielding ten randomly selected sets of spaxels which are not used in the training process to assess the performance on (the test samples). The final values on the performance statistics are given by the median of the ten runs, and the 1$\sigma$ error is taken as the dispersion across the ten runs or the difference between the performance of the testing and training sample (whichever is higher). The mean difference between training and testing samples in performance for both classification and regression is set to be $<$ 1\% (i.e. we vary the parameters in our ANN to achieve this threshold). This approach is qualitatively very similar to both Bluck et al. (2019) and Teimoorina et al. (2016).

Additionally, we adopt an early-stopping routine to combat the potential for over-fitting in the training phase. Specifically, we remove 30\% of the training sample in the first step and test the performance of the intermediate trained network after each iteration on this sample. If there is no improvement on the early-stopping sample, we abandon further iterations in the full training sample even if further improvements within the training sample are possible. This approach has been shown to be very effective at preventing the over-fitting of data (e.g. Bluck et al. 2019 and references therein). However, in our present analyses, as mentioned above, we are also largely protected from this issue by setting a robust upper limit to the difference in performance between training and testing samples. 

For the classification analysis, where we aim to identify the most important parameters for distinguishing between star forming and quenched spaxels, we take the fraction of spaxels correctly classified by the network in the independent test set as our primary performance statistic, also known as the  `accuracy' (as in Bluck et al. 2019). However, we note that  MLPC{\small LASSIFIER} formally optimises the log-loss (entropy) function, which is nonetheless very strongly correlated with the accuracy. We prefer to utilise the accuracy simply because it is more intuitive, but note that none of our results (i.e. rankings) depend critically on the choice of function. Additionally, we also consider the area under the true positive rate (TPR) - false positive rate (FPR) curve as an additional check of performance (referred to as the `AUC', as utilised in Teimoorina et al. 2016). The advantage of the AUC statistic is that it is insensitive to the fraction of each class provided to the network. Results from these statistics are identical in terms of ranking, but are different numerically, as expected. In our classification analysis we select a `balanced' sample of 50\% star forming and 50\% quenched spaxels for both training and testing, mitigating the need for more complex performance statistics like the AUC. Furthermore, this approach has been shown to improve the performance of classification (see Bluck et al. 2019 for a discussion on this point).

For the regression analysis, where we aim to predict actual $\Sigma_{\rm SFR}$ values for star forming spaxels, we compute the mean square error on the network predicted $\Sigma_{\rm SFR}$ values (compared to their measured values), which is used by the network as the loss function. However, interpreting the mean square error is difficult because even a random number will give rise to a certain level of predictivity, set by the structure of the distribution in $\Sigma_{\rm SFR}$. This is analogous to a random number obtaining a success rate of 50\% in binary classification (from an evenly sampled training and validation set), although in the regression case there is no single number which reflects a random level of performance. As such, it is the improvement over random which is really interesting in a regression analysis. Following Bluck et al. (2019), we define the improvement over random in regression as follows:

\begin{equation}
{\rm IoR \, (\%)} = \frac{{\rm RMSE}_i - {\rm{RMSE}_{\rm Rand}}}{0 - {\rm RMSE}_{\rm Rand}} \times 100\%
\end{equation}

\noindent where,

\begin{equation}
{\rm RMSE} \equiv \sqrt{\langle (\Sigma_{\rm SFR,p} - \Sigma_{\rm SFR,t})^2 \rangle}
\end{equation}

\noindent RMSE indicates the root mean squared error, the subscripts $i$ and Rand indicate the result for each variable (or set) in turn and the result for a random number, respectively. Zero is the maximum possible performance for regression (i.e. identical predictions to the test sample). In the second expression above, the subscript $p$ refers to the network predicted value, and the subscript $t$ refers to the truth value, which is our measured $\Sigma_{\rm SFR}$ for each spaxel (derived in Section 3). Thus the improvement statistic gives the percentage improvement over random, where 0\% indicates a performance indistinguishable from random (i.e. no relationship at all to star formation) and a performance of 100\% indicates that all $\Sigma_{\rm SFR}$ values for all spaxels in the test sample are predicted identically to their measured values.

In addition to ANN, we also utilise an RF analysis in the following sub-sections. The principal advantage of the RF approach is that it enables us to determine how effective a given parameter (or set of parameters) is in predicting quenching, or $\Sigma_{\mathrm{SFR}}$ values, in concert with other parameters. An RF treats multiple parameters as if they were in a competition, selecting the most useful for each decision fork (i.e. the parameter which minimises the entropy, or log-loss function, for classification and MSE for regression). By quantifying the increase in accuracy by each parameter in each fork within each tree of the random forest, the relative importance of each parameter (and group of parameters) is established. This competitive feature is especially useful when the data is highly inter-correlated with itself in a complex (and hence hard to counteract) manner. Our RF approach enables us to establish how useful each parameter is compared to each other available parameter, and to ascertain how effective global, environmental, and local parameters as a group perform in classification of spaxels and prediction of $\Sigma_{\rm SFR}$ values. 

For the RF analysis we utilise R{\small ANDOM}F{\small OREST}C{\small LASSIFIER} and R{\small ANDOM}F{\small OREST}R{\small EGRESSOR} from S{\small CIKIT}-L{\small EARN}, for classification and regression respectively. In both cases we utilise 100 estimators (independent decision trees) allowed to reach a maximum depth of 250. Varying either of these parameters does not significantly impact the final performance results, although continuously increasing the number of estimators does (slightly) improve the overall accuracy, at the price of longer run times. We control for over-fitting by varying the minimum number of samples permitted at the leaf-nodes (i.e. the minimum number of spaxels required in each output of a decision fork). We optimise the performance (MSE for regression and AUC for classification), requiring agreement between training and testing samples of 0.01 in both cases. It is possible to achieve the same tuning to the data with other parameters (e.g. maximum depth), but our experience with random forests indicate that this is the most straightforward route to avoiding over-fitting.

For classification, the minimum leaf-node sample is set to 50\,000, and for regression it is set to 5\,000. Increasing these values leads to over-fitting at fixed other parameters (greater than 0.01 discrepancy between training and testing samples), whereas decreasing these values leads to lower performance in both the training and testing samples. Hence, our random forest is optimised for our specific data set. Note that it is unsurprising that the minimum number of samples required in the leaf-nodes is lower for regression than classification, since the former is a more precision task than the latter. As with the ANN analyses, we run 10 independent training and testing runs, and take the median of the set as our final performance indicator, with the standard deviation of the set as the error. Unlike in the ANN analysis, for the RF we take the relative performance as our primary output statistic, which quantifies how effective each parameter in the set is for separating spaxels into star forming and quenched states (in classification), or predicting $\Sigma_{\mathrm{SFR}}$ values (in regression). This statistic directly tests how beneficial each parameter is to the regression or classification task in comparison to the rest of the available parameters.

Ultimately, the success of given parameters (and sets of parameters) in the classification analyses to predict whether spaxels are star forming or quenched may be taken as establishing how connected each parameter and set is to the process(es) of quenching. Alternatively, the success of given parameters (and sets of parameters) in the regression analyses to accurately estimate $\Sigma_{\rm SFR}$ in star forming regions may be taken as establishing how connected each parameter and set is to the process of ongoing star formation. As noted above, one may expect the performance of parameters to vary significantly between predicting quenching and rate of star formation.

\subsubsection{Artificial Neural Network Analysis: Single Parameter Performance}

\begin{figure*}
\includegraphics[width=1\textwidth]{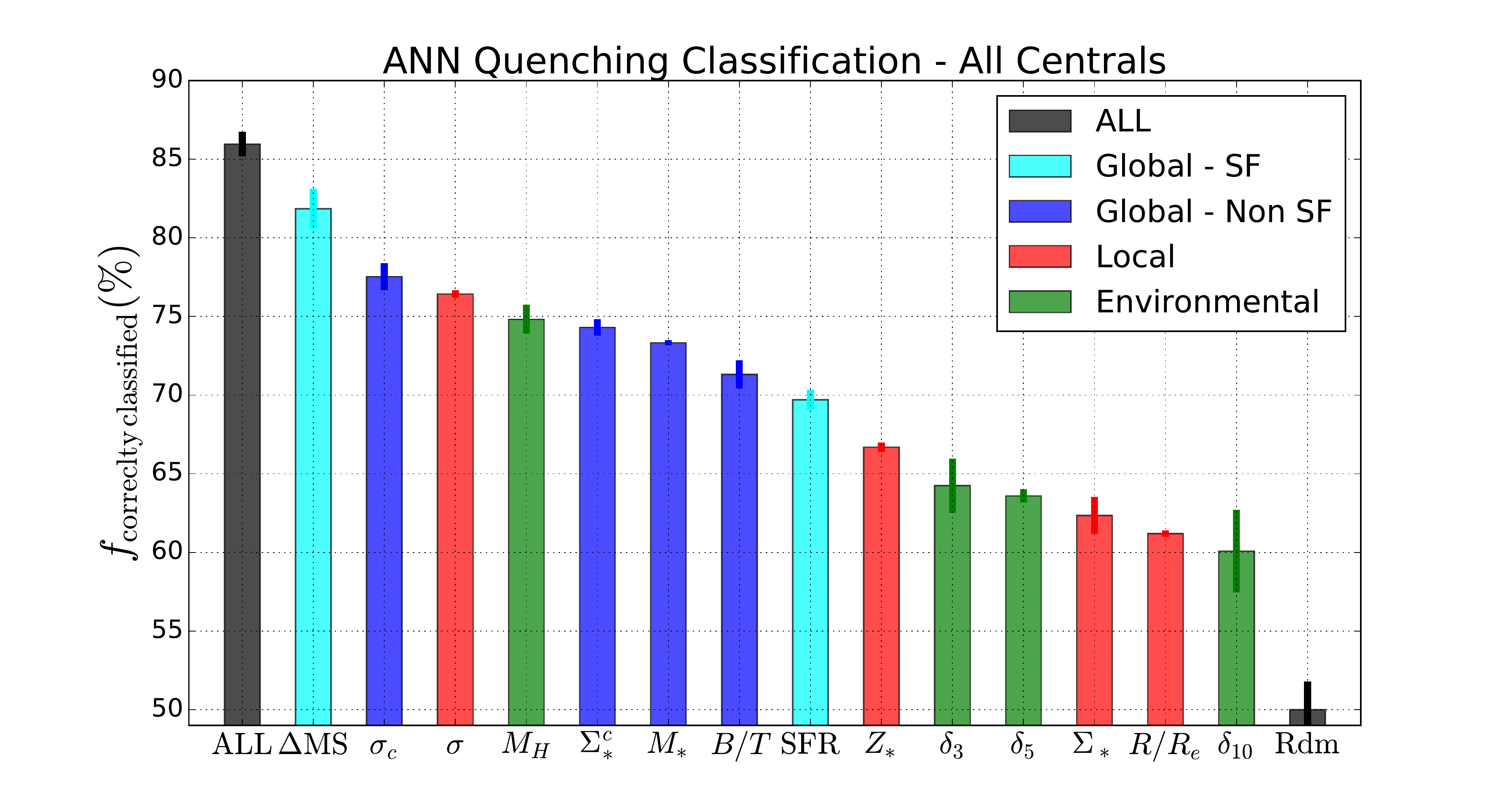}
\includegraphics[width=1\textwidth]{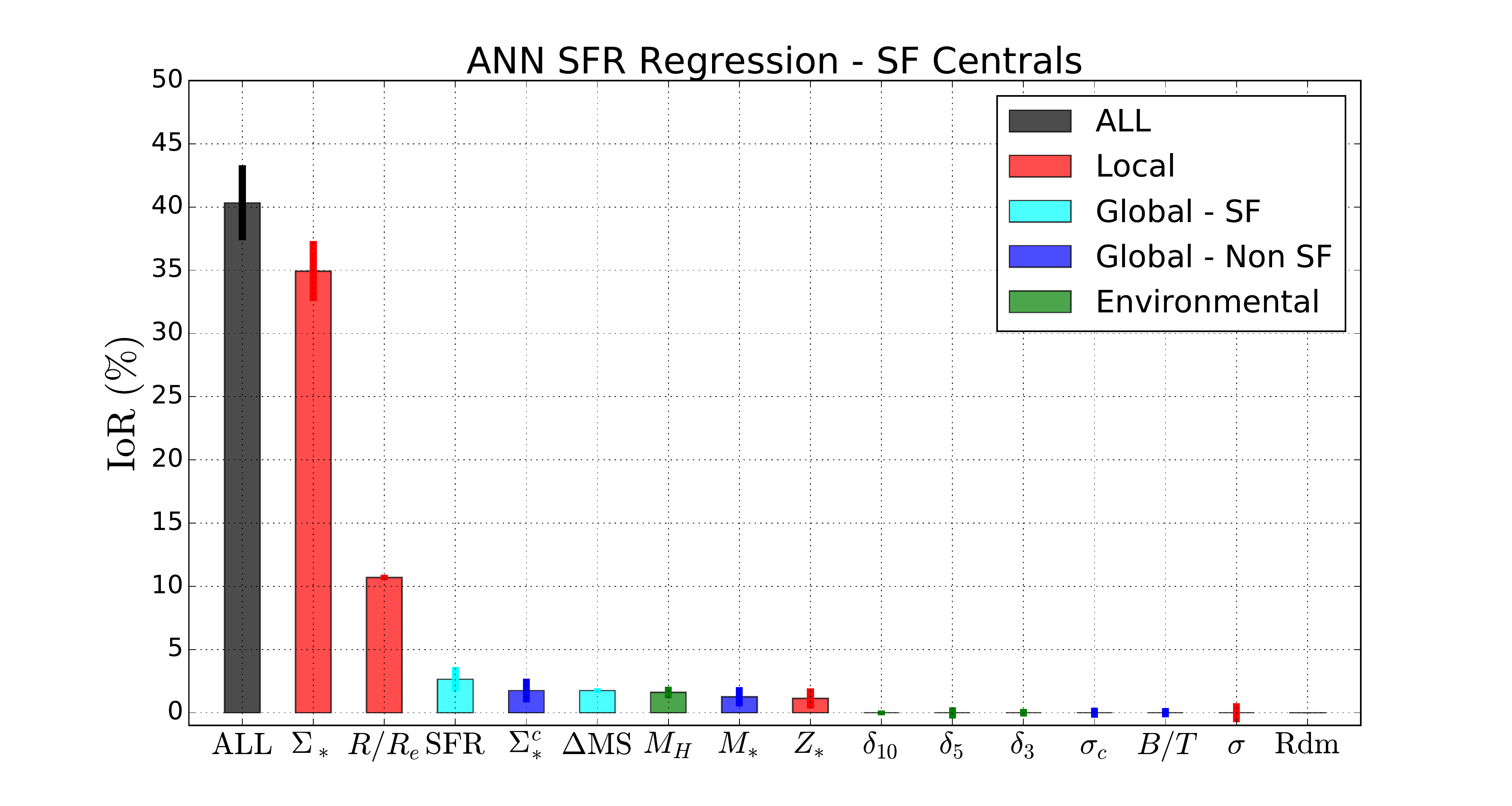}
\caption{{\it Top panel:} Results from an ANN classification analysis to predict whether spaxels will be star forming or quenched, based on single input parameters, and a multi-parameter set utilising all variables simultaneously. The y-axis shows the fraction of correctly classified spaxels by the network, while the x-axis lists the parameters made available to the network in each case (ordered from most to least predictive). {\it Bottom panel: } Results from an ANN regression analysis to predict actual $\Sigma_{\mathrm{SFR}}$ values in star forming regions. Here the y-axis shows the IoR statistic (defined in eq. 13), while the x-axis lists the parameters made available to the network in each case (ordered from most to least predictive). In both panels, the result utilising simultaneously all 14 variables is shown in black, and the result using a random set of numbers is shown in grey. These extremes indicate the maximum and minimum performance of the network possible with these data, and can be used as `yard sticks' to judge the individual parameter performances against. The type of parameter is indicated by the colour of the bar, as labelled by the legend. Error bars are given as the variance across ten independent training and testing runs, or as the mean difference in performance between training and testing (whichever is larger). For star formation (lower panel), it is clear that $\Sigma_*$ (a local/ spatially resolved parameter) is by far the most predictive parameter out of this set. For quenching (top panel), there is a much more even distribution in predictivity across the individual parameters. Nonetheless, global parameters generally perform better for predicting quenching than local or environmental parameters. Thus, there is a complete inversion in the dependence on parameters (and scales of measurement) as we change from predicting quenching to star formation rates.}
\end{figure*}

\noindent {\it QUENCHING:}\\
\noindent The goal of this part of the machine learning analysis is to train a network to predict whether spaxels will be star forming or quenched, given individual parameters for training and testing. For training, truth value labels (1 for quenched, 0 for star forming) are fed to the network based on the cuts to the $\Delta$SFR$_{\rm res}$ distribution (shown in Fig. 5). We exclude the $\sim$ 8\% of spaxels in the green valley region, with ambiguous levels of star formation. Crucially, many of the parameters used in the training process are not trivially connected to star formation, and hence their success or failure at predicting whether spaxels will be quenched reveals information on how connected those parameters are to the process of quenching. Additionally, we consider two parameters which are directly related to quenching at the global level: SFR and $\Delta$MS, the total star formation rate of the galaxy and the logarithmic distance the galaxy resides at from the {\it global} star forming main sequence (see Bluck et al. 2014, 2016), respectively. These parameters will help us to assess how well the global nature of quenching correlates with the local (spaxel-wise) nature of quenching. Ultimately, the most interesting aspect of this investigation is to explore how well the network can predict the star forming state of spaxels given this restrictive information.\\

We present our quenching classification results for central galaxies in the top panel of Fig. 8. The fraction of correctly classified spaxels is shown on the y-axis, and the parameters used to train the network are shown on the x-axis, with their parameter types indicated by the bar colour (as labelled by the legend). First, we train the network with all 14 global-SF, global(-non-SF), environmental, and local (spatially resolved) parameters used in this paper. The run with all parameters leads to an accuracy of 86.0$\pm$0.8\% of spaxels correctly classified, and an AUC = 0.926$\pm$0.006, which is formally classed as `outstanding' in the machine learning literature (see, e.g., Teimoorinia et al. 2019). The error on this value is estimated from the variance across the ten independent training and validation runs. The very small variation in the performance of the network when running on all available parameters is further confirmation that the network is stable and converged. As noted above, no significant improvement on this performance is seen from increasing the complexity of the network structure, and hence this level of performance represents the optimal result with these specific data. 

Astrophysically, the excellent performance of the network with these 14 parameters suggests that there are no significant other parameters needed in order to model spatially resolved quenching in galaxies effectively (at least at the level of spatial resolution in MaNGA). It is important to stress that this need not have been the case. Having said that, it is, of course, possible that other parameters would perform equally well or even slightly better. A particularly interesting parameter which is absent from our analysis is gas mass surface density (which is expected to set the level of ongoing star formation through the Kennicutt-Schmidt relation, modulo a potentially variable efficiency term, e.g. Kennicutt 1998, Piotrowska et al. 2019, Ellison et al. 2019). The excellent performance of these parameters for predicting quenching, combined with the assumed importance of gas content for quenching, may together imply that gas mass surface density must also be very well constrained by these data. However, to test this hypothesis directly one would need a large sample of spatially resolved gas mass measurements covering a wide range in star formation rates (which is well beyond the scope of this paper). As such, from this point on, we will concentrate our discussion on the {\it relative} performance of the individual parameters and groupings of parameters under investigation here. 

At the other extreme, we also present in the top panel of Fig. 8 the result for a random variable (denoted `Rand' on the x-axis). This run quantifies the minimum possible performance of the network, and has a median value of $\sim$50\%, as expected for a balanced sample. Given the two choices, if one were to flip a coin one would predict the star forming state of spaxels exactly as well as the random variable. Thus, each individual parameter and grouping must have a performance between random (50\% correctly classified) and the all variable run (86\% correctly classified). In this sense, the black bars in the top panel of Fig. 7 provide `yard sticks' to measure the performance of the other variables against, as well as to check that the network is behaving as expected. 

The fraction of correctly classified spaxels from a network run utilising each individual parameter alone is shown in the top panel Fig. 8 as narrow coloured bars (labelled by the x-axis). The class of parameter (i.e. global-SF, global-Non SF, local, environmental) is indicated by the colour of each bar, as labelled by the legend. For each parameter treated individually in Fig. 8, the global $\Delta$MS of the galaxy is the best single parameter. This is not terribly surprising since this parameter indicates the global star forming state of the galaxy, and hence its excellent performance at the spaxel level merely indicates that in star forming galaxies most spaxels are star forming and in quenched galaxies most spaxels are quenched. Nonetheless, the accuracy of $\Delta$MS in predicting the quenched state of individual spaxels is not perfect, yielding 81.9$\pm$1.2\% correct classifications. Thus, $\sim$18\% of spaxels must have a different star forming state to their host galaxies. This is an interesting sub-population for further exploration, which we will consider in detail in a forthcoming publication. For now, it is sufficient to emphasize that the quenching of galaxies is {\it to leading order} a global/ galaxy-wide process, with a non-negligible minority of contrary regions. It is important to stress that $\Delta$MS alone achieves 95\% of the accuracy of the all parameter run (82/86), which indicates that spatially resolved information is vastly sub-dominant to global information in parameterizing quenching\footnote{In one sense this result is highly intuitive. Given that it is common in the literature to define quenched and star forming galaxies (e.g. Baldry et al. 2006, Peng et al. 2010, 2012; Bluck et al. 2014, 2016) one might naturally expect that sub-galactic regions will simply trace the global state of the galaxy. However, the only way to directly test this hypothesis is to explore the sub-galactic regions, as done here with resolved spectroscopy from the MaNGA survey. Our conclusion is that the assumption of star forming state conformity within galaxies is correct to a $\sim$ 80\% level of accuracy.}.

Of the parameters which are not directly related to star formation, central velocity dispersion is the most predictive variable for classifying spatially resolved quenching in central galaxies (with 77.6$\pm$0.9\% of spaxels correctly classified). Of course, central velocity dispersion is also the most successful global (non-SF) parameter as well. By far the most successful environmental parameter is group halo mass (with 74.8$\pm$0.9\% of spaxels correctly classified). For local parameters, velocity dispersion at the location of the spaxel is by far the most effective parameter (with 76.4$\pm$0.2\% of spaxels correctly classified). It is intriguing that the environmental and local parameters which perform best (by a significant margin in both cases) are both highly correlated with global parameters for centrals. Halo mass is tightly correlated with stellar mass for centrals; and velocity dispersion profiles are quite flat (see later to Fig. 12), resulting in halo mass being an excellent predictor of stellar mass, and velocity dispersion at the spaxel location being an excellent predictor of the central velocity dispersion. All other local and environmental parameters achieve well under 70\% correct classifications, whereas for global parameters all of the variables considered have a predictivity of over 70\%. This is a very significant difference in performance, given the small errors from the stability of the independent runs (typically $\sim$1\%). 

Compared to Fig. 6, most of the single variables are ordered from most to least predictive for centrals in the same ordering as the difference between star forming and quenched spaxels in that parameter ($\Delta_{\rm SF|Q}$) . This is encouraging, since it offers a simple explanation for the predictivity of each variable in terms of how discrepant its values are between star forming and quenched spaxels. However, a variable having different values from star forming to quenched spaxels is only a very crude estimate of how predictive each parameter will be at discerning whether spaxels are star forming or quenched. For example, our ANN method does not rely on monotonicity or linearity in the data, both of which are implicitly assumed when taking the difference of an average. Moreover, the current ANN analysis offers a way to rank parameters and groups with well defined errors (from the variance of the network runs), and hence is far more robust and informative than merely quantifying the average difference in each parameter between star forming and quenched regions. On the other hand, the earlier analysis is still highly useful, because it offers a simple, visual and easy to intuit explanation of the single parameter rankings, as well as providing a basic consistency check on the machine learning. \\

\noindent {\it STAR FORMATION:}\\
\noindent The goal of this part of the ANN machine learning analysis is fundamentally different to the preceding paragraphs on quenching. Here we train our ANN to predict actual $\Sigma_{\rm SFR}$ values, rather than to predict whether given regions are star forming or quenched. Moreover, we restrict our analysis to the star forming sub-sample (blue shaded region in Fig. 5). Hence, there are no truly quenched spaxels at all in the current analysis. The primary reason for this selection is that robust $\Sigma_{\rm SFR}$ measurements do not extend into the quenched region of the resolved main sequence. The quenched population of spaxels are known to have low $\Sigma_{\rm SFR}$ relative to the resolved main sequence, but their specific values are completely unconstrained. Thus, the correct way to mitigate this uncertainty is to classify star forming and quenched regions, as done above. However, there is significant additional information contained in the star forming regions, where  $\Sigma_{\rm SFR}$ can be reliably constrained. Thus, the motivation of this part of the analysis is to answer a distinct question to the preceding section: {\it what parameters are most constraining for setting the level of star formation in star forming regions?}

In order to answer the above question, we now use regression (as opposed to classification), but otherwise utilise a very similar ANN architecture, and an identical methodology of utilising ten independent runs for both training and testing (explained in detail above). In keeping the analysis as similar as possible to the classification case, we maximise the reliability in comparing the performance of each parameter and group. In the bottom panel of Fig. 8 we show the results for our regression analysis to estimate $\Sigma_{\rm SFR}$ in star forming regions for central galaxies. As with the top panel of Fig. 8, the x-axis lists each individual parameter used in training and validation in turn, and the legend lists each group. Here the y-axis shows the improvement in performance over the result with a random variable (defined in eq. 13), unlike in the top panel of Fig. 8 where the y-axis shows the fraction of correctly classified spaxels. The reason for the use of the improvement statistic is to remove effects from variable distributions of $\Sigma_{\rm SFR}$ on the results (see above, and also Bluck et al. 2019 where this statistic is first defined). By definition, a random variable has 0\% improvement over random, and a perfect regression result (i.e. identical predicted values to the truth values) would yield a 100\% improvement.

It is striking how different the distribution in performances are for the quenching classification analysis (Fig. 8 top panel) and the star formation rate regression analysis (Fig. 8 bottom panel). In the star formation rate regression analysis, $\Sigma_*$ (measured in the spaxel) is {\it by far} the most effective single parameter for estimating $\Sigma_{\rm SFR}$ in star forming regions within galaxies. In fact, $\Sigma_*$ achieves an IoR $>$ 3 times that of the next most predictive variable (the spaxel's location within the galaxy: $R/R_e$). It is highly instructive to compare this result with the performance of $\Sigma_*$ for predicting whether a spaxel is star forming or quenched (in the top panel of Fig. 8). In classifying star forming and quenched regions, local $\Sigma_*$ is very poorly constraining, whereas for estimating actual values of $\Sigma_{\rm SFR}$ (in star forming regions) $\Sigma_*$ is highly effective. This result is critical to the narrative of this paper, so we will emphasise this point a little further here. 

If we know a given region in a galaxy is forming stars, measuring $\Sigma_*$ in that region enables a highly accurate prediction of $\Sigma_{\rm SFR}$ (ranked clear 1st for regression). However, if we know nothing about the region other than its $\Sigma_*$ value, whether the region is actively forming stars or not is almost entirely unconstrained (ranked 12th for classification). Conversely, knowing $\Delta$MS for central galaxies leads to a highly accurate classification of spaxels into star forming or quenched classes (ranked 1st); yet if we know a region is star forming, the galaxy total SFR gives virtually no constraint whatsoever on the level of star formation ensuing within a given region within the galaxy (with only $\sim$ 3\% improvement over random!). Thus, the parameters which are most effective for predicting the rate of star formation are very different to the parameters which are most effective for determining whether regions will be star forming or quenched. {\it Ultimately, quenching is governed by global parameters, yet star formation is governed by local/ spatially resolved parameters.}\\

\noindent {\it ALTERNATIVE SAMPLES:}\\
\noindent We repeat our entire classification and regression analyses for a wide variety of alternative data sets, sampling a variety of different methods. First we consider only spaxels with unique measurements of $\Sigma_{\rm SFR}$, and hence also $\Delta$SFR$_{\rm res}$ from H$\alpha$ and/or D4000 (i.e. completely removing the potential for trivial over-fitting). Second, we consider only unique binned spaxel regions within each galaxy, for the P{\small IPE}3D continuum fitting S/N thresholds (i.e. an analysis of voxels rather than spaxels). Both of these alternative runs lead to essentially identical results to the one for all spaxels presented in this section, with identical rankings provided for each individual parameter, within each group.  

Additionally, we reproduce the classification analysis for a segregation of quenched and star forming spaxels based on SSP derived $\Sigma_{\rm SFR}$ values. All of the main results and conclusions are identical to this section, where we utilise our hybrid H$\alpha$ - D4000 method (described in Section 3). We also test classification methods based purely on H$\alpha$ detection, on the age of the stellar population, and a combination of H$\alpha$ detection and age. The key results of this section (i.e. that quenching is a global process but star formation is a locally governed process) is recovered in every single alternate analysis we have explored. Therefore, the ANN results are extremely stable to sample variation and to the method used for assigning star forming state to spaxels. See Appendix A for further tests on the stability of the machine learning results.

\subsubsection{Random Forest Analysis: Multi-Parameter Performance}

\begin{figure*}
\includegraphics[width=1\textwidth]{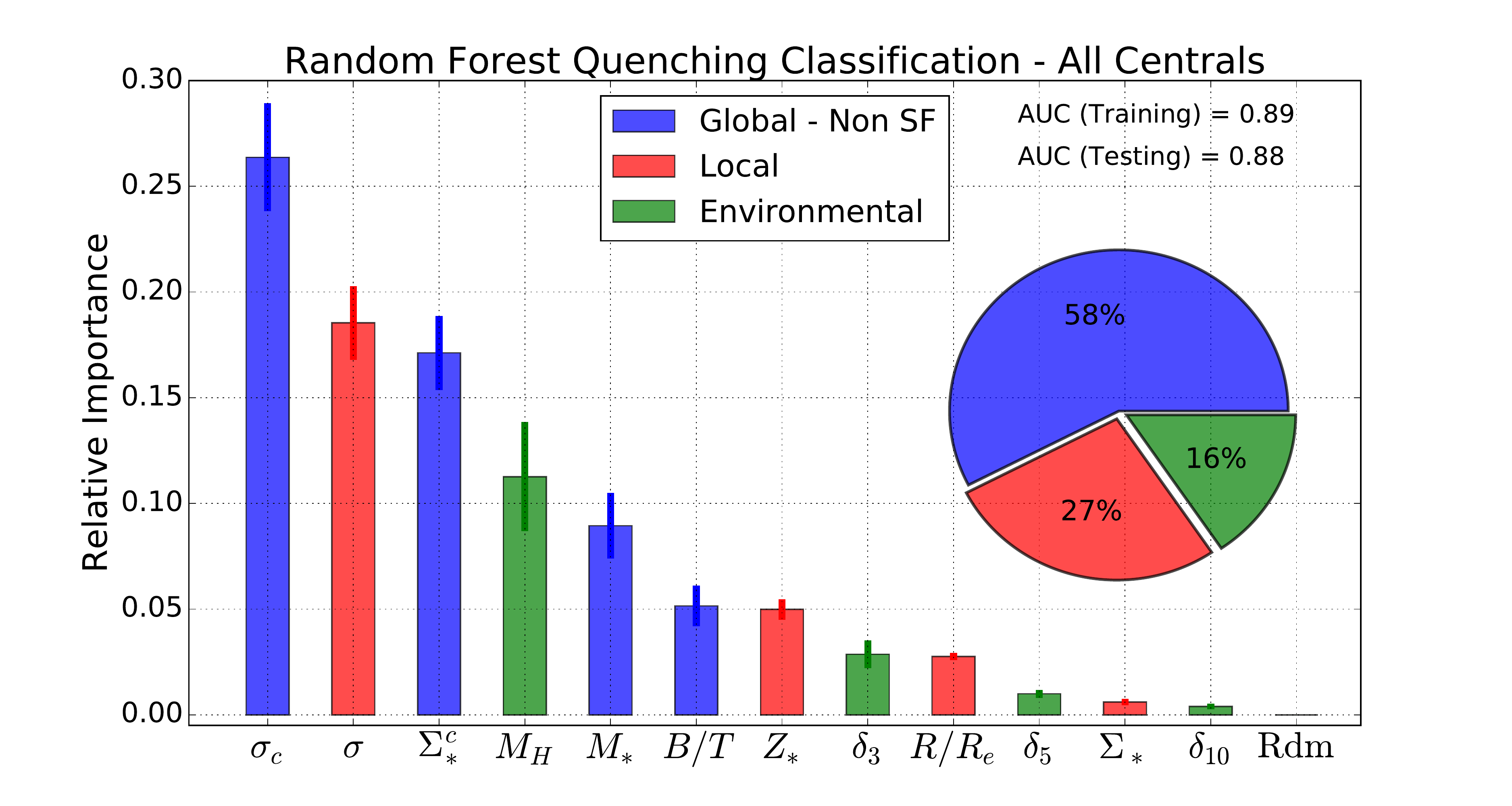}
\includegraphics[width=1\textwidth]{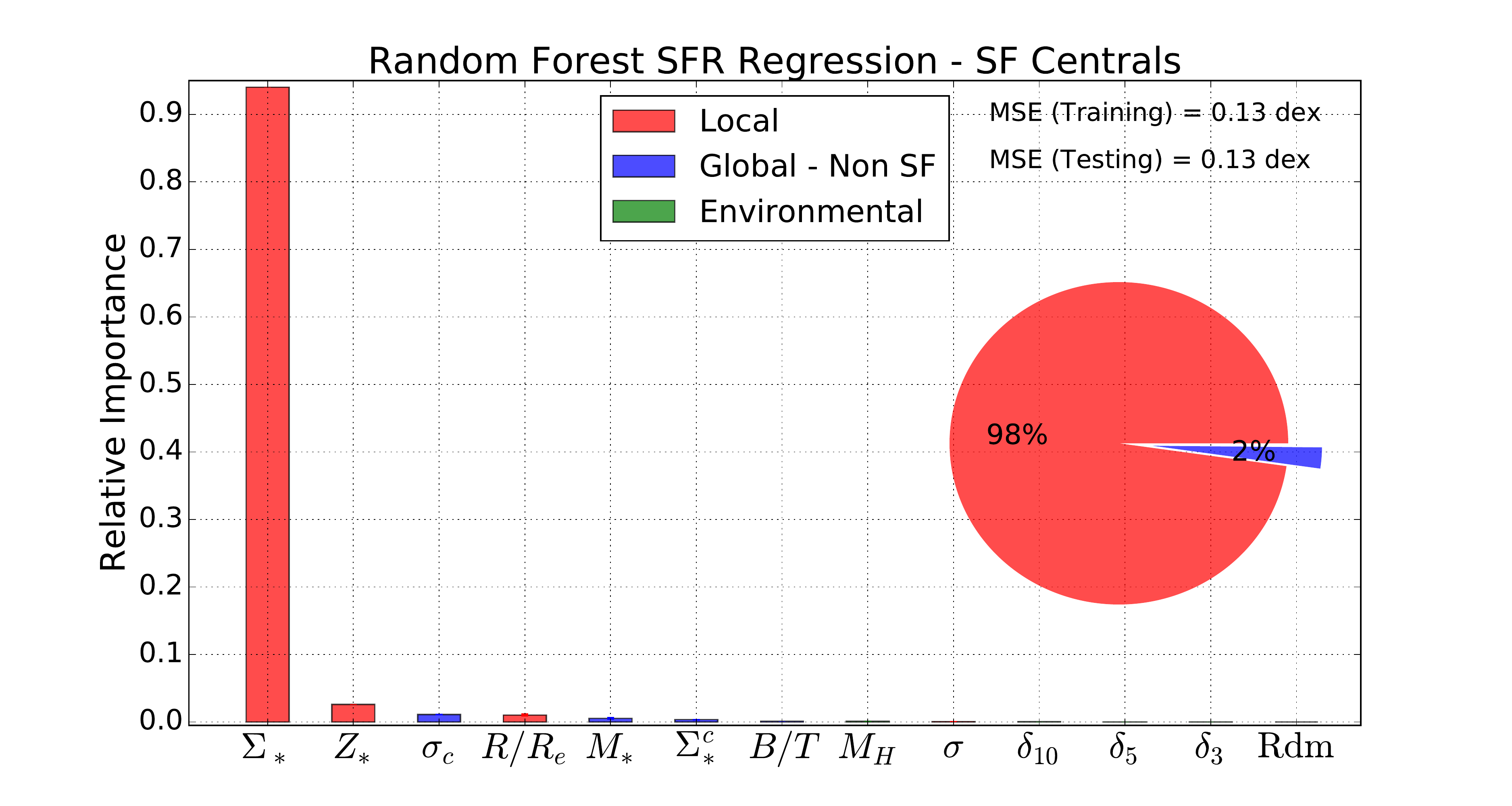}
\caption{{\it Top panel: } Results from an RF classification analysis to predict whether spaxels will be star forming or quenched, based on a run utilising all 12 parameters which are not trivially connected to star formation.  {\it Bottom panel: } Results from an RF regression analysis to predict $\Sigma_{\rm SFR}$ values in star forming regions, based on a run utilising the same 12 parameters. On both panels, the y-axis shows the relative importance of each parameter in the combined analysis, with each parameter labelled by the x-axis (ordered from most to least predictive). The colour of each bar indicates the class of the parameter under investigation, as labelled by the legend. Error bars show the variance of 10 independent training and testing runs; and the median performance metrics of the set of independent runs (AUC for classification and MSE for regression) are displayed for both training and testing samples on each panel. Additionally, we present a pie chart sub-plot on each panel showing the relative importance to the random forest analysis of each category of parameters, as labelled by the legend. For star formation, almost all of the predictive power in these data is contained by $\Sigma_*$ alone, with local parameters being overwhelmingly the best set. For quenching, there is a much flatter distribution of importance. Nonetheless, central velocity dispersion is the most important single variable for predicting quenching, and global parameters clearly perform as the most informative set.}
\end{figure*}

In the previous sub-section we explore the absolute predictive power of each of our 12 parameters which are not directly connected to star formation, in addition to SFR and $\Delta$MS, for both star formation and quenching. In this sub-section, we explore how important each parameter is to predicting quenching and star formation in concert with the other parameters. Specifically, we run two RF analyses, utilising all 12 parameters which are not trivially connected to star formation. The first is a classification analysis to predict the star forming state of spaxels (i.e. star forming or quenched) and the second is a regression analysis to predict actual $\Sigma_{\rm SFR}$ values in star forming regions. In this respect the analysis is very similar to the prior ANN analysis (discussed above). However, here we train only with the full set of parameters, and present the {\it relative importance} of each parameter to the RF. This statistic explicitly quantifies the informative power of each parameter in competition with the rest of the data set, exposing how parameters work in concert to predict star formation and quenching. 

More specifically, in the S{\small CIKIT}-L{\small EARN} implementation we adopt, the relative importance (more commonly referred to as the `feature importance') at each decision tree node is computed as the fraction of the data which reaches the node, weighted by the decrease in impurity after the split (defined via the Gini coefficient for classification and via the MSE for regression). The final value of the relative importance is taken as the sum over all nodes in a tree, and the average over the $N_{\rm est}$ independent decision trees (which we take as 100). The power of this approach is through averaging over relatively uncorrelated decision trees. The lack of correlation is ensured by two levels of randomisation: 1) for each tree, a bootstrapped sample is constructed, randomly selecting the total number of data points from the training sample, with return; and 2) for each node, only a randomly selected number of $\sqrt{N_{\rm features}}$ are considered. The upshot of this approach is that if two parameters have equal predictive power this will be easily exposed, and similarly if the predictive power of one parameter derives its strength solely from correlation with another parameter, this will also be exposed (although it may take several iterations, which is the logic behind utilising 100 estimators). As with the ANN approach, we run the RF analysis ten times over for ten randomly selected subsets of the data, selecting 50\% for training and 50\% for testing (separated on the galaxy level). In total our results for both RF classification and regression are based on averaging over 1000 independent decision trees, with differences ensured via sample and feature selection randomisation.  

In the top panel of Fig. 9, we show the results from an RF classification analysis to predict whether spaxels will be star forming or quenched. The y-axis shows the relative importance of each feature, and the x-axis labels each feature in turn. The class of the feature is indicated by the colour of the bar, as labelled by the legend. The overall performance of the random forest is given by the AUC statistic, as presented on the plot (AUC = 0.88 in the testing sample). In the case where all parameters are uncorrelated with each other, one would expect that the RF relative importances mimic the absolute fraction of correctly classified spaxels in the ANN analysis. This is clearly not the case here as there is a much stronger decline in the relative importance (in RF) than seen with the fraction of correctly classified spaxels (in ANN). This indicates that some of the predictive power of these individual parameters is derived from their correlations with one another.

For the individual parameters, the most important variable in our data set is clearly $\sigma_c$, with all other parameters showing a marked suppression relative to $\sigma_c$ when compared to the ANN analysis (see Fig. 8). This strongly suggests that much of the strength of correlation between, e.g., $M_*$ or $M_{H}$ and quenching arises simply from their correlation with $\sigma_c$. Interestingly, in the ANN analysis $\sigma_c$ (measured within 1kpc of the centre of the galaxy) performed only slightly better than local $\sigma$ (measured within each spaxel), whereas in the RF analysis here, $\sigma_c$ clearly outperforms $\sigma$. As such, if a single parameter is sought to parameterize central galaxy quenching with, $\sigma_c$ is the best choice out of the parameters we have considered (which is consistent with results from Wake et al. 2012, Bluck et al. 2016 and Teimoorinia et al. 2016 for global studies of galaxies). Nonetheless, it is evident that other parameters are still useful for predicting central galaxy quenching in conjunction with $\sigma_c$. Overall, the rankings in absolute predictivity (Fig. 8) and relative importance (Fig. 9) are similar for quenching, it is just that in the RF we see more marked separation between parameters, which is a result of the competitive nature of the RF approach.

\begin{figure*}
\includegraphics[width=1\textwidth]{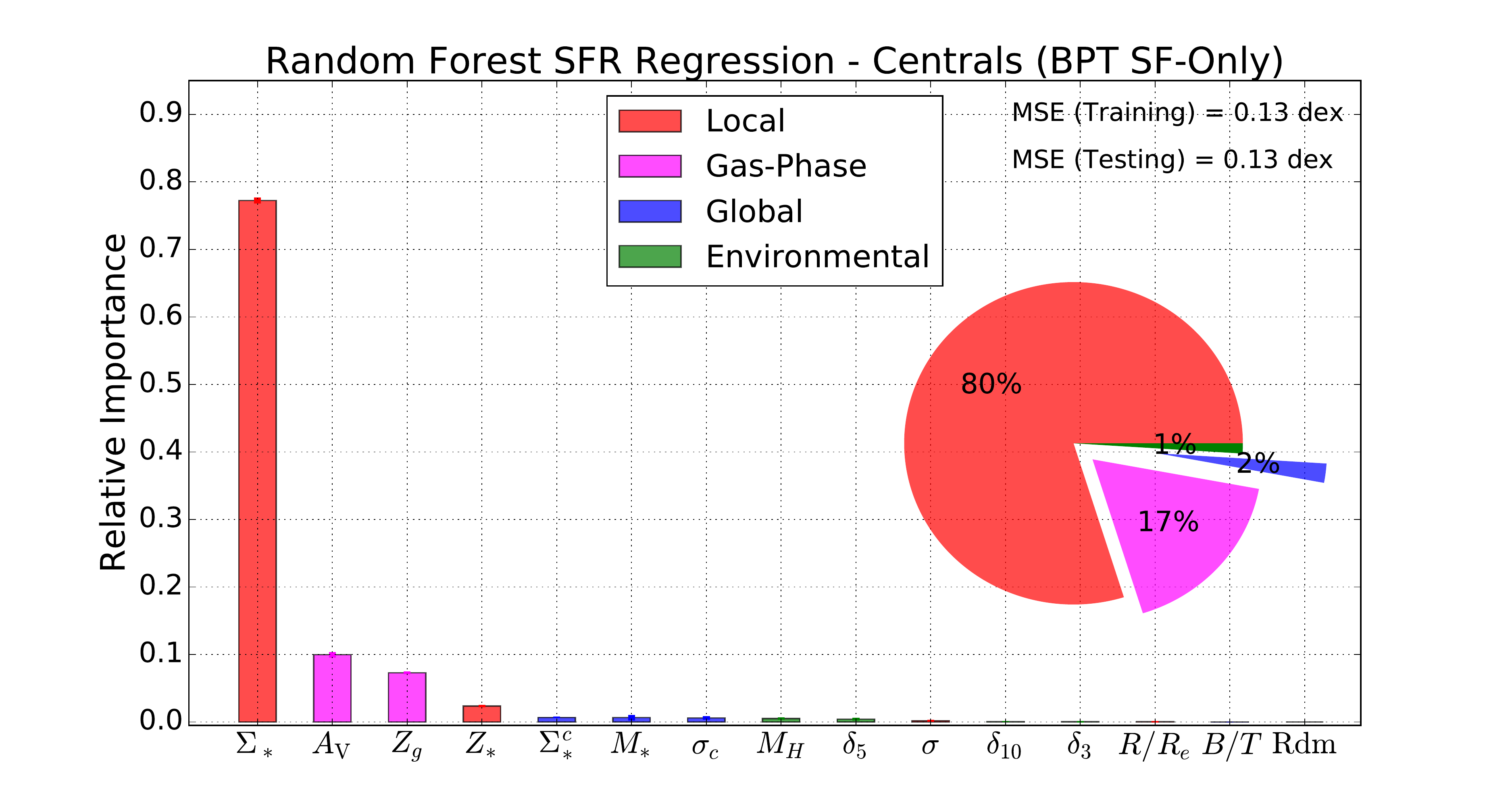}
\caption{Reproduction of Fig. 9 (bottom panel) showing the results from an RF regression analysis to predict $\Sigma_{\rm SFR}$ values in star forming regions. However, in this figure only spaxels with BPT emission lines with S/N $>$ 3, which are furthermore identified as `star forming' in the BPT diagram (see Fig. 2) are included. Additionally, we include two gas-phase parameters: the V-band extinction ($A_{\rm V}$) and the gas-phase metallicity ($Z_g$). Although local (non-gas phase) parameters still perform clearly as the best set, the gas-phase local parameters make a significant secondary contribution to the RF performance (as shown by the pie chart).}
\end{figure*}

In the RF analysis the relative importance quantifies the usefulness of each parameter in concert with the other parameters. Consequently, we may additionally explore the relative importance of different {\it classes} of parameters for quenching, which would be meaningless for the absolute predictivity in the ANN approach. In the top panel of Fig. 9 we show a pie chart visualising the relative importance of each class: global-Non SF, local and environmental. Working collectively, global parameters account for 58\% of the reduction in impurity in the random forest, with local parameters accounting for just 27\% and environmental parameters just 16\%. Thus, quenching is primarily governed by global parameters, out of the parameters considered in this work. Even more interestingly, it is parameters associated with the {\it centre of galaxies} ($\sigma_c$ and $\Sigma_*^{c}$) which are the most predictive of the global parameters. Hence, quenching is most accurately constrained by global parameters measured at the centre of galaxies. We will consider in detail the ramifications of this result, and possible theoretical explanations to it, in the discussion section (Section 5). 

In the bottom panel of Fig. 9, we show complementary results from an RF analysis to predict actual $\Sigma_{\rm SFR}$ values in star forming regions. The structure of this figure is identical to the one above. For star formation, $\Sigma_*$ is overwhelmingly the most important parameter out of the entire set. Consequently, local (spatially resolved) parameters are almost entirely predictive of ongoing rates of star formation, with a negligible contribution from global and environmental parameters (see the pie chart in the bottom panel of Fig. 9). Therefore, star formation is governed by local phenomena, yet quenching is governed by galaxy-wide global phenomena. Even a casual visual inspection of the two panels in Fig. 9 reveals that the dependence of star formation on these parameters is radically different to that of quenching. Quenching appears to be a much more complex problem than star formation, in that many parameters remain useful in the RF analysis for the former but the latter is reduced to a single parameter problem. 

Ultimately, the RF and ANN analyses yield highly consistent results, but there are a few interesting subtle differences between them. For example, for predicting $\Sigma_{\rm SFR}$ values in star forming regions, the location of the spaxel within the galaxy ($R/R_e$) is the second best variable in the ANN analysis, whereas in the RF analysis it performs notably worse, with an importance close to zero. The reason for this is that $\Sigma_*$ and $R/R_e$ are correlated, given that one finds the highest mass densities at the centre of galaxies. As such, the RF accounts for this correlation by preferentially weighting the most effective of the two parameters, whereas the ANN simply records the accuracy obtainable with each in isolation. This example highlights the main advantage, as well as a possible disadvantage, of the RF approach. In RF we consider the performance of each parameter {\it relative} to the rest of the set (hence our choice of performance label). If one changes the input variables one changes the relative importances for all parameters, not only the new ones. In the ANN analysis this is not the case. The performance of, say, $M_*$ is unrelated to the performance of every other parameter. As a result, there is unique information contained in the rankings of parameters with RF and ANN: the former treats the ranking as a contest, whereas the latter treats each contribution independently. For our data, the rankings are similar between the two approaches, but there is no requirement that this be the case.\\

\noindent {\it GAS-PHASE ANALYSIS:}

\noindent In the preceding part of this sub-section we investigate the full star forming sub-sample of spaxels. The advantage of the above sample is that it is complete. However, the disadvantage is that the majority of $\Sigma_{\rm SFR}$ values are inferred indirectly through the sSFR - D4000 relation (see Section 3.2). In this part, we consider the emission-line BPT star forming sub-sample (i.e. the sample shown by the magenta line in Fig. 5). For this sample all $\Sigma_{\rm SFR}$ values are computed via dust corrected H$\alpha$ flux. This enables us to make an important test on the rankings of the parameters in regression, and also allows us to explore some additional parameters in the gas-phase (which require emission line measurements to compute).

We present the results of our regression analysis for the BPT star forming sub-sample in Fig. 10. In comparison to the full star forming sub-sample (shown in the bottom panel of Fig. 9), most parameters are raked in a very similar manner. Importantly, local parameters are collectively still by far the best group, and $\Sigma_*$ is by far the best individual parameter. This is especially reassuring because in the prior analysis $\Sigma_*$ enters into the calculation of $\Sigma_{\rm SFR}$ via the conversion from sSFR(D4000). The fact that $\Sigma_*$  is still the best parameter in the sample where $\Sigma_{\rm SFR}$ is derived exclusively through H$\alpha$ implies that this calibration cannot be responsible for the high performance of $\Sigma_*$. As an additional check on the regression analysis, in Appendix A we perform an alternative analysis based on the  $\Sigma_{\rm SFR}$ values derived through full spectrum multi-SSP model fitting. The main results are all identical to the two versions presented here in the results sections.

For the high S/N emission line sample (analysed in this part of this sub-section), we can also add spatially resolved parameters related to the gas-phase properties of each spaxel. One parameter which is particularly interesting is gas-phase metallicity, or more specifically the oxygen abundance, defined as: $Z_g = 12 + \log_{10}(O/H)$. Here we construct the oxygen abundance from the $O3N2$ index, via the Marino et al. (2013) calibration for star forming regions within galaxies. Additionally, we include the level of gas extinction in V-band ($A_{\rm V}$), as inferred from the Balmer decrement, assuming a Cardelli et al. 1989 attenuation curve. As noted at the start of this section, the gas mass surface density ($\Sigma_g$) is a particularly important parameter to consider. However, accurate measurements of the spatially resolved gas content of galaxies (e.g. through HI 21cm or CO(1-0) rotational transition measurements, for the atomic and molecular phase, respectively) are not currently available for our large sample of galaxies (or even a significant fraction of them). Nonetheless, in several works (e.g., Barrera-Ballesteros et al. 2018, Concas \& Popesso 2019, Piotrowska et al. 2019) $\Sigma_g$ is found to correlate strongly with $A_{\rm V}$, or (essentially equivalently) the Balmer decrement. As such, we may use $A_{\rm V}$ as a conceptual proxy for $\Sigma_g$, assuming a constant gas-to-dust ratio and that $A_{\rm V}$ traces the dust mass along the line of sight.

In Fig. 10 we add the two gas-phase parameters ($A_{\rm V}$ and  $Z_g$) to the parameters which can be measured in all spaxels (star forming and quenched; emission line and non-emission line). Collectively, the gas-phase parameters perform second best, after the local (non gas-phase) parameters. Perhaps surprisingly (given the Kennicutt-Schmidt relation, e.g., Kennicutt 1998), $A_{\rm V}$ is not found to be the best single parameter for predicting $\Sigma_{\rm SFR}$ values. As with the full star forming sample, $\Sigma_*$  is still (by a very significant margin) the best single parameter. Of course, it is certainly possible that a more accurate and direct measurement of $\Sigma_g$ would yield a better result (indeed see Ellison et al. 2019 for evidence of this in a sample of $\sim$35 MaNGA galaxies observed with ALMA). Nonetheless, it is also possible that there is a significant variation in the efficiency of star formation (SFE $\equiv \Sigma_{\rm SFR} / \Sigma_g$ = $1/\tau_{\rm dep}$) throughout, and between, galaxies in our sample. Unfortunately, our present analysis is not constraining in this regard. This notwithstanding, the gas-phase parameters perform better at predicting $\Sigma_{\rm SFR}$ than all of the global and environmental parameters considered in this work, and all of the local parameters except $\Sigma_*$. This suggests that the local gas-phase properties within galaxies are important for setting ongoing levels of star formation.

Gas-phase metallicity ($Z_g$) performs as the third most predictive variable in our RF regression analysis, although it is far less informative that $\Sigma_*$. This is interesting because for global star formation in galaxies, Ellison et al. (2008) established that $Z_g$ is an important additional parameter to $M_*$ for setting total SFR, which has been confirmed by many other works (see, e.g., the review by Maiolino \& Mannucci 2019). This has become known as the fundamental metallicity relation (FMR). However, there has been much debate about whether there exists an FMR on resolved scales for spaxels (e.g., Barrera-Ballesteros et al. 2018, Maiolino \& Mannucci 2019 and references therein). Our RF analysis in Fig. 10 reveals that resolved star formation is set primarily by $\Sigma_*$, with only a very small (but still significant within the errors) additional effect from $Z_g$. Essentially, this result suggests that there is not a significant FMR on resolved kpc-scales (at least on average in the spaxel data), since the impact of gas-phase metallicity leads to only marginal improvement on the prediction of $\Sigma_{\rm SFR}$ over $\Sigma_*$ alone.

\subsection{Principal Component Analysis: Global vs. Local Test}

\begin{figure*}
\includegraphics[trim = 5mm 0mm 15mm 0mm, clip, width=0.49\textwidth]{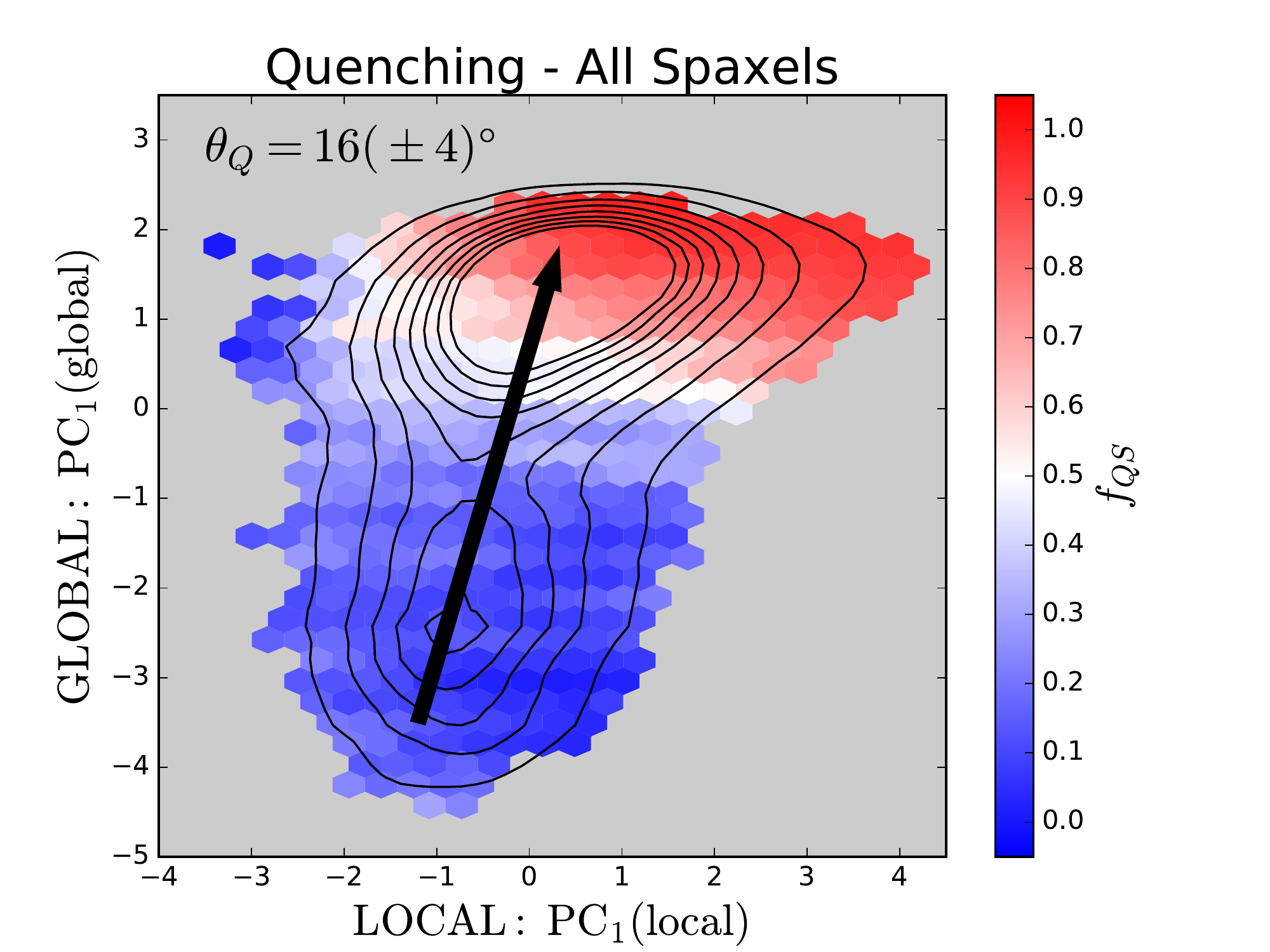}
\includegraphics[trim = 5mm 0mm 12mm 0mm, clip, width=0.49\textwidth]{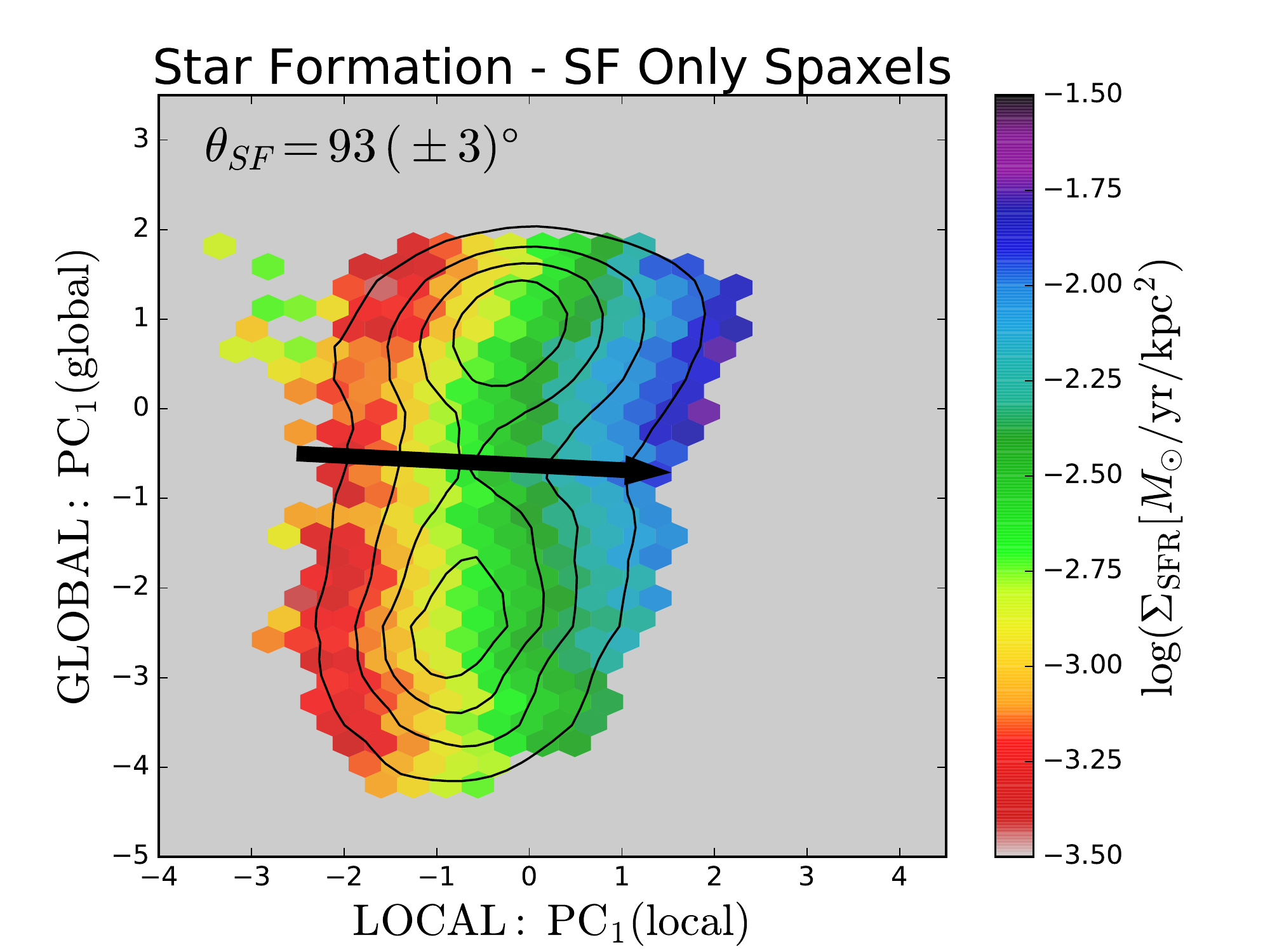}
\caption{Final test of global vs. local for star formation and quenching. Both panels plot the global hyper-parameter (PC$_{1}$(global)) against the local hyper-parameter (PC$_1$(local)), which represent the first principal component for the global and local data sets, respectively. In the left panel, the PC$_1$(global) - PC$_1$(local) plane is colour coded by the fraction of quenched spaxels ($f_{QS}$) in each hexagonal bin. Alternatively, in the right panel, the PC$_1$(global) - PC$_1$(local) plane is colour coded by mean star formation rate surface density ($\Sigma_{\rm SFR}$) in each hexagonal bin. From visual inspection, it is clear that quenching progresses predominantly vertically (i.e. as a function of global parameters); whereas star formation (probed in star forming regions) progresses predominantly horizontally (i.e. as a function of local parameters). We quantify this effect using the axis ratio statistic (see eqs. 16 \& 17). $\theta_Q$ shows the angle (clockwise from vertical) with the steepest increase in $f_{QS}$. Conversely, $\theta_{SF}$ shows the angle with the steepest increase in $\Sigma_{\rm SFR}$. There is a pronounced rotation from global to local dependence as we change from $\theta_Q$ to $\theta_{SF}$. Thus, {\it quenching is a global process but star formation is a local process}.}
\end{figure*}

The main conclusion from the machine learning analysis of the preceding section is that quenching is governed by global processes, yet star formation is governed by local processes. This is an important result because it sets the scale of interest for the two processes. In other words, in order to understand quenching one must look for global galaxy-wide physics, particularly with an origin associated with the centre of the galaxy. On the other hand, to understand ongoing star formation one must look for local spatially resolved physics, occurring within each region of a star forming galaxy. Of course, it is possible that we are missing some critical parameters which may ultimately change this conclusion. Nonetheless, the high level of classification and regression accuracy found by both the ANN and RF analyses suggest that we are capturing at least the majority of the information relevant for both star formation and quenching (from kpc to Mpc scales). 

In this section, we consider an alternative method to compare the importance of local and global parameters to quenching and star formation, which does not rely on machine learning. To this end we utilise a principal component analysis (PCA) technique. Specifically, we construct a local hyper-parameter (PC$_{1}$(local)) from the set of all local (spatially resolved) parameters for centrals: $\sigma$, $\Sigma_{*}$, $Z_{*}$ and $R/R_e$; and a global hyper-parameter (PC$_{1}$(global)) from the set of all global parameters for centrals, which are not trivially connected to star formation: $\sigma_c$, $\Sigma_*^c$, $M_*$, and $B/T$. First, we rescale each parameter by subtracting off the mean value, and normalising by the standard deviation. Thus, the data are first converted to a scale-free unitless form. We then apply a standard eigenvector/ eigenvalue decomposition of the combined multi-dimensional data cubes. We take as our hyper-parameter for each grouping the 1st principal component, which in all cases contains the highest variance of the full set of hyper-parameters. The local hyper-parameter acts as a proxy for the full local dataset, and the global hyper-parameter acts as a proxy for the full global dataset. Compressing the dimensionality of our data in this manner is useful for visualising the impact of each set of parameters on star formation and quenching, as well as for exploring how important local parameters are collectively at fixed global parameters, and vice versa, without relying on (arguably more obscure) machine learning techniques.

In Fig. 11, we present the PC$_{1}$(global) - PC$_{1}$(local) relationship, as indicated by density contours (shown as black lines), for the full central galaxy spaxel distribution (left panel) and the star forming only sub-sample (right panel). Before moving further, it is interesting to note that the global hyper-parameter distribution is highly bimodal for both samples, whereas the local hyper-parameter distribution is uni-modal for both samples. This is a direct consequence of several of the global parameters having bimodal distributions or pronounced power-law tails ($B/T$, $\sigma_c$, $\Sigma_*^c$), whereas none of the local parameters exhibit bimodality. We separate each panel in Fig. 11 into small hexagonal regions and display the fraction of quenched spaxels ($f_{QS}$, left) and the mean $\Sigma_{\rm SFR}$ (right) as the colour of each region, as labelled by the colour bars. Viewing Fig. 11 by eye, it is clear that quenching progresses predominantly vertically across the diagram, indicating that quenching is governed primarily by global variables in our data. Conversely, the rate of star formation in star forming regions progresses predominantly horizontally across the diagram, indicating that star formation is governed by local variables in our data. This is qualitatively in line with our conclusions from the RF analysis (above).

We construct a new statistic to quantify the dependence of quenching (or star formation) on two parameters simultaneously. We begin with the partial correlation coefficient (PCC), which quantifies the strength of correlation between two variables at a fixed third variable (see Bluck et al. 2019, Bait et al. 2017). Specifically, the PCC is defined as:

\begin{equation}
\rho_{AB,C} = \frac{\rho_{AB} - \rho_{AC} \cdot \rho_{BC}}{\sqrt{1 - \rho_{AC}^2} \sqrt{1 - \rho_{BC}^2}}
\end{equation}

\noindent where, e.g., $\rho_{AB}$ indicates the Spearman rank correlation coefficient between variables $A$ and $B$. The Spearman rank correlation expands on the Pearson correlation statistic by first rank ordering parameters, which essentially relaxes the assumption of linearity in favour of the more mild assumption of monotonicity (which is clearly supported by this data). Next, we utilise the PCC statistics of the colour bar variable in Fig. 11 with each of the x- and y-axis variables to construct the optimal route through each diagram to maximise either quenching (left panel) or star formation (right panel). We then construct the quenching axis by treating the two PCC values as components of a vector, explicitly calculating:

\begin{equation}
\theta_{Q} = {\rm tan}^{-1} \bigg( \frac{\rho_{Y f_{QS}, X}}{\rho_{X f_{QS}, Y}} \bigg)
\end{equation}

\noindent and

\begin{equation}
\theta_{SF} = {\rm tan}^{-1} \bigg( \frac{\rho_{Y \Sigma_{\rm SFR}, X}}{\rho_{X \Sigma_{\rm SFR}, Y}} \bigg)
\end{equation}

\noindent where, e.g., $\rho_{Y f_{QS}, X}$ indicates the partial correlation of $f_{QS}$ with the y-axis variable, at fixed values of the x-axis variable. Thus, $\theta_{Q}$ gives the angle clockwise from vertical (aligned with the y-axis) which indicates the optimal direction through the two dimensional plane for maximising the fraction of quenched spaxels. If  $\theta_{Q}$ = 90$^{\circ}$ this indicates that the quenched fraction is entirely correlated with the x-axis parameter, and entirely uncorrelated with the y-axis parameter. Conversely, if $\theta_{Q}$ = 0$^{\circ}$ this indicates that the quenched fraction is entirely correlated with the y-axis parameter, and entirely uncorrelated with the x-axis parameter. A value of $\theta_{Q}$ = 45$^{\circ}$ indicates an even split in correlation between the x- and y-axes, with quenched fraction depending equally on both axes. A negative value of $\theta_{Q}$ indicates that the x-axis variable is negatively correlated with the quenched fraction, at fixed values of the y-axis variable; and a value of $\theta_{Q} > 90^\circ$ indicates that the y-axis variable is negatively correlated with the quenched fraction at fixed values of the x-axis variable. As such, the quenching axis acts like a compass which points in the direction of maximal quenching.

In exact analogy to $\theta_{Q}$, $\theta_{SF}$  acts as a compass which points in the direction of maximal increase in $\Sigma_{\rm SFR}$. Note that if the two processes were equivalent (i.e. if quenching was a result solely of a lack of star formation drivers in these data), one would expect $\theta_{Q} = \theta_{SF} - 180^{\circ}$ (accounting for quenching and increased star formation being opposites). In both cases, errors are determined from a Monte Carlo (MC) simulation, varying each parameter within its respective errors and recomputing the correlation, partial correlation, and quenching axis statistics in each case. The 1$\sigma$ error is then taken as the variance across the multiple MC runs.

We find that $\theta_{Q} = 16(\pm 4)^{\circ}$ and $\theta_{SF} = 93(\pm 3)^{\circ}$. Thus, star formation is governed almost entirely by local processes, i.e. it is a function solely of local variables in our data. Conversely, quenching is much more dependent on global parameters than local parameters, although there is a weak correlation between quenching and local parameters, even at fixed global values. These results are in precise accord with our prior machine learning analysis, and hence lend further confidence to our overarching conclusion: that quenching is a global galaxy-wide process, yet star formation is a locally dependent phenomenon. Finally, it is important to highlight that the difference $\theta_{SF}$ - $\theta_{Q}$ (= 77$^{\circ}$) is significantly offset from the expectation value of 180$^{\circ}$ (for the simple assumption of quenching merely being an extension of the star formation process). Therefore, we conclude that star formation and quenching are distinct physical processes, requiring different physical explanations which must operate on very different physical scales. As noted before, this conclusion is, of course, subject to change with the inclusion of new parameters. However, the high degree of accuracy with which we can predict star formation rates and the presence of quenching with these data strongly suggests that the addition of new parameters is unlikely to significantly change our conclusions.

\subsection{Further Insights from Parameter Profiles}

\begin{figure*}
\includegraphics[width=0.49\textwidth]{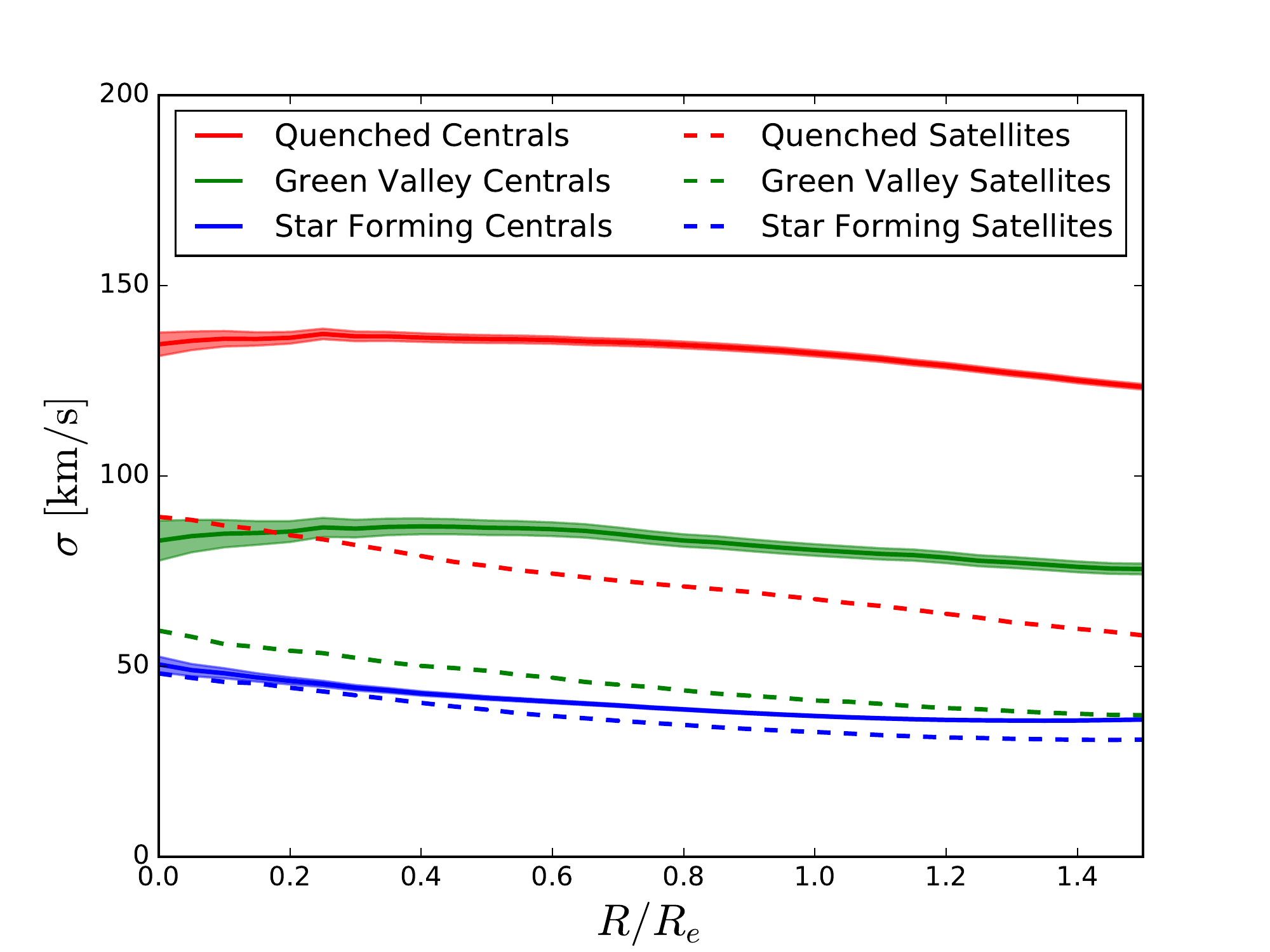}
\includegraphics[width=0.49\textwidth]{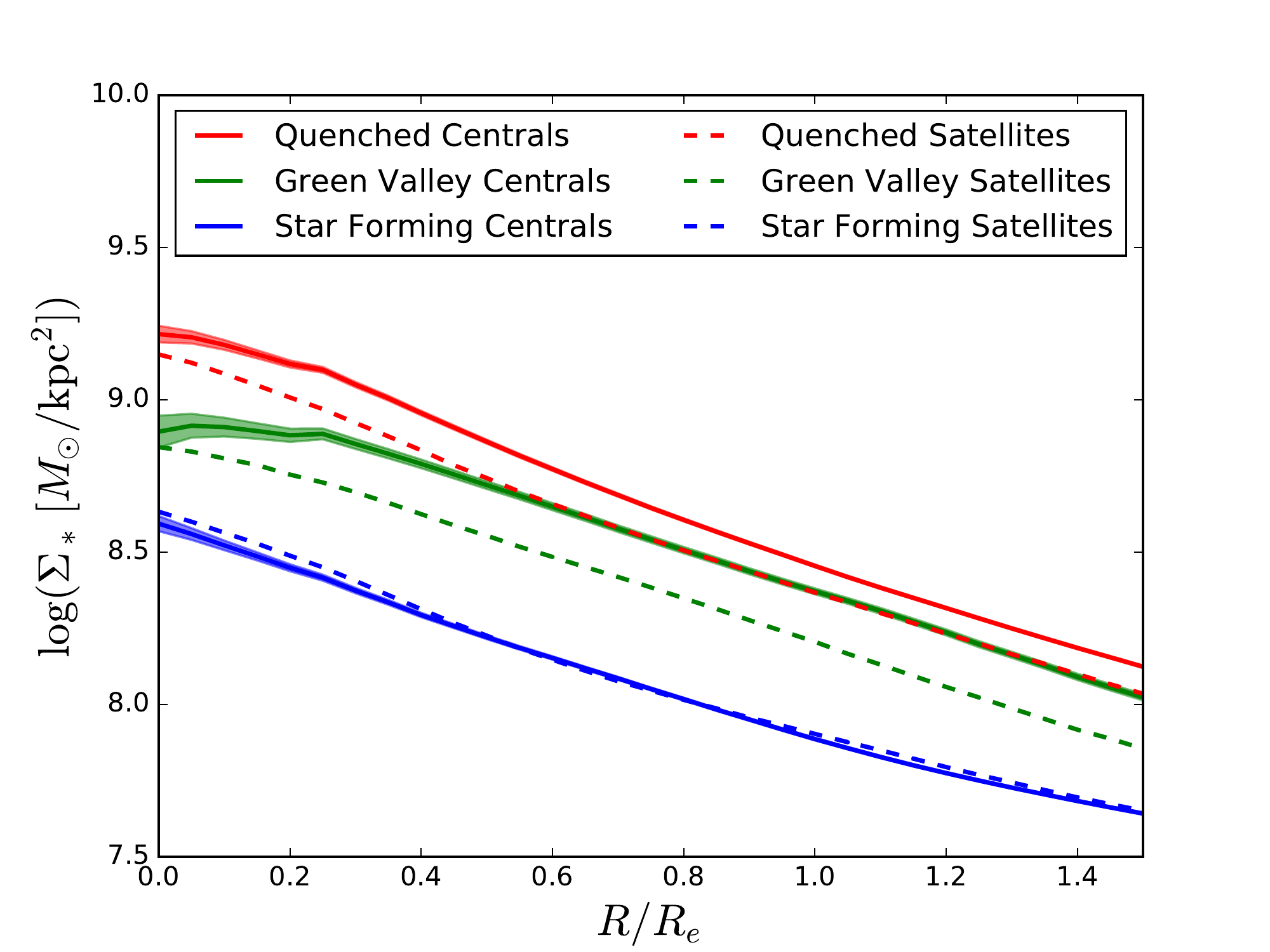}
\caption{Velocity dispersion profiles (left panel) and stellar mass surface density profiles (right panel), shown separately for quenched (red), green valley (green) and star forming (blue) spaxels (as defined in Fig. 5). The profiles are shown separately for centrals (solid lines with shaded regions) and satellites (dashed lines). Spaxels are selected only from galaxies which are well resolved (with $R_e > 2 \times$FWHM, 81\% of the spaxel sample). The velocity dispersion profiles are relatively flat, whereas the mass surface density profiles steeply decline with radius. Interestingly, the offset in velocity dispersion between quenched and star forming centrals is far larger than the offset between quenched and star forming satellites. For stellar mass surface density, the differences between centrals and satellites are smaller. }
\end{figure*}

In Sections 4.2.2 \& 4.2.3, we find that $\sigma_c$ (evaluated within 1 kpc) is the best performing single parameter for classifying spaxels into `star forming' or `quenched' categories in central galaxies. Curiously, $\sigma$ measured at the spaxel location also performs very well, and is in fact only slightly less constraining than $\sigma_c$ in the ANN analysis. In the RF analysis, we do see that  $\sigma_c$ is superior to $\sigma$, yet the latter still performs very well (and is indeed the second most useful parameter in the RF). Alternatively, $\Sigma^c_*$ (evaluated within 1kpc) is much more constraining of quenching than $\Sigma_*$ at the spaxel location (in both the ANN and RF approach). In both cases the central value is (significantly) more constraining than the value at the spaxel location, but for $\Sigma_*$ this difference is much larger than for $\sigma$. In this sub-section, we look at the mean profiles of $\sigma$ and $\Sigma_*$ to gain further insight into the results of the machine learning analyses.

In Fig. 12 we show the mean profiles for $\sigma$ (left panel) and $\Sigma_*$ (right panel), each plot as a function of the elliptical semi-major axis, normalised by the effective radius of the galaxy from which it is drawn (measured in the r-band SDSS image). Results are shown for centrals (solid lines) and satellites (dashed lines, for comparison), and each panel is subdivided into `quenched', `green valley' and `star forming' subsamples, based on the cuts in $\Delta$SFR$_{\rm res}$ (shown in Fig. 5). For this analysis we select only well resolved galaxies, with effective radii greater than twice the full width half maximum (FWHM) of the point spread function (PSF) on the IFU reconstructed images (81\% of the spaxel data).

All of the $\sigma$ profiles are very flat compared to the equivalent $\Sigma_*$ profiles, which all exhibit a steep decline with radius. This general trend is similar in the individual galaxy profiles as well, but there is much more diversity in shape. However, for the machine learning analysis it is the average properties of the spaxel data which matters, since the network is not given the ability to determine which spaxels come from which galaxy. For centrals, there is a very large separation in average $\sigma$ between the star forming classes. Quenched spaxels reside in galaxies with very high $\sigma$, whereas star forming spaxels reside in galaxies with very low $\sigma$. Green valley spaxels are drawn from galaxies with intermediate $\sigma$, on average. These differences are clear throughout the galaxy, up to 1.5 times the effective radius (beyond this distance MaNGA has sparse coverage and hence we cannot draw robust conclusions). The flatness of the $\sigma$ profiles further implies that knowing $\sigma$ in {\it any} observed region allows one to estimate $\sigma_c$ (within 1 kpc), with a high degree of accuracy on average.

For centrals, varying the star forming state of spaxels impacts the $\Sigma_*$ profiles in a similar manner to the $\sigma$ profiles. Quenched spaxels reside in regions with higher $\Sigma_*$ on average than star forming spaxels, with green valley spaxels having intermediate values. However, the magnitude of the difference is much less for $\Sigma_*$ than for $\sigma$. Moreover, the steep gradient in $\Sigma_*$ as a function of radius implies that one cannot infer the central $\Sigma^c_*$ (within 1 kpc) simply from knowing the values of $\Sigma_*$ at a random location within the galaxy. Thus, there is a high degree of variability within galaxies with respect to $\Sigma_*$, and a much smaller variability with respect to $\sigma$. Ultimately, the extent of variation in $\Sigma_*$ as a function of radius is greater than the variation in $\Sigma_*$ as a function of star forming state. Essentially, this is the opposite result as to $\sigma$. This fact explains why the ANN is similarly constraining in its predictive power with $\sigma$ and $\sigma_c$, but is very different in its predictive power for $\Sigma_*$ and $\Sigma^c_*$. The RF analysis, however, is more capable of separating the correlated variables, revealing that it is indeed the central $\sigma$ which matters most for quenching. The fact that the central values of each of these parameters is superior to the local value suggests that quenching is governed by processes originating at the centre of galaxies. Local parameters are only effective at predicting quenching if they correlate strongly with the central region. This is the case for $\sigma$, but not for $\Sigma_*$.

For satellites, variation in the star forming state affects $\sigma$ far less than for centrals. Hence, spaxels from quenched satellites have significantly lower values of $\sigma$, both at the spaxel location and in the central regions of their galaxies, than quenched centrals. This result is interesting because it suggests that $\sigma_c$ (and by extension $\sigma$ in the spaxel) is deeply connected to quenching in centrals, but much less so for satellites. For $\Sigma_*$, the variation with star forming class is similar between centrals and satellites, comparable in magnitude to the satellite case with $\sigma$, but much less pronounced than for $\sigma$ in central galaxies. There is a natural interpretation of this result through evoking the $M_{BH} - \sigma_c$ relationship (e.g., Ferrarese \& Merritt 2000, Gebhardt et al. 2000, Saglia et al. 2016) which we present in full in the Discussion section. 

In summary of this sub-section, $\sigma$ varies far more with star forming state than with location within the galaxy (up to 1.5 $R_e$), which explains why $\sigma$ and $\sigma_c$ perform so similarly in their classification of spaxels. Conversely, $\Sigma_*$ varies significantly more as a function of radius than as a function of star forming state, which explains why the predictive power of $\Sigma_*$ and $\Sigma^c_*$ differ significantly in predicting star forming classes. For both $\Sigma_*$ and $\sigma$, it is the value of each parameter at the centre of the galaxy which is most effective at predicting quenching (see Fig. 9). This fact is simply more obvious for $\Sigma_*$ for the reasons outlined above. Thus, as concluded in the previous three sub-sections, quenching is governed by global parameters, particularly those associated with the central most regions of central galaxies.

%
%   DISCUSSION
%

\section{Discussion - What Quenches Central Galaxies?}

Before discussing in detail which physical mechanisms for quenching galaxies are plausible in light of our machine learning results and rankings, we first pause briefly to consider what we have learned in terms of star formation. For star forming regions within galaxies, $\Sigma_{\rm SFR}$ is most accurately constrained by the stellar mass surface density of the region in question. Moreover, we find that local (spatially resolved) parameters collectively perform far better than global or environmental parameters in concert at estimating $\Sigma_{\rm SFR}$ within a given region of a star forming galaxy. Thus, star formation is a local process. Another way to put this is that $\Sigma_{\rm SFR}$ varies significantly within a given galaxy, in such a way that it is highly correlated with other spatially resolved parameters, but poorly constrained by global or environmental parameters. This result is fascinating because it clearly demonstrates that a global (galaxy-wide) view of star formation is necessarily incomplete. Furthermore, from the local nature of star formation in galaxies we conclude that the observed global trends with star formation, particularly the star forming main sequence (e.g. Brinchmann et al. 2004), is an emergent phenomenon arising from more fundamental local processes (as also argued for in Cano Diaz et al. 2016). 

It would be entirely reasonable to conjecture that the same emergence of global correlation from local processes may be true for quenching as well as for star formation. However, in this work we demonstrate that this is emphatically not the case. Despite the considerably greater information held in the millions of spaxels with unique local information, global parameters collectively predict whether spaxels will be star forming or quenched with significantly higher accuracy than local or environmental parameters. This result is clearly seen in an ANN analysis (see Fig. 8), an RF analysis (see Fig. 9) and in a PCA analysis (see Fig. 11).  For central galaxies, the best single parameter (not trivially connected to star formation) for predicting whether a region will be star forming or quenched is central velocity dispersion (see Fig. 9). This fact is remarkable for two reasons: 1) no parameter at the location of the spaxel is as effective at predicting quenching as a parameter measured at the centre of the galaxy; and 2) for the parameters under investigation in this work, a single number reflecting the central kinematics of a galaxy is superior to thousands of local measurements for constraining quenching in central galaxies. Consequently, we conclude that quenching is irreducibly a global process, at least in terms of the parameters we have studied in this work and for the galaxy sample under investigation here. As such, quenching affects galaxies as a whole, and is best modelled as a function of global parameters. This result immediately suggests that a stifling of gas supply into galaxies (i.e. `starvation' or `strangulation') may be the dominant quenching mode for centrals, since this mechanisms has the capacity to engender galaxy-wide quenching (see also Peng et al. 2015 \& Trussler et al. 2019 for very similar conclusions via independent reasoning).

The success of central velocity dispersion in predicting the star forming state of central galaxies is in agreement with previous work (especially Wake et al. 2012, Teimoorinia, Bluck \& Ellison 2016, Bluck et al. 2016). However, crucially, these prior studies did not have access to resolved star formation rates in galaxies (only the global total), and furthermore did not have spatially resolved parameters to model quenching with. As such, we significantly expand on these prior works by investigating quenching on a resolved, as well as global and environmental, scale. Nonetheless, it is highly reassuring that central velocity dispersion is also found to out-predict other global and environmental parameters in previous studies, albeit without knowledge of the spatially resolved dimension.

In the remainder of this discussion we consider a number of possible quenching mechanisms, and carefully assess whether or not they are consistent with our observational results.

\subsection{The Heating Solution: Three Quenching Paradigms}

\begin{figure*}
\includegraphics[width=0.49\textwidth]{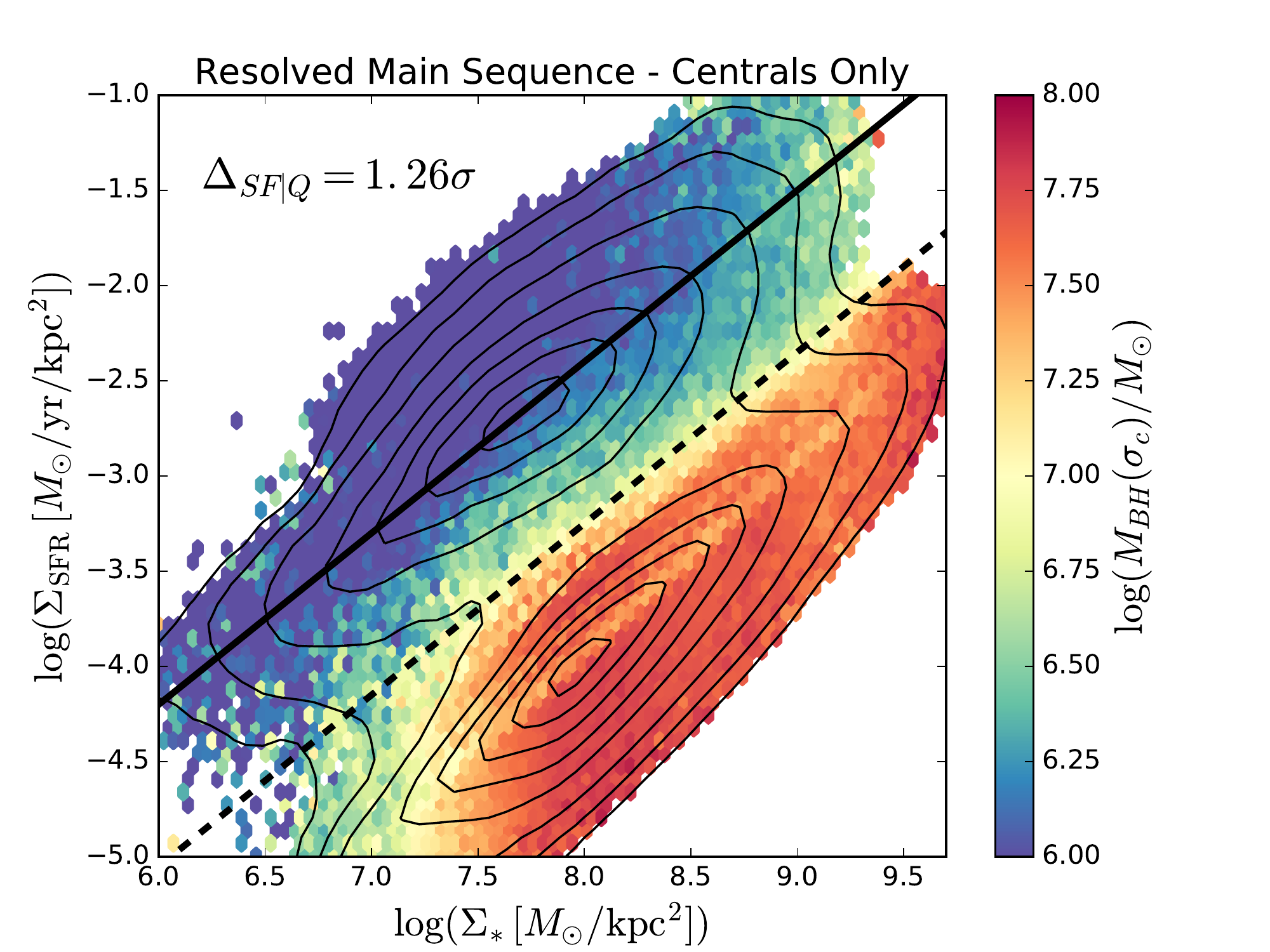}
\includegraphics[width=0.49\textwidth]{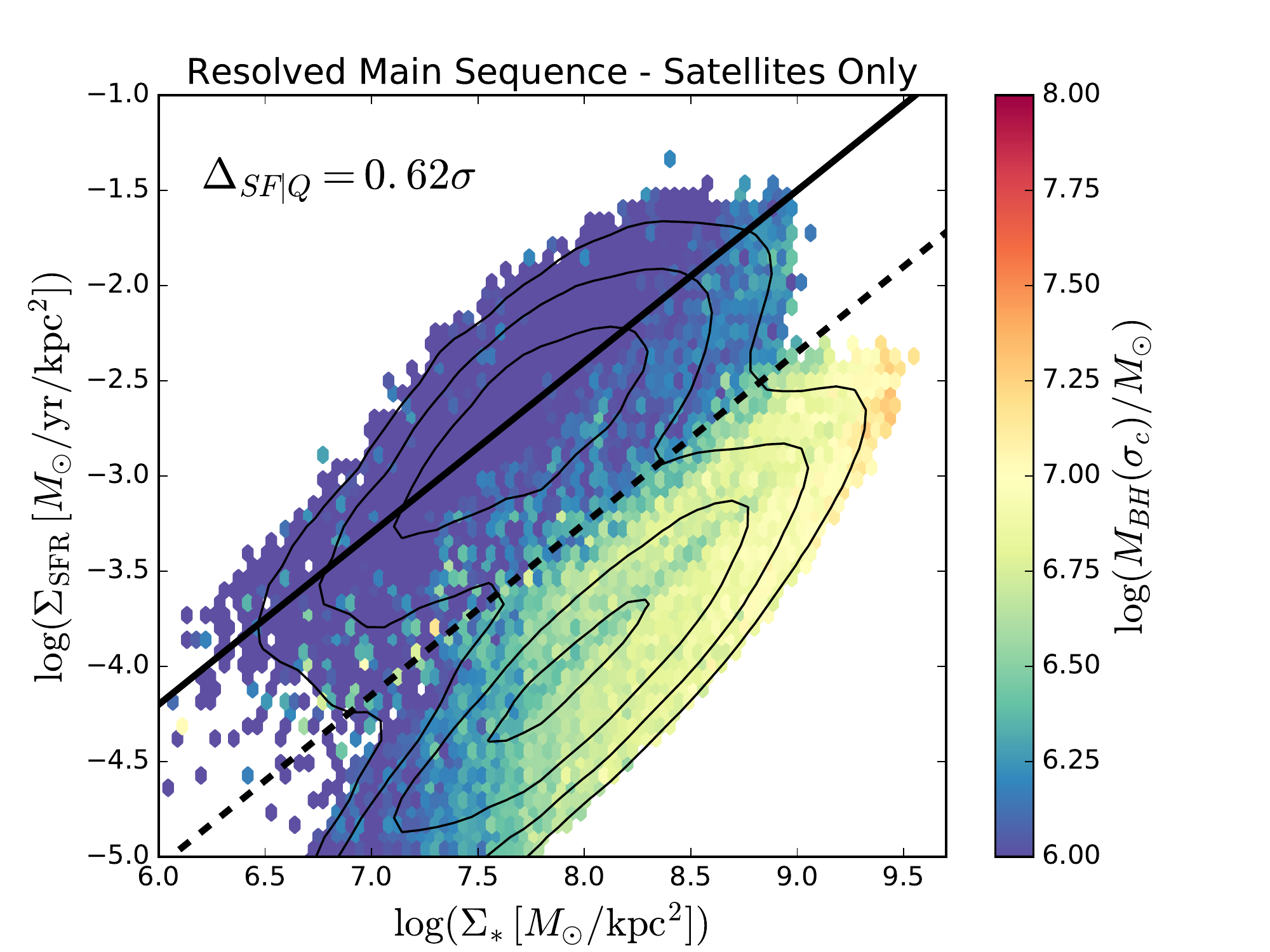}
\caption{The resolved star forming main sequenced colour coded by estimated supermassive black hole mass for central galaxies (left panel) and satellite galaxies (right panel). Black hole mass is inferred from the $M_{BH} - \sigma_c$ relationship for all galaxy types taken from Saglia et al. (2016). For central galaxies, it is clear that quenched spaxels reside preferentially in galaxies hosting significantly more massive black holes than the host galaxies of star forming spaxels. This result is much weaker for satellite galaxies. We quantify the variation in average host $M_{BH}$ between star forming and quenched regions using the $\Delta_{\rm SF|Q}$ statistic (defined in eq. 12). }
\end{figure*}

As discussed in the Introduction, in order for a galaxy to maintain quiescence over a sustained period of time, cooling of the hot gas halo must be prevented. In turn, this shuts down the inflow of cold gas into a galaxy, which is needed as fuel for further star formation. The most obvious solution to this cooling problem is via heating. In the heating solution, energy is input into the system in order to offset the expected cooling of the hot gas halo. The source of the energy required to offset cooling will reveal the physical process(es) which give rise to quenching in the heating scenario. In this discussion, we consider three broad possibilities for the source of heating and hence central galaxy quenching: 1) AGN feedback; 2) stellar and supernova feedback; 3) gravitationally driven virial shocks. In selecting these three possible avenues for quenching we have drawn upon a considerable body of prior research (including, Croton et al. 2006, Dekel \& Birnboim 2006, Bower et al. 2006, 2008, Henriques et al. 2015; Vogelsberger et al 2014a,b, Schaye et al. 2015, Bluck et al. 2014, 2016). Although there are other possibilities for the source of the energy needed to quench galaxies (e.g. magnetic fields and cosmic rays), they have received much less attention to date in the literature due to severe limitations involved in accurately modelling these potential solutions.

In the following sub-sections of this Discussion we explore in detail the $M_{BH}$ - $M_*$ - $M_{\rm Halo}$ parameter space for central galaxies in MaNGA and in the SDSS\footnote{Note that black hole mass and halo mass are on the same conceptual footing since they are both inferred indirectly through calibrations with observational data. Halo mass is inferred from abundance matching, applied to the total stellar mass of the group or cluster. Black hole mass is inferred from the $M_{BH} - \sigma_c$ relation. Even stellar mass (one might be tempted to argue) is ultimately inferred from SED model fitting to optical waveband measurements. The final uncertainties on these three measurements are actually quite comparable ($\sim$ 0.2 - 0.5 dex on average). We consider in detail in Appendix A2 whether differential measurement errors could drive any of our results (the answer is no).}. To aid in interpreting the following results, in Appendix B we construct a simple analytic argument to link these three observable parameters to theoretically motivated sources of late-time heating and quenching in central galaxies. Our main conclusion from Appendix B is that $M_{BH}$ traces the total integrated energy released from AGN feedback over the lifetime of a galaxy; $M_*$ traces the total integrated energy released from supernova (Type I \& II) feedback over the lifetime of a galaxy; and $M_{\rm Halo}$ traces the total integrated energy released from virial shocks into the hot gas halo. 

In our detailed analytic derivation, we find a direct proportionality between black hole mass and total integrated AGN feedback energy (as originally derived in Soltan et al. 1982), and an analogous direct proportionality between stellar mass and integrated supernovae energy (e.g. Henriques et al. 2019). For virial shocks, we find a power-law dependence of energy on halo mass, with an exponent: $\beta$ $\sim$ 1.5 (broadly consistent with Dekel \& Birnboim 2006 and Dekel et al. 2019). Consequently, quenching scaling primarily with $M_{BH}$ indicates AGN feedback; quenching scaling primarily with  $M_*$ indicates supernova and stellar feedback; and quenching scaling primarily with $M_{\rm Halo}$ indicates virial shock heating. The above is exactly as one might intuit, but it is still highly useful to make this case explicit.

\begin{figure*}
\includegraphics[width=1\textwidth]{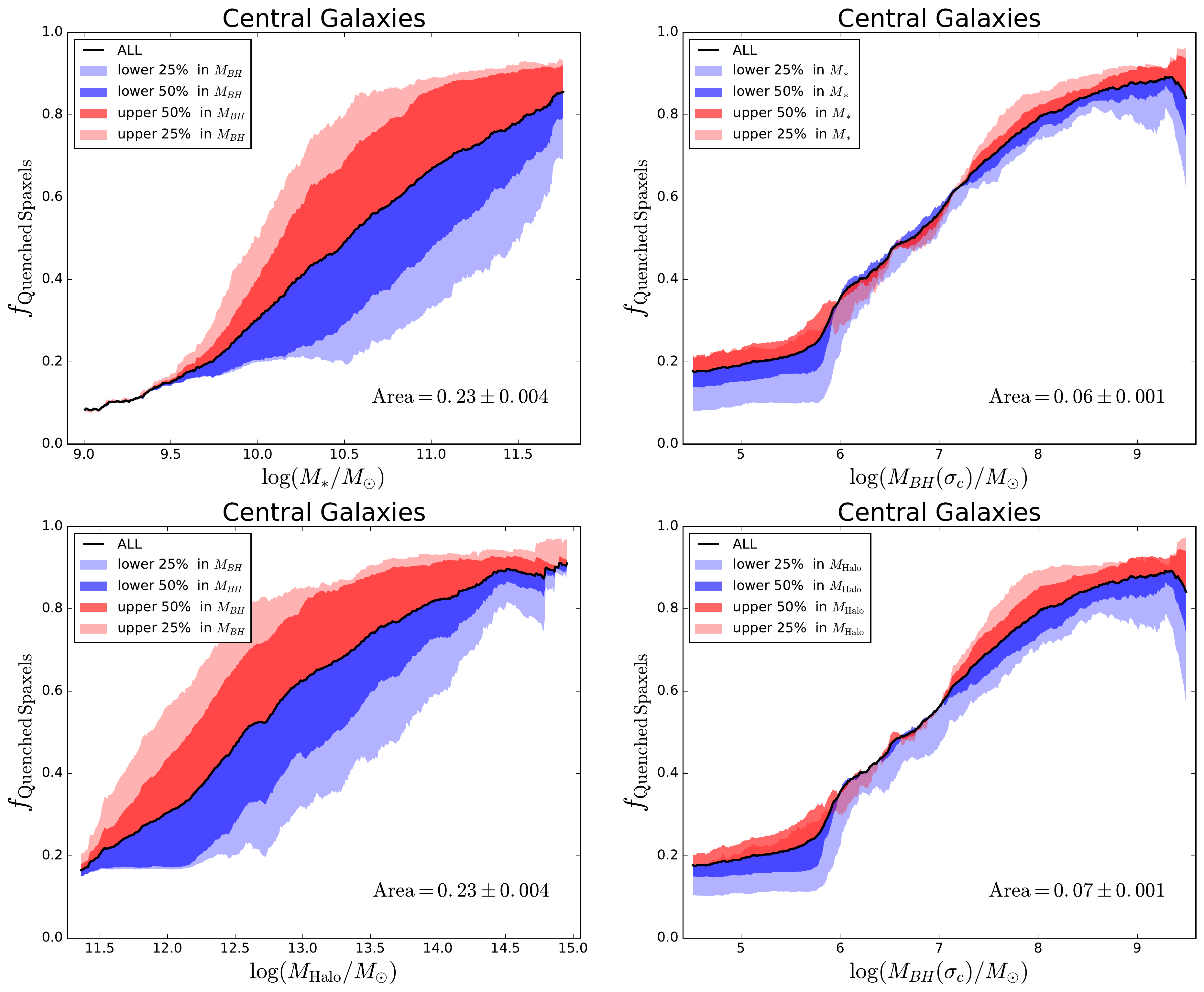}
\caption{The fraction of quenched spaxels in central galaxies plot as a function of: a) stellar mass split into ranges of black hole mass (top left panel); b) black hole mass split into ranges of stellar mass; c) group dark matter halo mass split into ranges of black hole mass; and d) black hole mass split into ranges of halo mass. The panels in this figure should be compared row-wise. It is clear that varying black hole mass at a fixed stellar or halo mass engenders a much greater impact on the fraction of quenched spaxels than varying stellar or halo mass at a fixed black hole mass. We quantify this effect by measuring the area contained by the upper 50 percentile to the lower 50 percentile (defined in eq. 19). We find that varying black hole mass at a fixed stellar or halo mass impacts the quenched fraction by greater than a factor a three times more than varying stellar or halo mass at a fixed black holes mass. [Note: black hole masses are inferred from the $M_{BH} - \sigma_c$ relationship (Saglia et al. 2016), and halo masses are inferred from an abundance matching technique applied to the total stellar mass of the group or cluster (Yang et al. 2009).] }
\end{figure*}

Following our rationale in Bluck et al. (2016), we estimate supermassive black hole mass from central velocity dispersion via the tight $M_{BH} - \sigma_c$ relation (e.g. Ferrarese \& Merritt 2000, McConnell \& Ma 2013, Saglia et al. 2016). In order to estimate supermassive black hole masses for our MaNGA sample, we utilise the relationship from Saglia et al. (2016) for all galaxy types, specifically computing:

\begin{equation}
\log(M_{BH}[M_{\odot}]) = 5.25 \times \log(\sigma_c[\mathrm{km/s}]) - 3.77
\end{equation}

\noindent which gives a formal scatter against 96 dynamically measured black hole masses of 0.46 dex. Detailed investigations in the appendices of Bluck et al. (2016) demonstrate that varying the functional form of the $M_{BH} - \sigma_c$ relation for early and late types, or pseudo and classical bulges, does not significantly affect the results or conclusions of the following analyses. It is important to note that these tests were performed on the SDSS DR7, which is the parent sample for MaNGA.

The main advantage of re-phrasing our results with $\sigma_c$ to estimated black hole mass ($M_{BH}(\sigma_c)$) is this allows us to link our observed results much more clearly to theoretical predictions (as explored in detail in Appendix B). For example, recent work from Illustris (Bluck et al. 2016), Eagle (Davies et al. 2019) and Illustris-TNG (Terrazas et al. 2019) make the direct prediction that supermassive black hole mass is the key observable regulating quenching in central galaxies. The reason for this is that black hole mass traces the total integrated feedback energy released by AGN in these simulations (exactly as we analytically derive in Appendix B from highly general theoretical arguments). In order to test this prediction, it is essential to estimate black hole masses for large samples of galaxies, which necessitates the use of a proxy. Central velocity dispersion is perhaps the best known correlator to central black hole mass, and more importantly is frequently found to be the tightest correlator, out of commonly available observables (e.g. Saglia et al. 2016 and references therein), which justifies our chosen approach. None of our conclusions are sensitive to the exact form of the $M_{BH} - \sigma_c$ relation used, or whether or not different relationships are used for different galaxy populations.

In Fig. 13 we present the full resolved main sequence relationship colour coded by black hole mass estimated from central velocity dispersion, for centrals (left panel) and satellites (right panel). For centrals, we find that the mean black hole mass varies significantly between the star forming and quenched density peaks in the resolved main sequence. Hence, quenched spaxels are much more likely to be drawn from host galaxies with higher mass black holes than star forming spaxels (in line with key predictions from cosmological simulations; see Bluck et al. 2016, Davies et al. 2019, Terrazas et al. 2019). This segregation in black hole mass is larger than for any other local, environmental or global parameter considered in this work for central galaxies (which is a direct consequence of the high performance of $\sigma_c$ in Section 4). For satellite galaxies (right panel of Fig. 13), the variation in black hole mass between star forming and quenched spaxels is much less pronounced than for centrals. Thus, it is highly unlikely that black hole mass can be the primary driver of quenching in satellites. Ultimately, the difference between centrals and satellites in terms of their spaxels' dependence on black hole mass for quenching suggests that different mechanisms must be sought to account for quenching in the two cases (as argued for previously in Peng et al. 2010, 2012, Woo et al. 2013, Bluck et al. 2014, 2016). We will explore in detail the quenching of satellites in MaNGA in an upcoming publication (Bluck et al. in prep.).

\subsection{Central Galaxy Quenching in the $M_{BH} - M_* - M_{\rm Halo}$ Parameter Space}

\subsubsection{Fraction of Quenched Spaxels Analysis}

The rankings shown in the machine learning sections are based on the absolute number of correct classifications in ANN (Fig. 8), and the usefulness of each parameter in conjunction with the other parameters in an RF analysis (Fig. 9). Yet, perhaps even more interesting is the question of how varying one parameter at fixed other parameters affects the probability of a region within a galaxy being quenched. To show this, in Fig. 14 we present the fraction of quenched spaxels as a function of stellar mass, halo mass and black hole mass (derived and motivated in the previous sub-section). Critically, we subdivide each parameter into ranges of the other parameter, exposing the impact on resolved quenching of varying one parameter at a fixed value of another parameter. This plot was first devised in Bluck et al. (2016), where the same parameters were investigated in terms of the global quenching of galaxies, i.e. the fraction of quenched galaxies replaces the fraction of quenched spaxels shown here. Whilst we may expect the performance of these parameters to be similar to the global analysis, this is not guaranteed given the potential for sub-galactic complexity in terms of star formation and quenching.

In Fig. 14 we show (from left to right, top to bottom) the fraction of quenched spaxels as a function of: a) stellar mass subdivided by black hole mass, b) black hole mass subdivided by stellar mass, c) halo mass subdivided by black hole mass, and d) black hole mass subdivided by halo mass. Comparing first the top row, we see clearly that varying black hole mass at a fixed stellar mass leads to a much greater impact on the fraction of quenched spaxels than varying stellar mass at a fixed black hole mass. Similarly, in the bottom row, varying black hole mass at a fixed halo mass leads to a much greater impact on the fraction of quenched spaxels than varying halo mass at a fixed black hole mass. Thus, black hole mass is more fundamentally linked to the quenching of centrals than either stellar or halo mass. This is the same conclusion as in Bluck et al. (2016) but now for a resolved analysis of star forming and quenched regions within galaxies.

To quantify the above effect, we modify the novel area statistic of Bluck et al. (2014, 2016) to apply to the fraction of quenched spaxels, as opposed to quenched galaxies. We define the spatially resolved area statistic as:

\begin{equation}
{\rm Area} = \frac{1}{\Delta \alpha} \int{ \big( f_{QS}(\alpha|_{\beta_{\rm upp}}) - f_{QS}(\alpha|_{\beta_{\rm low}}) \big) d\alpha }
\end{equation}

\noindent where $f_{QS}$ indicates the fraction of quenched spaxels at the x-axis parameter $\alpha$. This fraction is evaluated for a fixed range in parameter $\beta$, where in this work we take $\beta_{\rm upp}$ as the top 50\% in parameter $\beta$ and $\beta_{\rm low}$ as the bottom 50\% in parameter $\beta$. To give an example, in the top left panel of Fig. 14, $\alpha = M_{*}$ and $\beta = M_{BH}$, the top right panel switches these around. Higher values of the area statistic indicate a greater impact on quenching of varying the $\beta$-parameter at fixed values of the $\alpha$-parameter. 

The advantage of sorting and using percentile ranges (as opposed to using a fixed binning) is twofold. First, given that in general the parameters under investigation are inter-correlated with each other, a fixed binning would necessarily only extend for part of the range of the x-axis. Second, splitting into a percentile range allows the relative importance of parameter $\beta$ at fixed $\alpha$ throughout each individual value of $\alpha$. The error on the area statistic is taken as the Poisson counting error, and we have checked that this gives reasonable values from a full Monte Carlo simulation. See Bluck et al. (2016) for more details and tests on this statistical method, particularly as applied to galaxy quenching.

Using the area statistic to quantify the impact of each parameter at fixed values of the other parameters, we find that varying black hole mass at a fixed stellar mass engenders more than three times the variation in the quenched fraction of spaxels than the other way around (top panels of Fig. 14). We find a similar result for black hole mass at a fixed group halo mass as well (bottom panels of Fig. 14). Thus, a difference of a few per cent in classification accuracy (in Section 4) leads to a profound difference in the constraining power at fixed values of the other parameters. As such, we conclude that black hole mass (as inferred from central velocity dispersion) is {\it much} more constraining of central galaxy quenching than either stellar or halo mass. This result is very similar to what we found for the fraction of quenched galaxies in Bluck et al. (2016). It is highly interesting, however, that our present resolved study leads to conclusions which are so similar to our prior galaxy-wide study. This fact may lend further qualitative support to the overarching conclusion of this paper - that quenching is fundamentally a global process, i.e. not reducible to a set of more fundamental local phenomena.

\subsubsection{The Quenching Axis \& Comparison to the SDSS}

\begin{figure*}
\includegraphics[width=0.49\textwidth]{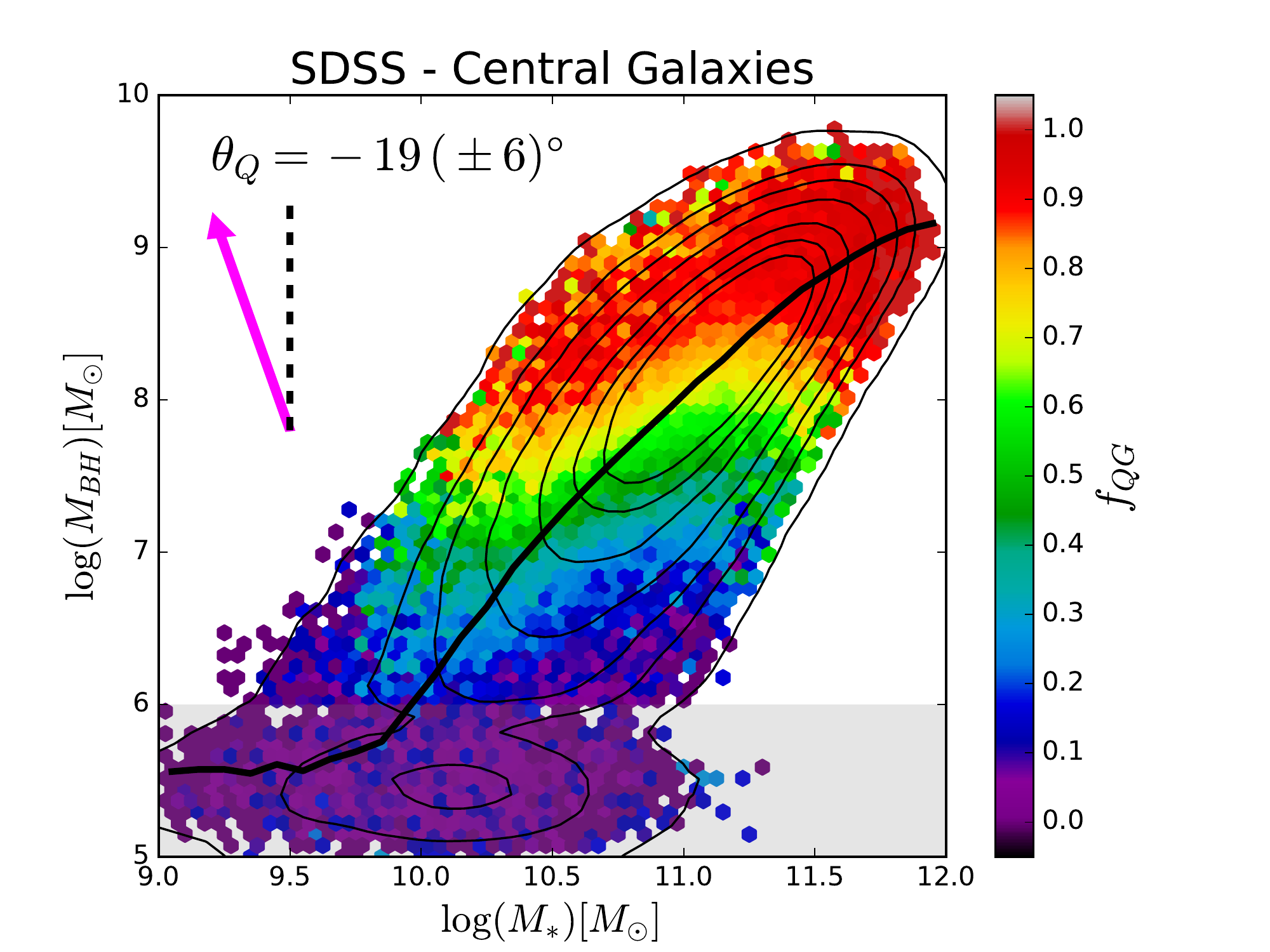}
\includegraphics[width=0.49\textwidth]{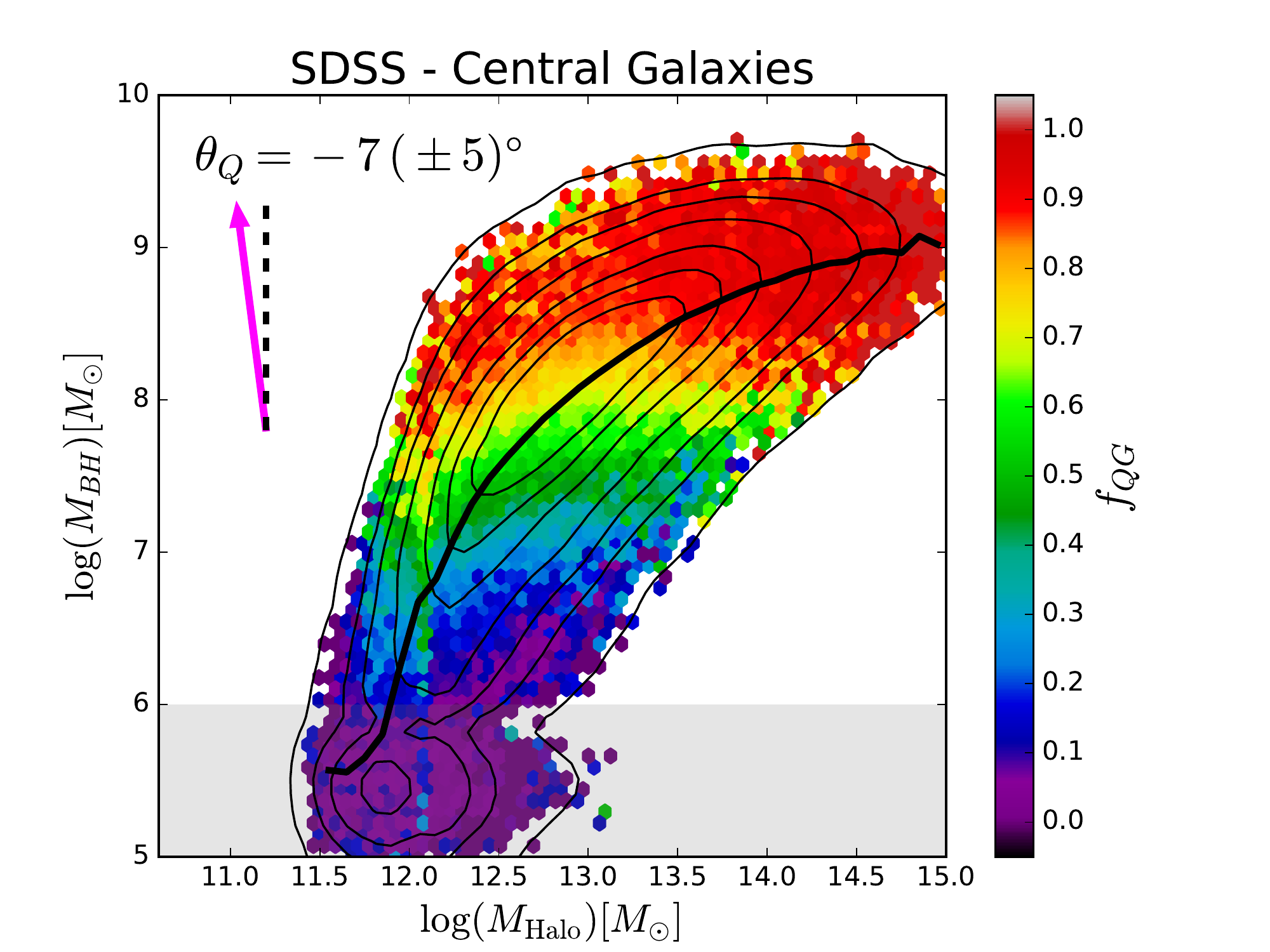}
\includegraphics[width=0.49\textwidth]{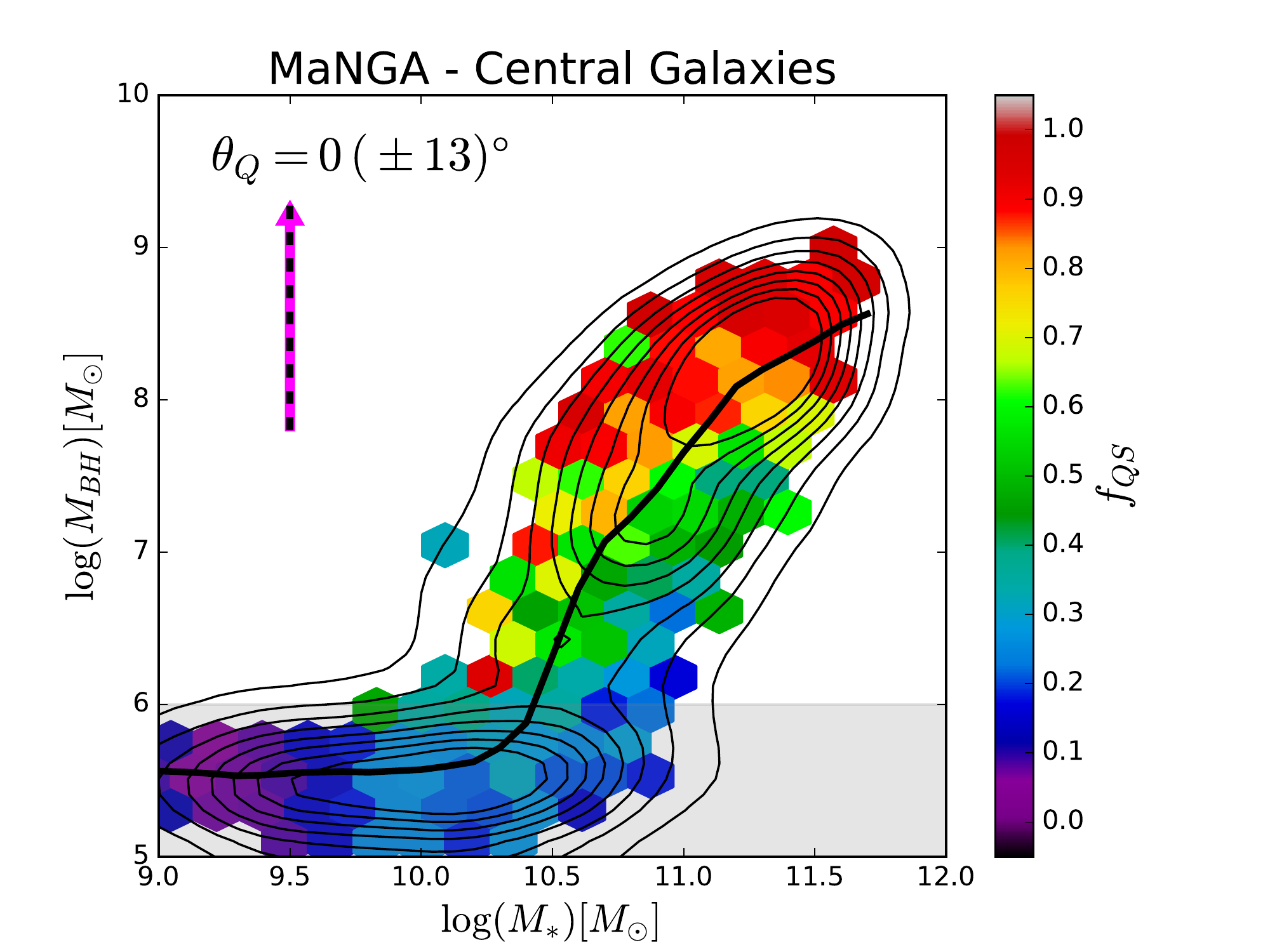}
\includegraphics[width=0.49\textwidth]{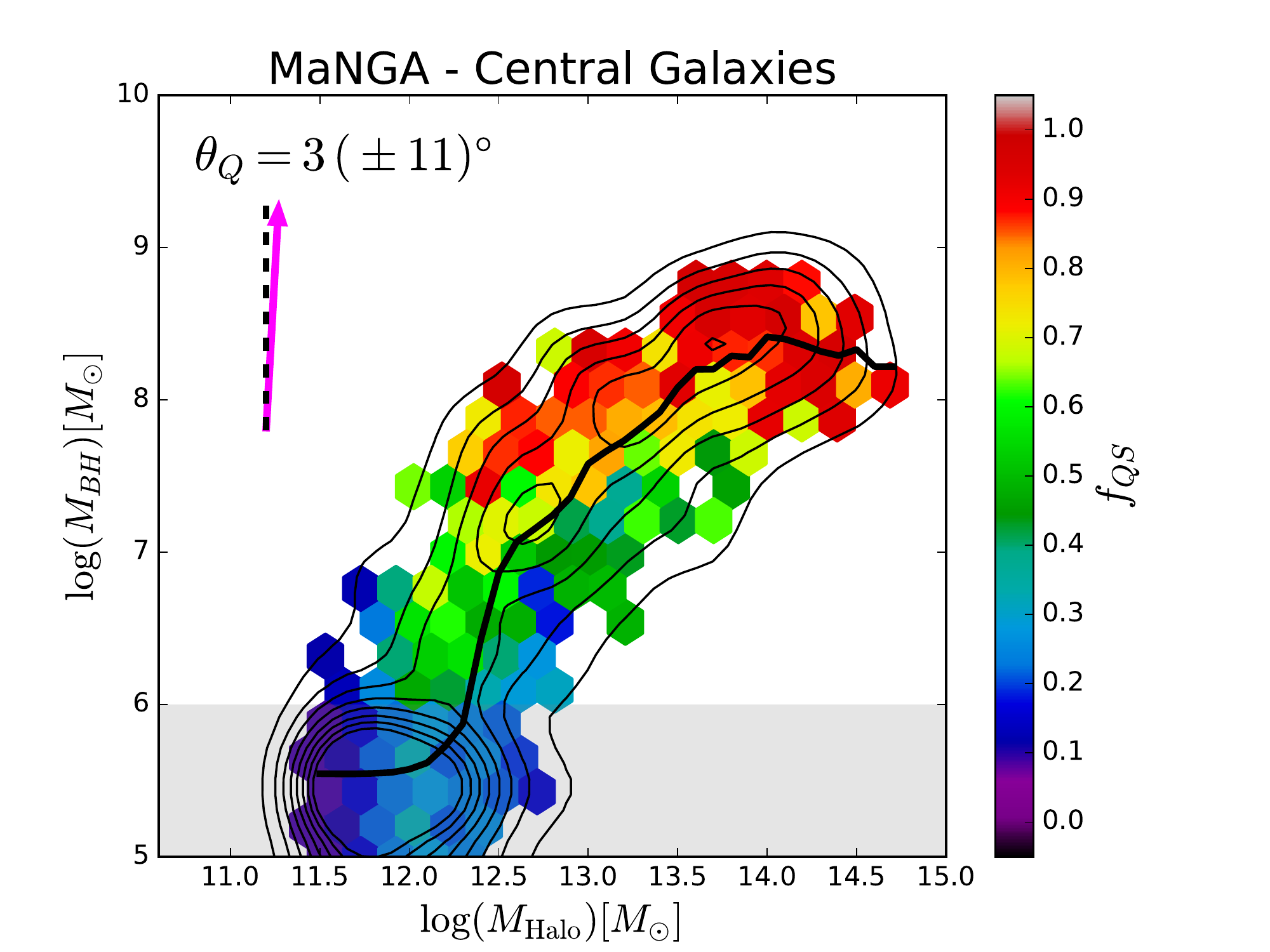}
\caption{Quenching comparison for central galaxies between the SDSS (top row) and MaNGA (bottom row). Specifically, this figure plots: a) estimated black hole mass vs. stellar mass, colour coded by the fraction of quenched galaxies ($f_{QG}$) within each hexagonal bin (for the SDSS); b) black hole mass vs. group halo mass, also coloured by $f_{QG}$ (for the SDSS); c) the same as a) but for MaNGA using the fraction of quenched spaxels ($f_{QS}$); d) the same as b) but for MaNGA using $f_{QS}$ in the colour bar. In all panels, density contours are shown as light black lines and the median relationships are shown as a thick black lines. Hexagonal bins display the quenched fraction statistics, and bin sizes are chosen to achieve a minimum of 10 galaxies per bin (for both MaNGA and the SDSS). On all panels it is clear that the fraction of quenched galaxies (or spaxels) varies primarily as a function of black hole mass, not stellar or halo mass. We quantify this effect with the quenching axis ($\theta_{Q}$, explained in eq. 16 and associated text.) }
\end{figure*}

Since we have established that quenching is a global process (see Section 4), the spatially resolved dimension is, to leading order, unimportant for investigating quenching (unlike for star formation). Hence, the SDSS DR7 parent sample of MaNGA provides a rich data source with a hundred times more galaxies than currently observed with MaNGA, from which we can meaningfully analyse galaxy quenching. In this sub-section, we make a brief comparison to the essential quenching results in the SDSS.

In Fig. 15 we show the relationship between black hole mass (estimated from central velocity dispersion) and stellar and halo mass in the SDSS (top row) and in MaNGA (bottom row). For the SDSS sample, we restrict to face-on objects in order to utilise the measured velocity dispersion in the aperture as a tracer of the actual velocity dispersion in the stellar population, rather than a noisy estimator of rotation. We construct a weighting to counteract the systematic loss of galaxies at low masses (and disc-dominated morphologies) from our sample as in Bluck et al. (2016). In the SDSS sample, quenching is defined via the logarithmic distance from the (global) star forming main sequence, whereby systems with $\Delta$MS $<$ -1 are deemed quenched, and systems with $\Delta$MS $>$ -1 are deemed star forming (see Bluck et al. 2016 for full details). For the spaxel sample, quenching is defined as in Fig. 5 and throughout this paper. On all panels of Fig. 15 the relationships are split into hexagonal bins, with a bin size chosen to achieve at least ten galaxies per bin (note in the SDSS the number of galaxies per bin is far higher than this). The fraction of quenched galaxies (for the SDSS) and quenched spaxels (for MaNGA) is indicated by the colour of each bin, and labelled by the colour bar in each panel.

Viewing Fig. 15 by eye, it is clear that the fraction of quenched galaxies and spaxels varies primarily as a function of black hole mass for centrals, and indeed there is very little variation in either $f_{QG}$ or $f_{QS}$ at a fixed black hole mass with either stellar or halo mass. Qualitatively at least, the general trends between the resolved and global plots of Fig. 15 are similar, albeit with the MaNGA data being much noisier.

We apply our quenching axis statistic (defined in Section 4.3, see eq. 16) to the SDSS in the top panels of Fig. 15, and to MaNGA in the bottom panels of Fig. 15. We find that the quenching axis is much closer to 0$^{\circ}$ than to $\pm 90^\circ$ in all cases, indicating that black hole mass is much more strongly correlated with quenching, at fixed values of stellar and halo mass, than the other way around. In this sense, Fig. 15 is in qualitative agreement with Fig. 14 for MaNGA, and the results of Bluck et al. (2016) for the SDSS. 

Interestingly, for both stellar mass and halo mass in the SDSS we find that $\theta_{Q} < 0^{\circ}$, which indicates that increasing stellar or halo mass at a fixed black hole mass actually {\it decreases} the probability of a galaxy being quenched. This result is significant for stellar mass, but is marginal for halo mass. Ultimately, finding negative values of the quenching axis is very interesting because it reveals that the strong positive correlations between stellar and halo mass and the fraction of quenched galaxies (e.g., Baldry et al. 2006, Peng et al. 2010, 2012, Woo et al. 2013) are entirely removed, and even inverted, at a fixed black hole mass. Therefore, the relationships between stellar and halo mass and the quenching of galaxies must be explained as originating from a deeper connection between quenching and black hole mass (see also Bluck et al. 2016, Teimoorinia et al. 2016, Terrazas et al. 2016, 2017 for similar conclusions). Thus, phenomenologically at least, {\it mass quenching is black hole mass quenching}.

For MaNGA, we find that the quenching axis is consistent with 0$^\circ$ for both stellar and halo mass, and hence there is no significant effect on quenching from either stellar or halo mass at a fixed black hole mass. However, the errors on the fraction of quenched spaxels in MaNGA are higher than in the SDSS, due to a lack of distinct galaxies (2500 vs. 400 000 centrals). Nonetheless, the general conclusions from these analyses are the same in both surveys: quenching is significantly more correlated with black hole mass than with stellar or halo mass, especially at fixed values of the other variables. 

Given the theoretical arguments presented in Appendix B (and summarised above), the clear dependence of central galaxy quenching on black hole mass (as estimated from central velocity dispersion) is consistent with quenching via AGN feedback. Furthermore, the lack of correlation between quenching and stellar or halo mass (at a fixed black hole mass) strongly disfavours models which quench central galaxies through stellar, supernova or virial shock feedback processes. See Appendix B for a much more detailed consideration of the theoretical implications of these observational results. It is important to emphasise here that the lack of correlation between stellar or halo mass and quenching, at fixed black hole mass, reveals that the strong positive correlations between these parameters cannot be causal in nature, as has been argued for in many prior works (e.g., Dekel \& Birnboim 2006, Peng et al. 2010, 2012, Woo et al. 2013, 2015, Dekel et al. 2019). The fact that the strong correlation between quiescence and black hole mass remains even at fixed stellar or halo mass is highly encouraging for the AGN feedback model, but, of course, it remains possible that some even deeper connection will in the future be discovered. We consider some important caveats in the final sub-section of this Discussion section.

\subsection{Alternative Explanations to Heating}

AGN feedback emerges from our analysis as the most probable heating source for quenching central galaxies (see the preceding sub-sections), and is furthermore currently the favoured model in the literature (see the Introduction, and also - Croton et al. 2006, Bower et al. 2006, 2008, Vogelsberger et al. 2014a,b, Schaye et al. 2015, Somerville \& Dave 2015, Brennan et al. 2017, Nelson et al. 2018, Henriques et al. 2019). However, given the lack of a fully realised model for the coupling of AGN feedback to galaxy hot haloes, and the relative lack of direct observational evidence (although see Fabian 2012, Hlavacek-Larrondo et al. 2012, 2013, 2018), alternatives to the AGN feedback paradigm (and indeed heating itself) exist in the literature. In this final sub-section we consider the leading alternatives to AGN heating, and determine whether they may offer plausible explanations to our observational results or not.

The major alternative to AGN feedback for explaining the tight dependence of central galaxy quenching on central velocity dispersion is morphological quenching (Martig et al. 2009). In this scenario, galaxies quench not due to a lack of cool gas accreted onto the galaxy but due to tidal toques preventing the gravitational collapse of giant molecular clouds. There are two serious issues with this explanation though: 1) a natural prediction of the morphological quenching model is that the gas fraction of massive quenched galaxies would be similar to the gas fraction of massive star forming galaxies, which is observed not to be true (e.g., Saintonge et al. 2011, 2016, 2017, Piotrowska et al. 2019); and 2) morphological quenching is an unstable solution. Small perturbations in the form of minor mergers or internal galaxy relaxation would disturb gas clumps enough to allow periodic bursts of star formation in giant ellipticals, which are not observed. At the very least, if one wants to employ morphological quenching to explain our observational results, additional mechanisms to reduce the gas fraction in galaxies will also be required. This is not a small effect, since the (molecular and atomic) gas fraction is found to drop by up to a factor of ten (Saintonge et al. 2016, 2017), and yet cooling models without heating through feedback predict a significant {\it increase} in the gas fraction of high mass galaxies. Thus, feedback mechanisms would still be needed (most likely through AGN heating) in order to prevent excessive cooling of gas from the hot gas halo into galaxies, required to match observations. As such, we conclude that, given the alternatives currently discussed in the literature, AGN feedback is the most efficient and probable explanation of our observational results.

We also consider the possibility that the close connection between quenched fraction (of spaxels and galaxies) and central velocity dispersion is {\it a causal}. First, we emphasize again that correlation does not imply causation, and hence a tight correlation does not allow us to infer causation directly. Nonetheless, if a model predicts a tight correlation between two observables and observations rule this out, rejection or improvement of the model is required. In this sense, we have found {\it consistency} between the predictions of a generalised AGN feedback model and observations on both global and spatially resolved scales. Thus, we need not reject the AGN feedback paradigm on the basis of the observational evidence provided in this paper. It remains possible, however, that some future model would perform as well in some respects and better in others, and should this occur we would then adopt that model as the most probable explanation to the observations.

Lilly \& Carollo (2016) argue that the observed size evolution in galaxies leads to an {\it a causal} explanation of the tight correlation between the fraction of quenched systems and measures of central density (e.g., bulge mass, central density and perhaps central velocity dispersion). Galaxies which quench earlier in the history of the Universe will be smaller and denser for their mass than galaxies which quench later, or indeed are still star forming today. This argument is often referred to as `progenitor bias' because it highlights that single epoch studies are incapable of comparing quenched objects to the type of star forming objects that were their progenitors. It is important to note that these types of argument offer no explanation as to {\it why} galaxies quench, but simply urge caution in interpreting the significance of the correlation between mass density and quenching. As a field, we must still come up with a plausible quenching mechanism, we just cannot rely solely on contemporary observations to constrain this accurately. 

There are three results which strongly suggest that progenitor bias is not the dominant reason behind the strong correlation between quenched fraction and central velocity dispersion: 1) The predictivity of central velocity dispersion and central mass density is far less for satellites than for centrals (see Fig. 13), and yet a simple interpretation of the {\it a causal} model would leave all quenched objects denser than star forming objects, because they quenched at a time in the Universe's history when galaxies were smaller for their mass. 2) In Bluck et al. (2016) we find that central velocity dispersion is still the most effective global parameter for predicting the presence of centrals in the green valley. Hence, assuming that green valley centrals are undergoing quenching now, the difference in their inner structures cannot be attributed to progenitor bias. 3) Finally, and most importantly, it is not clear how size evolution alone can give rise to the strong correlation between central velocity dispersion and quenched fraction. Though smaller for their mass (and hence denser), galaxies at higher redshifts do not have significantly higher central velocity dispersions than low redshift galaxies selected at the same stellar mass (e.g., Forster Schreiber et al. 2006, 2009, 2011). In fact, galaxies at z$\sim$2  are found to be more disc dominated and more frequently rotationally supported than similar mass galaxies at z$\sim$0 (e.g., Forster Schreiber et al. 2009, Bluck et al. 2012, Mortlock et al. 2013, Buitrago et al. 2013, 2014). Note also that the observed increase in disc velocity dispersion is significantly smaller than the required increase in central velocity dispersion, by up to an order of magnitude. Therefore, progenitor bias does not appear to be a serious concern for central velocity dispersion, even if it may be for mass density. 

In conclusion, none of the current alternatives to the heating solution for quenching central galaxies are fully in line with observations. More specifically, morphological quenching fails to explain the extensive reduction in gas fraction of high mass quenched systems; and the progenitor bias argument fails to account for why quenched centrals have substantially higher central velocity dispersions than star forming centrals (especially when satellites do not). Thus, as discussed in the Introduction, heating has become the favoured approach for explaining both the cooling problem and quenching in massive galaxies. Within the heating paradigm, the observational results presented in this paper provide strong support for AGN feedback as the underlying heating source (particularly in the radio/ maintenance-mode). Furthermore, we also find strong evidence to disfavour supernova feedback and virial shocks as significant quenching mechanisms in high mass galaxies.

%
%   SUMMARY
%

\section{Summary}

We present an analysis of over 5 million spaxels from $\sim$3500 galaxies taken from the SDSS-IV MaNGA survey (DR15). We estimate star formation rate surface densities ($\Sigma_{\rm SFR}$) for all spaxels belonging to galaxies via a two stage approach. If the spaxel has high S/N ($>$3) in all of the BPT emission lines, and is furthermore identified as `star forming' on the BPT diagram, we utilise dust corrected $H\alpha$ flux to infer its $\Sigma_{\rm SFR}$. If the spaxel has low S/N ($<$3) in any of the BPT emission lines, or else if it is identified as `AGN' or `composite' on the BPT diagram, we estimate $\Sigma_{\rm SFR}$ from the observed resolved correlation between sSFR and the strength of the 4000 \AA \hspace{0.05cm} break (D4000). In this manner we obtain an indicator of star formation for every galaxy spaxel in the sample ($>$99\%), unlike in the majority of publications to date utilising IFU spectroscopy. We validate our hybrid H$\alpha$ - D4000 approach for estimating $\Sigma_{\rm SFR}$ against alternative methods in Appendix A (from SSP model fitting and EW(H$\alpha$)). None of the results or conclusions presented in this paper depend critically on the choice of SFR method. We present the {\it complete} resolved star forming main sequence in Fig. 3 for the first time.

We develop a machine learning approach utilising a multi-layered ANN to classify spaxels into `star forming' and `quenched' categories based on information from individual, and strategic groupings, of parameters measured at various scales. Specifically, we consider parameters from three broad ranges: local (spatially resolved, measured at the spaxel location), global (one value per galaxy, pertaining to the galaxy as a whole), and environmental (one value per galaxy, pertaining to the environment in which the galaxy resides). Additionally, we train our ANN architecture to predict actual $\Sigma_{\rm SFR}$ values in star forming regions, via regression. We also utilise a random forest analysis to ascertain how effective parameters are in concert for predicting star formation rates and quenching. Our primary results from the machine learning analyses are as follows:

\begin{enumerate}

\item We find that global parameters (acting in concert) are far superior to local and environmental parameters at predicting whether regions within central galaxies will be star forming or quenched (see Figs. 8 \& 9, upper panels). Thus, quenching is a global process not reducible to a set of more fundamental local processes, at least at the $\sim$kpc resolution scale of MaNGA.\\

\item For central galaxies, we find that central velocity dispersion is the most predictive single parameter for determining whether regions within galaxies will be star forming or quenched (see Fig. 9, upper panel). This result is in qualitative agreement with Bluck et al. (2016) \& Teimoorinia et al. (2016), which utilise single aperture spectroscopy from the SDSS. However, we significantly expand on those earlier works by including spatially resolved parameters in our analysis, and by considering star formation and quenching as spatially resolved phenomena, which may vary across a galaxy.\\

\item For predicting actual $\Sigma_{\rm SFR}$ values in star forming regions (via regression), we find that local parameters are by far the most effective group for all galaxy types. Furthermore, the stellar mass surface density ($\Sigma_*$) at the spaxel location is by far the most predictive single parameter for estimating $\Sigma_{\rm SFR}$ in star forming regions (see Figs. 8 \& 9, lower panels). Thus, star formation is a local process, governed by spatially resolved phenomena at the spaxel location.\\

\item We present an additional test on the machine learning results utilising a principal component analysis. Once again we confirm that quenching is governed by global parameters, yet star formation is governed by local (spatially resolved) parameters (see Fig. 11).\\

\item In Appendix A we provide extensive testing of the stability of the main results presented above. We demonstrate that the ranking of parameters is highly stable to sample selection, SFR method, and the impact of measurement error and noise. 

\end{enumerate}

In Appendix B we demonstrate through analytic derivation that $M_{BH}$ traces the total integrated energy released from AGN feedback, $M_*$ traces the total integrated energy released through supernovae (of all types), and  $M_{\rm Halo}$ traces the total energy released from virial shocks. Armed with these theoretical insights, in Section 5 we explore quenching throughout the $M_{BH}$ - $M_*$ - $M_{\rm Halo}$ parameter space for central galaxies. We estimate $M_{BH}$ from central velocity dispersion and $M_{\rm Halo}$  from an abundance matching technique applied to the total stellar mass of the group. Through a variety of statistical tests (see Figs. 14 \& 15, and associated text) we determine that black hole mass is the main driver of quenching in central galaxies. Furthermore, stellar and halo mass are both very poor tracers of quenching, at fixed black hole mass. Therefore, we conclude that black hole mass (as estimated via central velocity dispersion) is the key observable parameter driving quenching, consistent with modern predictions for AGN feedback (particularly in the radio mode). Additionally, we conclude that supernova feedback and virial shocks are not significant quenching mechanisms for central galaxies.

In summary, we find that quenching is fundamentally a global process, governed for centrals by the properties of the inner most regions within galaxies. Conversely, star formation is a local process, governed by the physical conditions at the spaxel location.

\section*{Acknowledgments}

We thank the anonymous referee for a highly insightful and constructive report which has helped to improve the presentation of this work. We are very grateful for several stimulating discussions on this work during its preparation phase, especially to Alice Concas, Christopher Conselice, Emma Curtis-Lake, Stephen Eales and Robert Gallagher. AFLB and RM gratefully acknowledge ERC Advanced Grant 695671 "QUENCH", and support from the Science and Technology Facilities Council (STFC). SFS gratefully acknowledges the support of CONACYT grant CB-285080 and FC-2016-01-1916, and funding from the PAPIIT-DGAPA-IA101217 (UNAM) project. SLE gratefully acknowledges funding from an NSERC Discovery Grant. JMP gratefully acknowledges funding from the Merac Foundation. 

This work makes use
of data from SDSS-I \& SDSS-IV. Funding for the SDSS has been provided by the Alfred P. Sloan Foundation, the Participating Institutions, the National Science Foundation, the U.S. Department of Energy, the National Aeronautics and Space Administration, the
Japanese Monbukagakusho, the Max Planck Society, and the
Higher Education Funding Council for England. Additional funding towards SDSS-IV has been
provided by the U.S. Department of Energy Office of Science. SDSS-IV acknowledges support and resources from
the Center for High-Performance Computing at the University of Utah. The SDSS website is: 
www.sdss.org

The SDSS is managed by the Astrophysical Research Consortium for the Participating Institutions of the SDSS Collaboration. For SDSS-IV this includes the Brazilian
Participation Group, the Carnegie Institution for Science,
Carnegie Mellon University, the Chilean Participation Group,
the French Participation Group, Harvard-Smithsonian Center
for Astrophysics, Instituto de Astrofísica de Canarias, The
Johns Hopkins University, Kavli Institute for the Physics
and Mathematics of the Universe (IPMU) / University of
Tokyo, Lawrence Berkeley National Laboratory, Leibniz Institut fur Astrophysik Potsdam (AIP), Max-Planck-Institut fur
Astronomie (MPIA Heidelberg), Max-Planck-Institut fur Astrophysik (MPA Garching), Max-Planck-Institut fur Extraterrestrische Physik (MPE), National Astronomical Observatory
of China, New Mexico State University, New York University,
University of Notre Dame, Observatario Nacional / MCTI,
The Ohio State University, Pennsylvania State University,
Shanghai Astronomical Observatory, United Kingdom Participation Group, Universidad Nacional Autonoma de Mexico, University of Arizona, University of Colorado Boulder,
University of Oxford, University of Portsmouth, University of
Utah, University of Virginia, University of Washington, University of Wisconsin, Vanderbilt University, and Yale University.

The MaNGA data used in this work is publicly available at: 
http://www.sdss.org/dr15/manga/manga-data/

The Participating
Institutions of SDSS-I \& II are the American Museum of Natural History, Astrophysical Institute Potsdam, University of Basel, University of Cambridge, Case Western Reserve University, University of Chicago, Drexel University, Fermilab, the Institute for Advanced Study, the Japan Participation Group, Johns
Hopkins University, the Joint Institute for Nuclear Astrophysics, the Kavli Institute for Particle Astrophysics and
Cosmology, the Korean Scientist Group, the Chinese Academy
of Sciences (LAMOST), Los Alamos National Laboratory,
the Max-Planck-Institute for Astronomy (MPIA), the Max-Planck-Institute for Astrophysics (MPA), New Mexico State
University, Ohio State University, University of Pittsburgh, University of Portsmouth, Princeton University, the United States Naval Observatory, and the University of Washington.

%\begin{thebibliography} {99}

%
%   APPENDIX - Data/ Method Stability Tests
%

\appendix

\section{Tests on the Stability of the Main Results}

\begin{figure*}
\includegraphics[trim = 20mm 0mm 20mm 0mm, clip, width=1\textwidth]{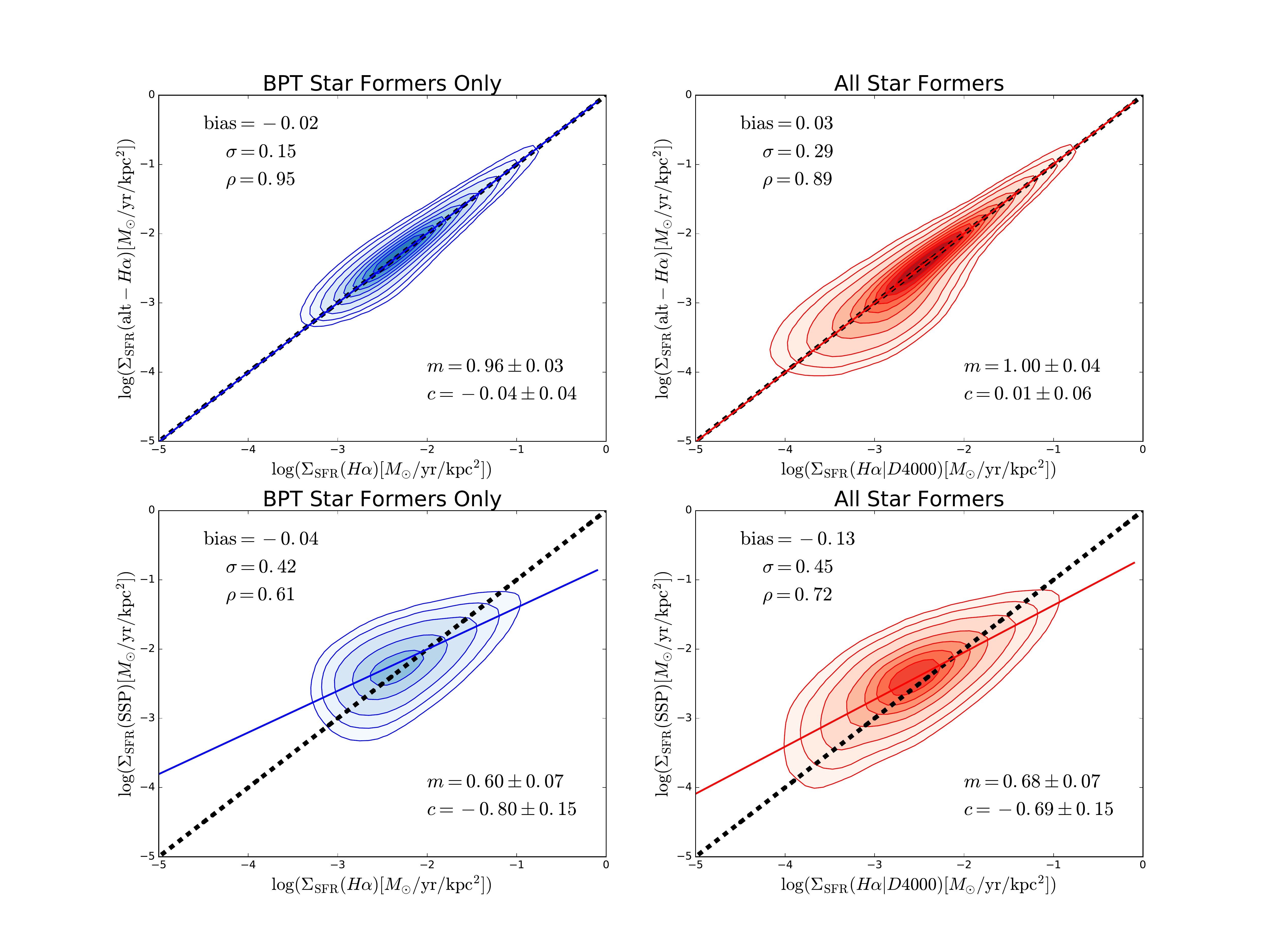}
\caption{Direct comparison of $\Sigma_{\rm SFR}$ methods, with all axes showing a distinct measurement of $\Sigma_{\rm SFR}$. The x-axes show the $\Sigma_{\rm SFR}$ values utilised throughout the main body of the paper (left panels from $H\alpha$ only for BPT star forming regions, shown in blue; and right panels from the H$\alpha$ - D4000 hybrid method for all star forming regions, shown in red). The top panel y-axes show an alternative H$\alpha$ based method (utilising an EW(H$\alpha$) threshold to identify star forming regions, and an average Balmer decrement at constant radius to incorporate more spaxels into the analysis). The bottom panel y-axes show photometric $\Sigma_{\rm SFR}$ values derived from SSP fitting. On each panel the bias, dispersion ($\sigma$), correlation strength ($\rho$), gradient ($m$) and offset ($c$) of a linear regression fit are all displayed. Light coloured lines and shaded regions indicate linearly spaced density contours. Solid coloured lines indicate the best fit lines from linear regression, and the black dashed lines indicate the 1:1 relation. There are strong correlations between all methods in all samples. The H$\alpha$ based methods are particularly highly correlated, and exhibit a near 1:1 relation. Comparing to the photometric SFRs, the relationships are clearly not 1:1 but there remains a reasonably tight relationship.}
\end{figure*}

In this appendix we show a variety of tests and checks on the primary results and conclusions of this paper. We start with an analysis of two alternative prescriptions for estimating $\Sigma_{\rm SFR}$ (via an alternative H$\alpha$ method and photometric SFRs derived from SSP model fitting). We then consider the impact of measurement error on the main results. All of the tests we perform indicate that the results presented in the main body of the paper are highly stable to the above potential issues.

\subsection{Alternative Approaches for Estimating $\Sigma_{\rm{SFR}}$ \& Classifying Spaxels}

Our two stage approach for estimating $\Sigma_{\rm SFR}$ from dust corrected H$\alpha$ flux or the sSFR - D4000 relation (used throughout the main body of the paper, see Section 3) is a simple, empirically motivated method, which passes all of our intrinsic tests on its performance. However, it is important to ascertain whether or not the adoption of alternative methodologies for estimating $\Sigma_{\rm SFR}$ would significantly impact any of our results or conclusions. In this part of the appendix we consider two such alternatives.

First we consider a qualitatively similar approach by inferring $\Sigma_{\rm SFR}$ from dust corrected H$\alpha$ flux. However, in this approach we do not require that the H$\beta$ line is detected at S/N $>$ 3 in the spaxel. Instead, we take the average value of the Balmer decrement measured at an approximately constant galacto-centric radius. The advantage of this approach is that it allows us to recover significantly more spaxels from which an approximate dust correction can be made ($\sim$60\% vs. 20\%). The disadvantage of this approach is that the dust correction is more approximate, missing variation as a function of azimuthal angle (e.g., as a result of dust lanes and spiral features). The key advantage of recovering more spaxels would be lost if we still categorized spaxels solely through their location on the BPT diagram (as in Fig. 2), since an accurate H$\beta$ flux in the spaxel would still be required. To combat this, we utilise an identical approach to Cano-Diaz et al. (2016) and S\'anchez et al. (2017) to identify star forming regions approximately by location on the BPT diagram (lower than the Kewley et al. 2001 line, without S/N cuts) and additionally though a cut in EW(H$\alpha$) $>$ 6\AA \hspace{0.05cm} (which has been shown to be highly effective at identifying star forming regions). We also correct for the impact of AGB-ionised diffuse gas on the SFR measurement, by subtracting off EW(H$\alpha$) $\sim$ 1\AA \hspace{0.05cm} (following Binette et al. 1994). Note that the EW(H$\alpha$) cut and correction are largely redundant in our original analysis because our imposed S/N limit on H$\beta$ leads to very high H$\alpha$ fluxes in all cases, which almost invariably leads to high EW(H$\alpha$) values as well.

\begin{figure*}
\includegraphics[width=0.85\textwidth]{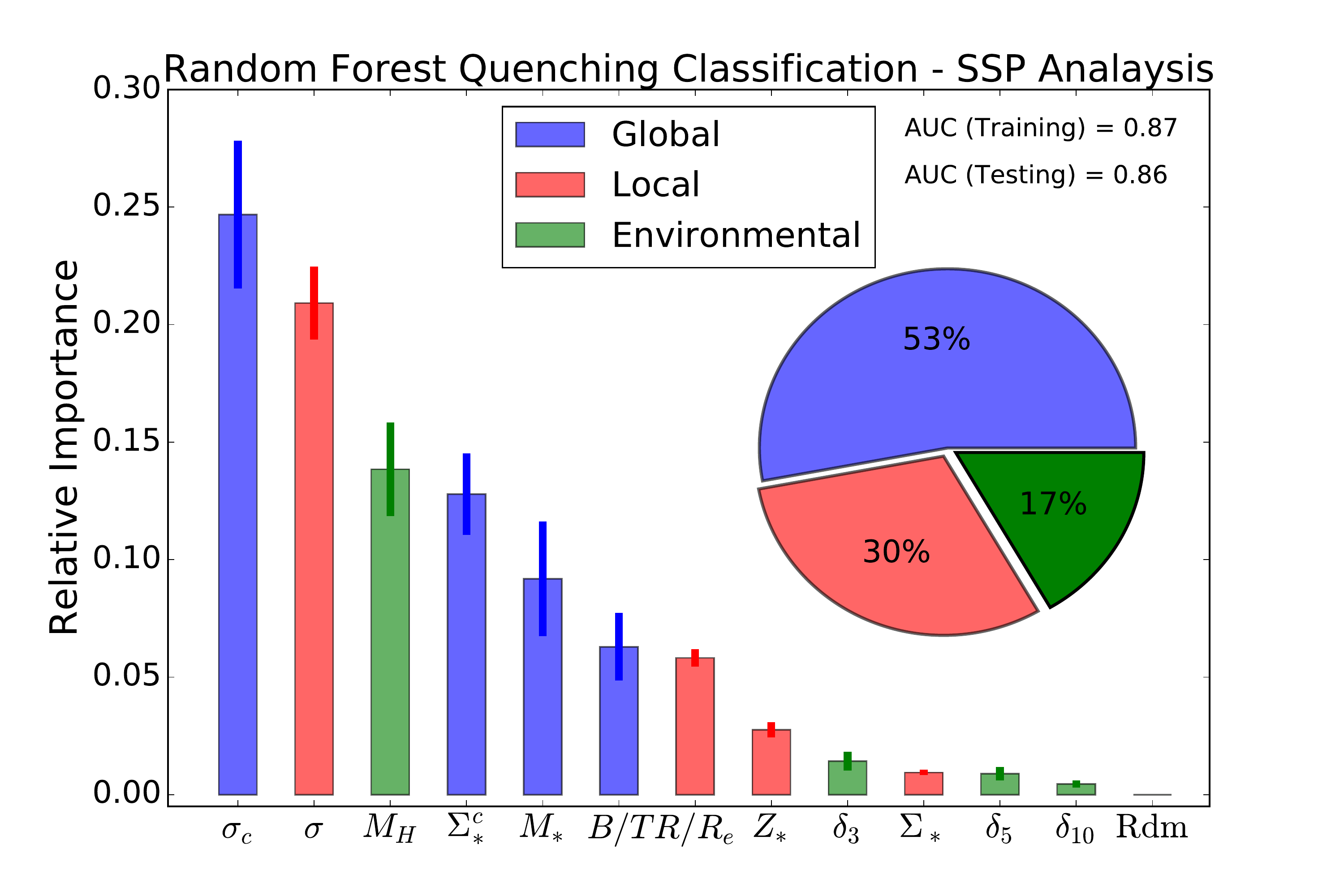}
\includegraphics[width=0.85\textwidth]{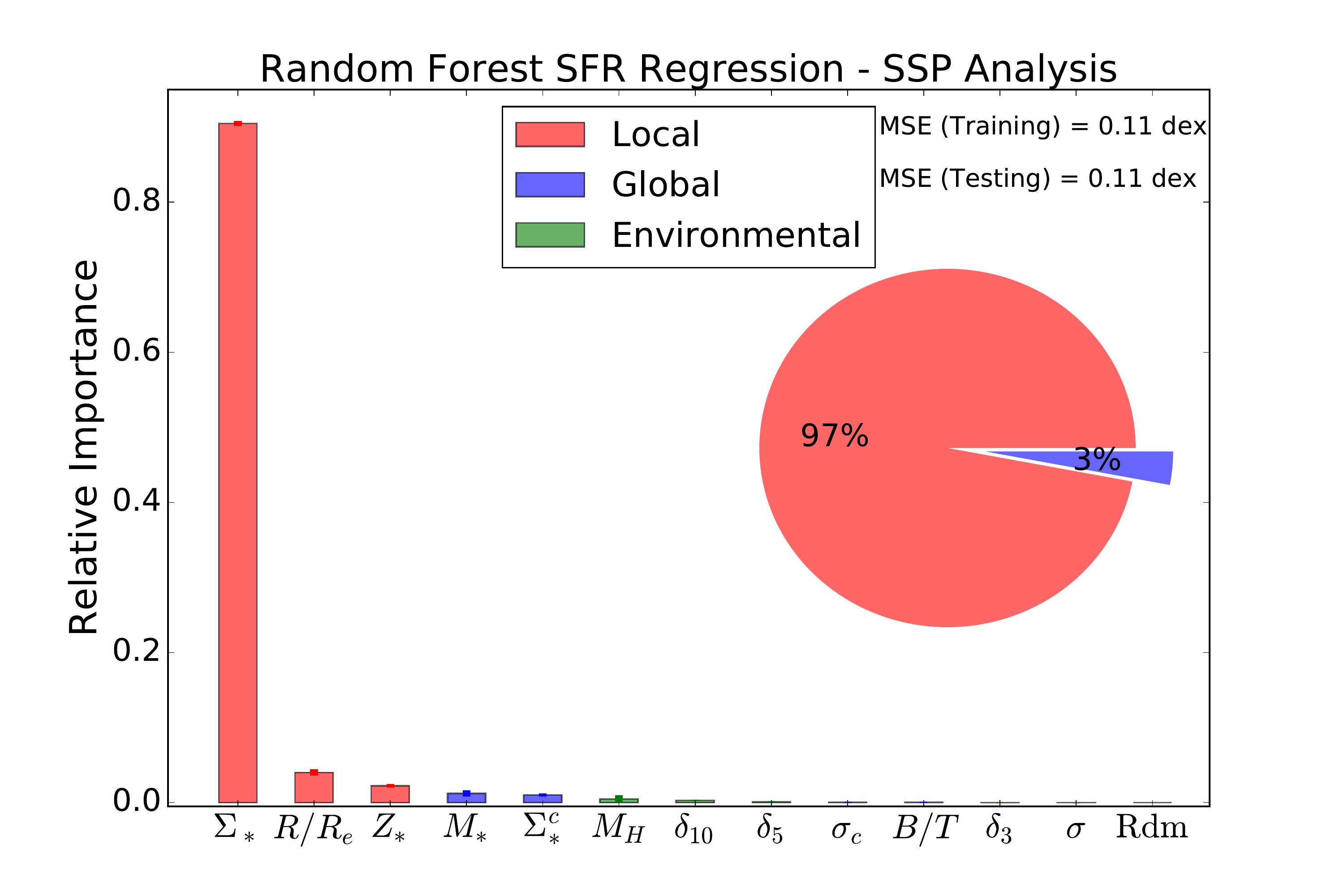}
\caption{ {\it Top panel:} Reproduction of the RF classification analysis to predict quenching in central galaxies (originally presented in Fig. 9, top panel), here using a quenched threshold set by photometric sSFR(SSP) values to train the network. {\it Bottom Panel:} Reproduction of the RF regression analysis to predict $\Sigma_{\rm SFR}$ values in star forming regions for central galaxies (originally presented in Fig. 9, bottom panel), here using photometric SFR values derived from SSP fitting to train the network. Note that the D4000 index is not used in any way in this figure. Both analyses lead to very similar results and identical conclusions to the original versions.} 
\end{figure*}

All considered, $\sim$50\% of galaxy spaxels are given a $\Sigma_{\rm SFR}$ value from this alternative technique, a significant improvement over the original approach in terms of completeness. Nonetheless, that still leaves around half of the dataset without a $\Sigma_{\rm SFR}$ measurement (due to either low S/N in H$\alpha$ or low EW(H$\alpha$)). This necessitates either the assumption that all of those regions are quenched (which is premature, given the possibility of low mass surface densities and/or high extinction), or the use of an alternate method (e.g. through the sSFR-D4000 relation or SSP model fitting) to determine their star forming state. This is especially important to do carefully if the goal is to probe quenching (as in this paper). As with all of the $\Sigma_{\rm SFR}$ values used in this paper, we adopt the Kennicutt 1998 calibration assuming a Salpeter IMF and utilise a flat $\Lambda$CDM cosmology (with parameters given at the end of Section 1). Full details on this methodology are provided in S\'anchez et al. (2017, 2018), and references therein.

In the top panels of Fig. A1 we present a direct comparison of our $\Sigma_{\rm SFR}$ values against the version outlined above. On the left panel we show the comparison just for our BPT star forming sample (comparing H$\alpha$ with H$\alpha$). In the right panel we show the results for our complete hybrid sample. There is excellent agreement between the two methods, with a very high correlation strength, low offset and small scatter. Furthermore, the best fit from linear regression shows a gradient very close one in all samples, indicating essentially no systematic bias between these two approaches. In one sense this may be unsurprising, since both methods use dust corrected H$\alpha$ flux to infer $\Sigma_{\rm SFR}$. However, in the alternative approach (described above) the D4000 index is not used at all, whereas in our approach it actually sets the majority of $\Sigma_{\rm SFR}$ values (in the top right panel of Fig. A1). Furthermore, the different approaches utilise different prescriptions for identifying star forming regions, which given the similarity in their final $\Sigma_{\rm SFR}$ values lends further confidence to both approaches. Ultimately, the excellent agreement between these two methods strongly implies that our results for star forming regions would be essentially identical if we implemented these alternative $\Sigma_{\rm SFR}$ measurements into our analysis. Indeed, we have explicitly tested this and it is the case.

The chief disadvantage of utilising H$\alpha$ exclusively in the estimation of $\Sigma_{\rm SFR}$ is that it is not detected in $>$ 40\% of galaxy spaxels. Although it might be tempting to assume that all of these spaxels with null detections in H$\alpha$ are quenched, this could also be due to low surface mass densities and/or high levels of extinction (as noted above). To combat this issue, we adopt the sSFR - D4000 calibration throughout the main body of this paper, but there are other possible approaches to consider. 

The stellar decomposition of observed spectra by {\sc Pipe3D} allows us to derive the amount of dust-corrected light (and mass) that originates from stars with different ages (within a given sampling defined by our chosen SSP model library).  Based on that decomposition, it is possible to estimate the cumulative assembly of stellar mass along the look back time (taking into account the redshift of the object). More specifically, we co-add the observed masses at each age across all cosmological epochs to the present. Then, assuming that all assembled mass is due to star formation, the SFR at each time is derived simply as the differential cumulative stellar mass in each time step with respect to the adjacent one, divided by the time range (see S\'anchez et al. 2019b for a more complete discussion of this method). Finally, we estimate the actual SFR by combining the stellar mass assembled in the last 32 Myr (divided by this time range), following precisely the prescription outlined in detail in Gonz\'alez Delgado et al. 2016 and S\'anchez et al. 2018. This gives a photometric estimate of $\Sigma_{\rm SFR}$ obtained via fitting to SSP models. Finally, if there are no stars formed within 32 Myr, as ascertained by the ages of the SSPs, we assume the star formation rate is zero and give this a nominal low value (analogous to in our D4000 method). It is important to stress that the timescales of $\Sigma_{\rm SFR}(H\alpha)$ and  $\Sigma_{\rm SFR}({\rm SSP})$ are different (probing the average rate of star formation over $\sim$4 Myr and 32 Myr, respectively). Nonetheless, both prescriptions lead to a meaningful estimate of spatially resolved star formation, which can be used to validate each approach.

\begin{figure*}
\includegraphics[width=1\textwidth]{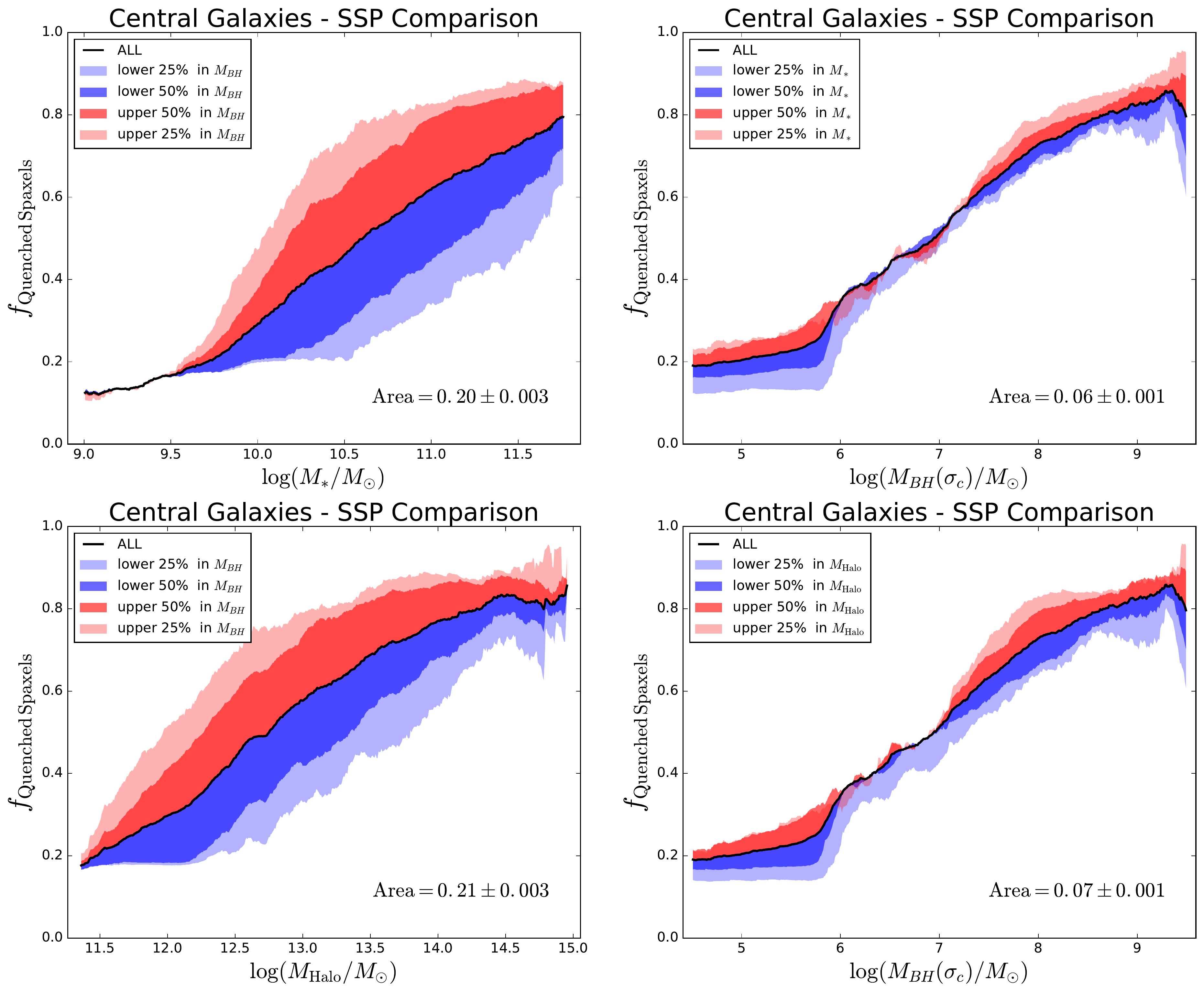}
\caption{Reproduction of Fig. 14, classifying spaxels into star forming and quenched sets via the SSP photometric SFR values, with a threshold at sSFR(SSP) = $10^{-11.5} {\rm yr}^{-1}$. Varying this threshold by $\pm$0.5 dex leads to no significant difference in these results. Note that this method is entirely independent of the D4000 index. It is clear that black hole mass (as estimated from central velocity dispersion) impacts the quenched fraction of spaxels significantly more at fixed values of stellar and halo mass than the other way around (exactly as seen in Fig. 14).}
\end{figure*}

In the bottom panels of Fig. A1 we present a direct comparison of our estimated $\Sigma_{\rm SFR}$ from H$\alpha$ in BPT star forming regions (left) and for all spaxels via our hybrid H$\alpha$ \& D4000 approach (right) with the $\Sigma_{\rm SFR}({\rm SSP})$ values described above. As seen before in S\'anchez et al. (2018, 2019) there is a strong correlation between these $\Sigma_{\rm SFR}$ approaches; however, the relationship is manifestly not one-to-one. This is expected since the two methods probe different timescales, and, moreover, use entirely different methodologies. Nonetheless, the strong correlation between the measurements (especially in the full-sample case, bottom right panel of Fig. A1) is encouraging.

Since the two H$\alpha$ methods give such similar results we do not consider them further here. The SSP method, on the other hand, is worth exploring further since the $\Sigma_{\rm SFR}$ values show significant differences with respect to the H$\alpha$ methods. To this end, we reproduce our entire machine learning analysis (see Section 4.2) utilising $\Sigma_{\rm SFR}({\rm SSP})$ instead of $\Sigma_{\rm SFR}({\rm H}\alpha | {\rm D4000})$. For the sake of brevity, we focus only on the RF analyses here.

In the bottom panel of Fig. A2 we show the results from an RF regression analysis to predict $\Sigma_{\rm SFR}({\rm SSP})$ values in star forming regions (here defined as sSFR(SSP) $>$ $10^{11} {\rm yr}^{-1}$), for central galaxies. All of the main features of our original analysis with $\Sigma_{\rm SFR}({\rm H}\alpha | {\rm D4000})$ (shown in Fig. 9) are recovered exactly. Local parameters, acting in concert, are still by far the most predictive group, and stellar mass surface density (at the spaxel location) is still overwhelmingly the most predictive single parameter. The rankings of all other parameters are also very similar to the original case, and are largely negligible in importance. Hence, we have demonstrated that the use of very different $\Sigma_{\rm SFR}$ measurements leads to essentially identical results in our star forming analysis.

For comparison to our quenching classification analysis, we adopt an analogous approach defining star forming regions as having sSFR(SSP) $>$ $10^{11} {\rm yr}^{-1}$, and quenched regions as having sSFR(SSP) $<$ $10^{11.5} {\rm yr}^{-1}$, removing $<$ 10\% percent of green valley spaxels which lie between these limits. In total 87\% of spaxels are classified identically in both methods, and of the misclassifications a significant fraction lie in the green valley of one or both methods. We show the results for an RF classification analysis based on SSP measurements in the top panel of Fig. A2, for central galaxies. As with the star forming regression analysis, all of the main features of the quenching classification analysis are identical to the hybrid H$\alpha$ - D4000 method (shown in Fig. 9). More specifically, global parameters still perform as the best group; central velocity dispersion is still the most predictive single variable (and of course the most predictive global variable); halo mass is still the most predictive environmental parameter; and velocity dispersion at the spaxel location is still the most predictive local parameter. All other rankings are very similar to the original analysis as well, with the only notable difference being that $\Sigma_*^c$ and $M_{H}$ exchange places. Thus, we have established that the SFR methodology is not critical for establishing the results of this paper, since a very different approach leads to almost identical results.

Additionally, we have experimented with even more methods for classifying spaxels (based on the age of the stellar population, pure H$\alpha$ detection, and combinations of the above); and incorporated random noise into the input parameters at various levels (based on their error estimates). All of this extensive testing yields essentially identical results to that presented in the main body of the paper: {\it quenching is fundamentally a global process; whereas star formation is governed locally. The best parameter for predicting the level of star formation in star forming regions is $\Sigma_*$, and the best parameter for predicting quenching in centrals is $\sigma_c$.}

Finally, we test the relative performance of $M_{BH}(\sigma_c)$, $M_{\rm Halo}$ and $M_{*}$ for predicting quenching in the SSP analysis, to compare to our hybrid H$\alpha$ - D4000 analysis (shown in the main body of this paper). To achieve this, we reproduce the fraction of quenched spaxels analysis of Fig. 14 here in Fig. A3 for the SSP SFRs, defining quenched and star forming regions as above. We recover an almost identical result, whereby the impact on the quenched fraction of varying black hole mass (estimated from central velocity dispersion) at fixed stellar or halo mass is $\sim$three times greater than the other way around. Therefore, the primacy of black hole mass in governing central galaxy quenching is recovered in the SSP analysis, which strongly suggests that this important result is not dependent on the method used to classify spaxels into star forming and quenched categories.

\subsection{The Impact of Measurement Errors on the Primary Results of this Paper}

\begin{figure*}
\includegraphics[width=0.49\textwidth]{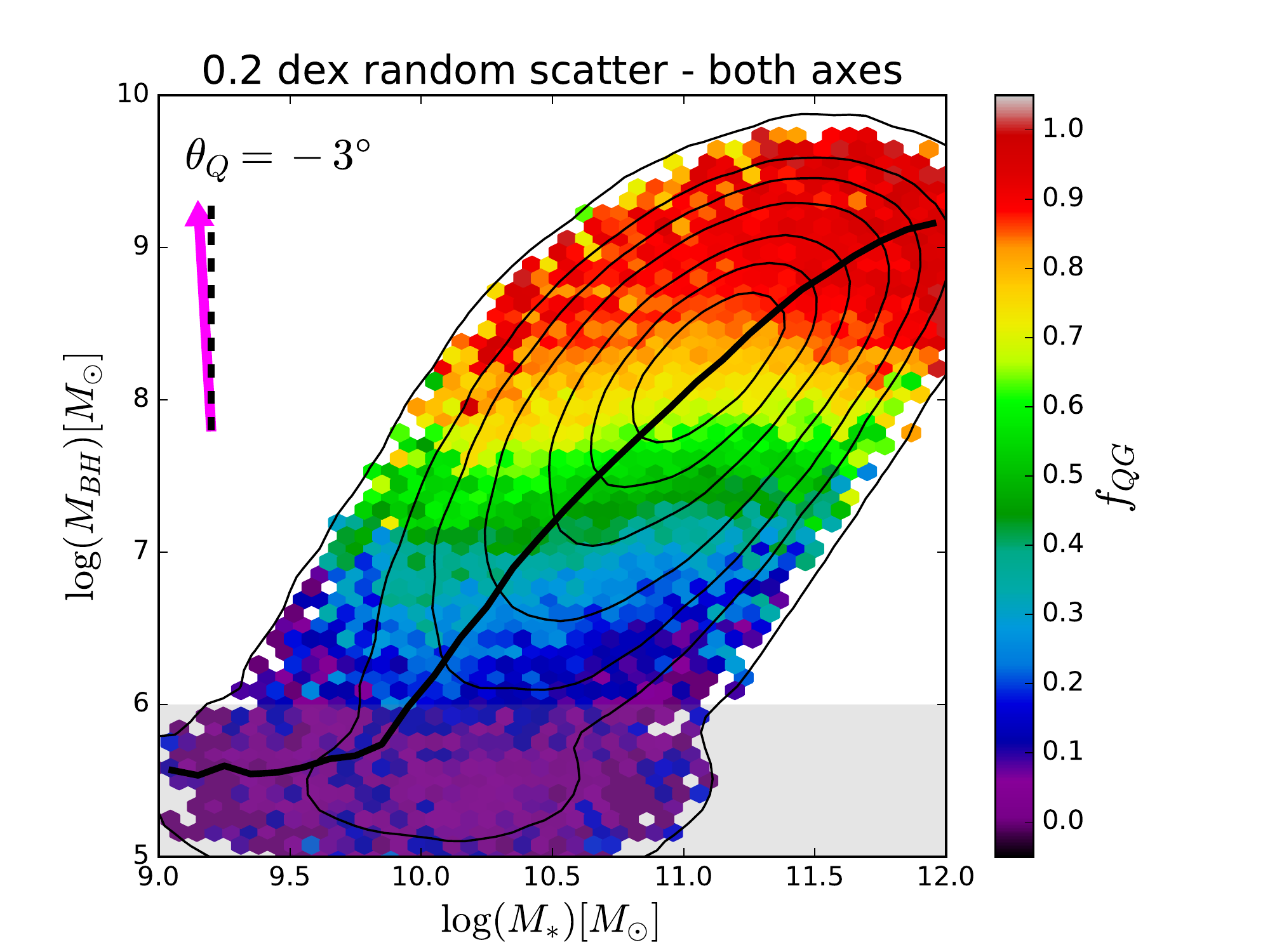}
\includegraphics[width=0.49\textwidth]{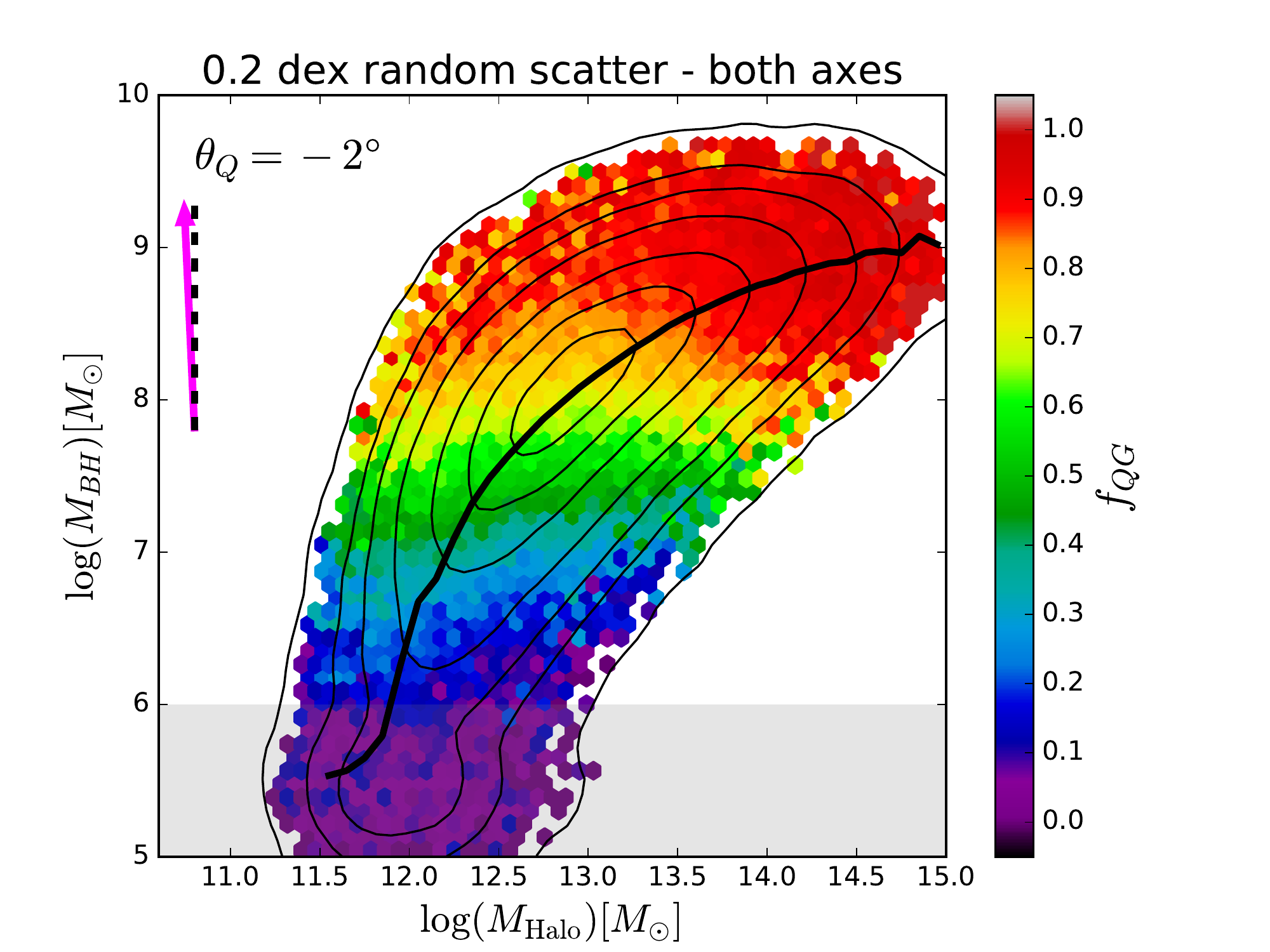}
\includegraphics[width=0.49\textwidth]{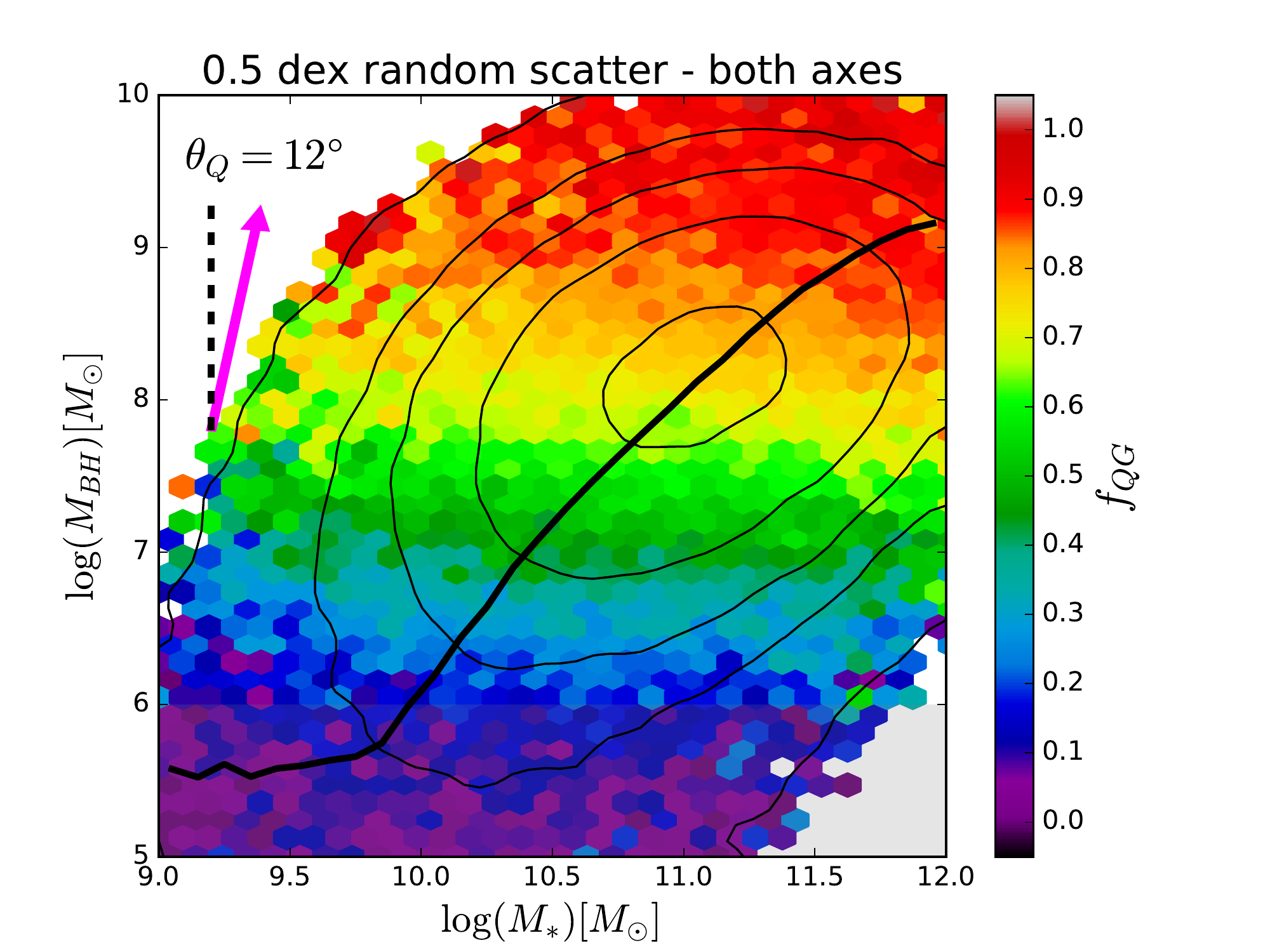}
\includegraphics[width=0.49\textwidth]{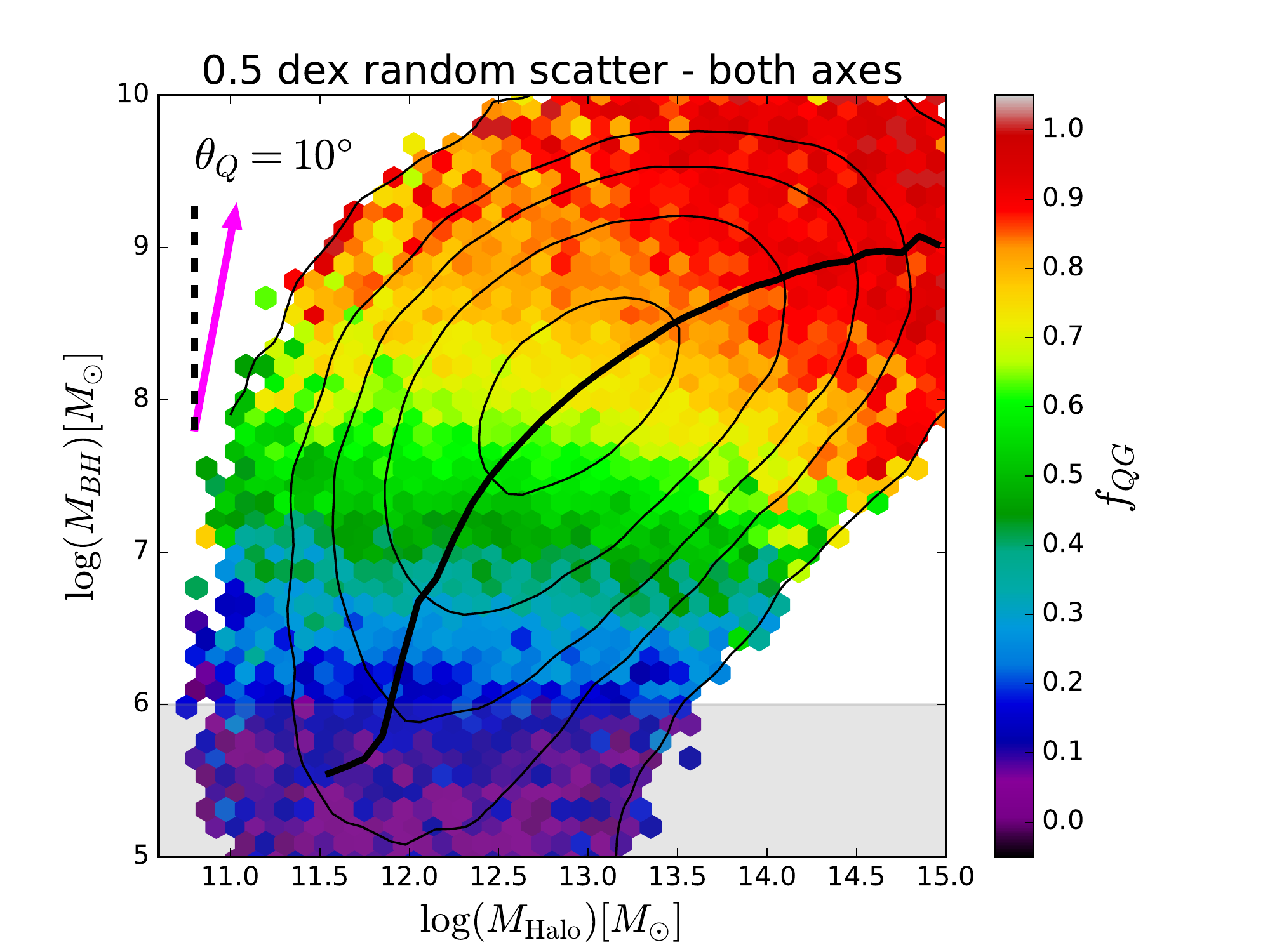}
\caption{Reproduction of Fig. 15 SDSS results with random noise added to both axes.}
\end{figure*}

\begin{figure*}
\includegraphics[width=0.49\textwidth]{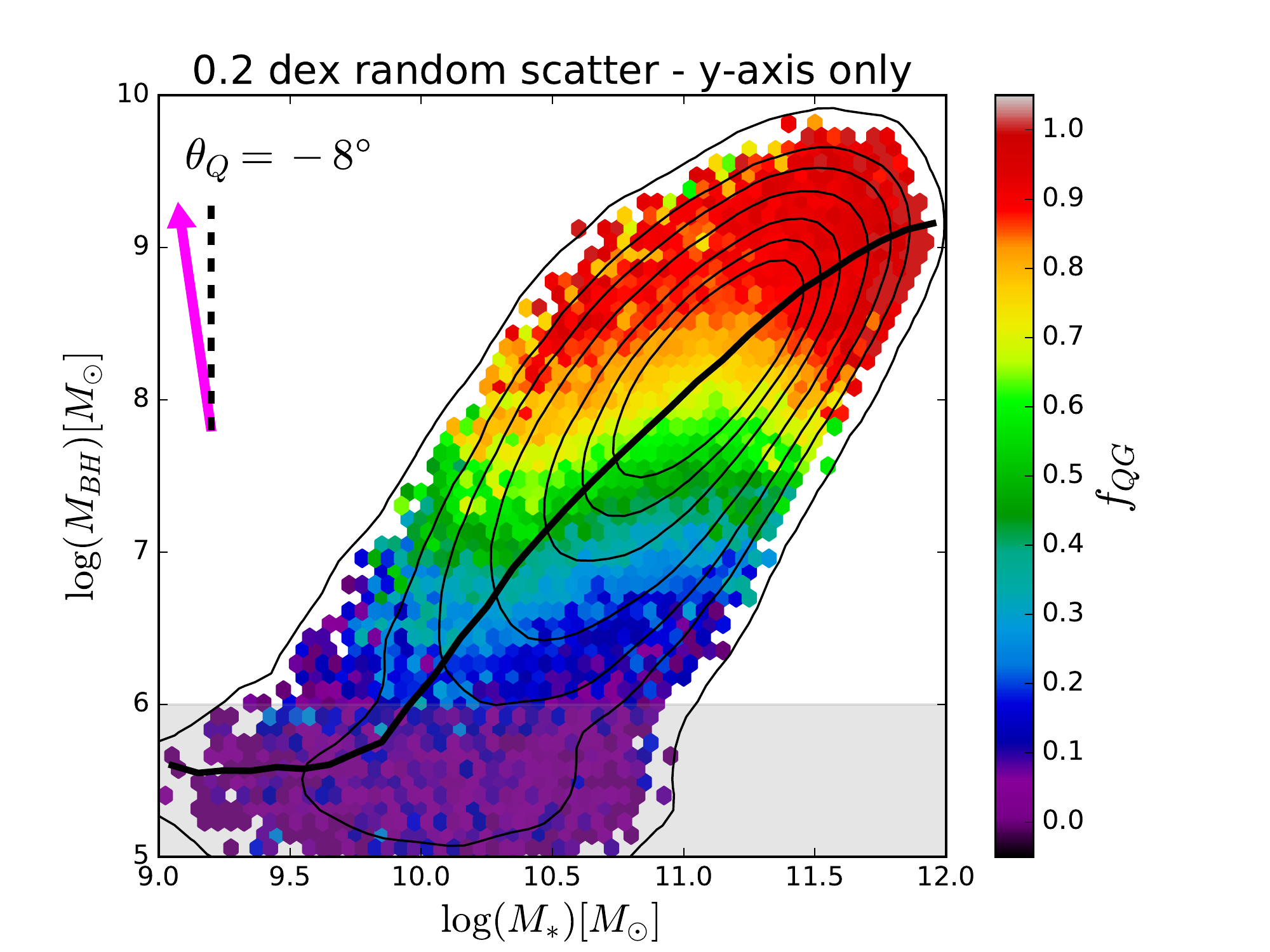}
\includegraphics[width=0.49\textwidth]{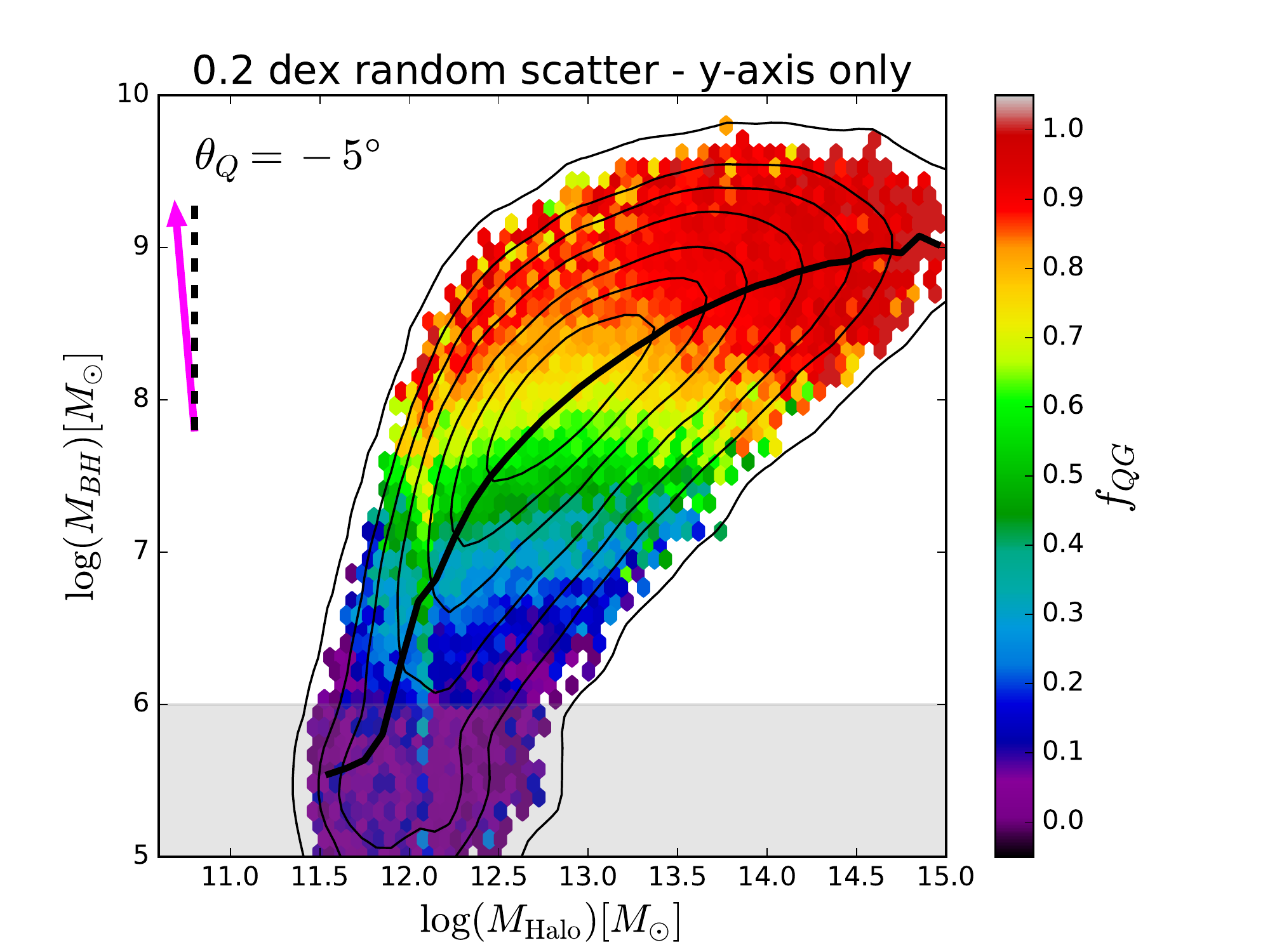}
\includegraphics[width=0.49\textwidth]{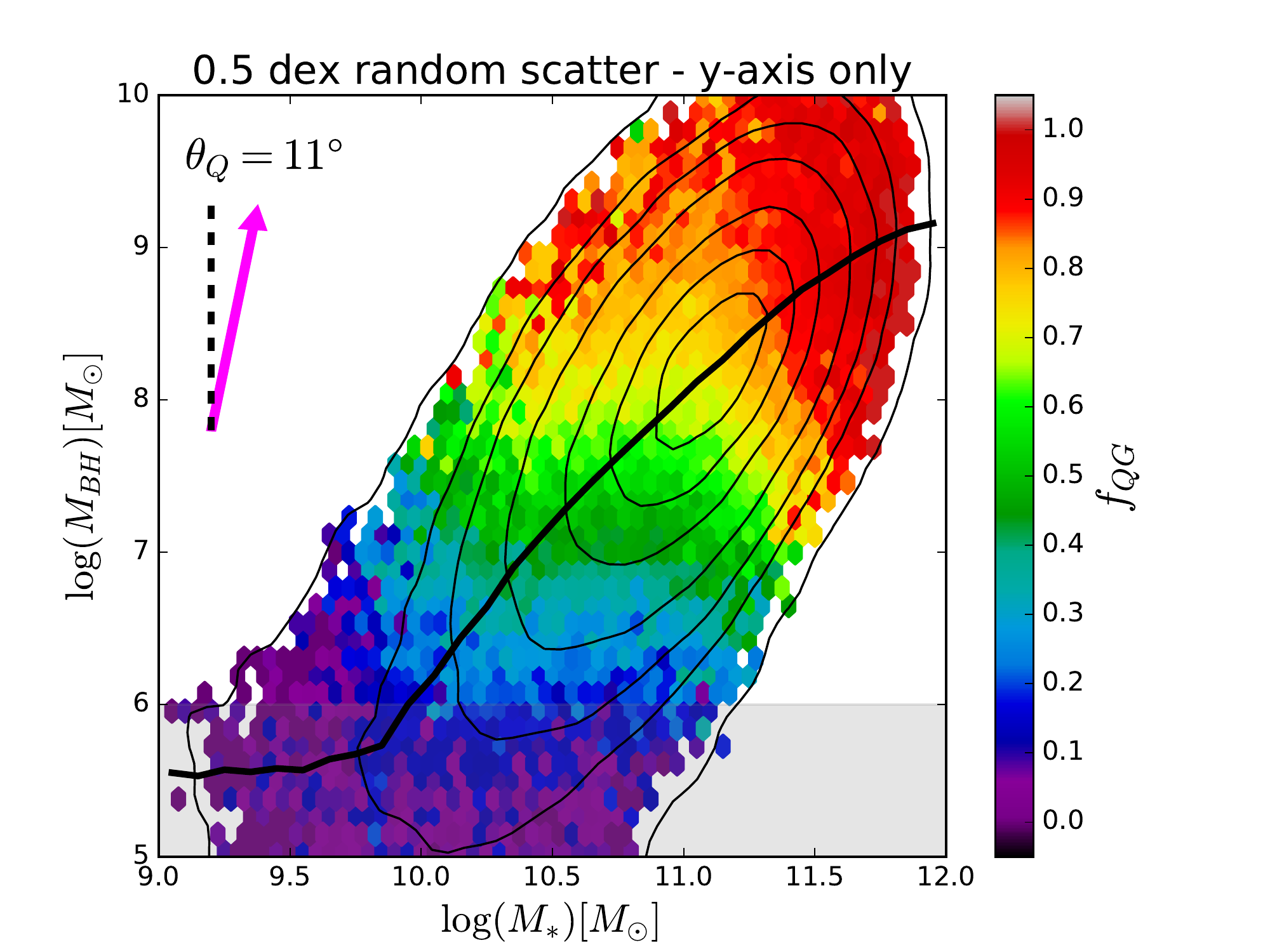}
\includegraphics[width=0.49\textwidth]{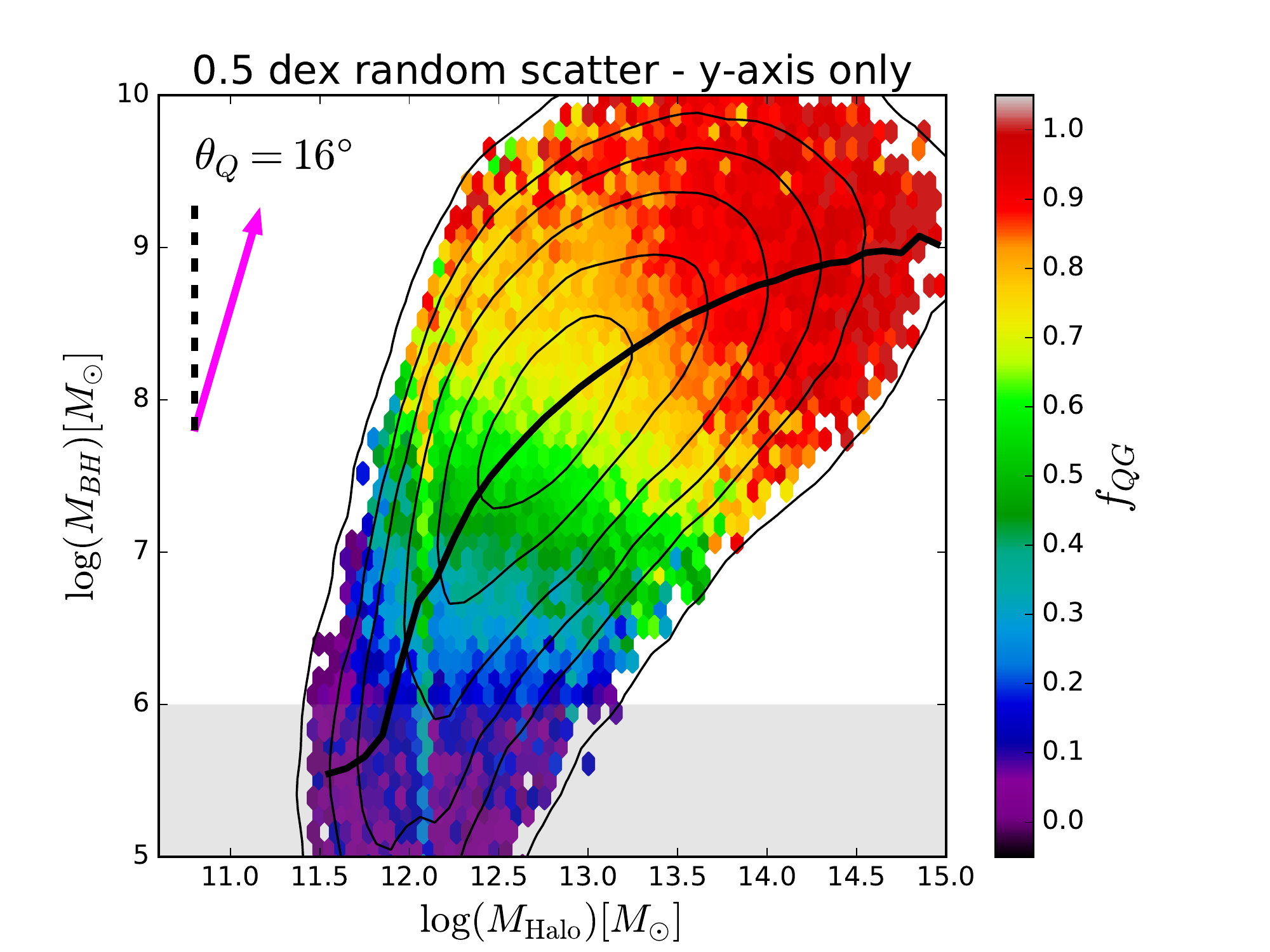}
\caption{Reproduction of Fig. 15 SDSS results with random noise added to the y-axis only.}
\end{figure*}

Throughout this paper we have investigated the connection of a host of global, local and environmental parameters to star formation and quenching. In our analysis we have used the largest and most accurate catalogues of these measurements currently available. Nonetheless, the performance of a given parameter (or group of parameters) is limited by the accuracy with which each parameter is measured. Naturally, increasing the measurement error on a given parameter will decrease its accuracy in predicting the level of star formation in a spaxel. Parameters with similar relative errors on their measurements may be compared freely, but parameters with a higher level of discrepancy in their error must be compared with caution.

The majority of parameters studied in this work have similar relative errors on their measurements. For example, global stellar mass, mass surface density in the spaxel, and mass surface density in the central 1kpc all have a typical error of $\sim$0.2-0.3 dex. Hence, the significant difference in performance between global and local mass measurements cannot be attributed to differences in their errors. Moreover, given that local parameters perform best at predicting $\Sigma_{\rm SFR}$ values in star forming regions, yet perform worst at predicting whether spaxels will be quenched or star forming, these differences cannot be attributed to measurement error. In the case where, for example, local parameters were measured with significantly less accuracy than global parameters, one would expect the latter always to outperform the former. This is clearly not the case. As such, the general conclusions presented throughout this paper (i.e. that star formation is a local process and quenching is a global process) cannot be explained by evoking the possibility of different measurement uncertainties between the parameters, and groups. Furthermore, the ranking results from the machine learning analyses of Section 4 are all highly stable to sample variation, SFR method, and to the addition of random noise into the data (as discussed in the previous sub-section). Thus, for the vast majority of parameters, we can safely conclude that our results are not dominated by errors in the measurements. 

On the other hand, some parameters which performed quite similarly in terms of their predictive power (e.g. halo mass and central velocity dispersion) may indeed have different levels of uncertainty on their measurements, which can potentially impact their rankings. The most important parameters to consider in this respect are $M_{*}$, $M_{\rm Halo}$ and $M_{BH}(\sigma_{c})$, which have proved very important in our interpretation (see Section 5 \& Appendix B). Given that we have found quenching to be a global process and hence that the spatially resolved dimension is largely unimportant (to leading order at least), we concentrate in this appendix on the MaNGA parent sample, SDSS DR7. The principal advantage of this approach is to utilise a much larger sample of galaxies ($\sim$500 000 vs. 3500) in order to investigate the potential impact of measurement errors on the rankings. In general, statistical noise tests are much more reliable on large samples.

In Fig. A4 we show the $M_{BH}$ - $M_*$ relation (left panels) and the  $M_{BH}$ - $M_{\rm Halo}$ relation (right panels), each colour coded by the fraction of quenched galaxies. In the top panel we introduce a moderate amount of random noise to each parameter, at the level of 0.2 dex. In the bottom panel we introduce a high level of random noise to each parameter, at a level of 0.5 dex. The low noise level represents an approximate lower limit to the error on each parameter, and the high noise level represents a conservative upper limit to the error on each parameter. These plots may be compared with the originals in Fig. 15.

Qualitatively, in Fig. A4 we see that progression from star forming to quenched (transitioning from blue to red in colour) occurs primarily in the vertical direction, for all realisations of the data considered. This is the same as seen for the unperturbed data in Fig. 15. We quantify this effect by using the quenching axis statistic (see eq. 16, and associated text). As before, we find that the quenching axis points far more in the vertical than horizontal direction. Thus, increasing the random noise equally to both axes does not significantly impact the main conclusion from this analysis: that quenching is more closely dependent on black hole mass than halo or stellar mass.

However, if one of the masses were measured much less accurately than the others this could impact the results in a way that would not be mimicked by adding random noise to all axes. This type of differential error can be modelled by adding noise to just one axis. The essential idea is to add random noise to the parameter one expects to be measured more accurately in order to level the playing field with the other parameters. 

In Fig. A5 we reproduce Fig. A4 but this time applying the 0.2 and 0.5 dex random noise only to the vertical axis (i.e. to $M_{BH}$). The reason for this is to simulate the hypothetical effect of halo mass and stellar mass being measured less accurately than black hole mass by a level of 0.2 and 0.5 dex, respectively. In all cases, the quenching axis still points more vertically than horizontally, and hence the conclusion that black hole mass is more connected with quenching in centrals than either stellar or halo mass is robust to both random error and differential random error up to a level of at least 0.5 dex. Therefore, even if the black hole masses were measured three times more accurately than the halo and stellar masses, we can still robustly conclude that it is black hole mass, not stellar or halo mass, which drives quenching. In order to achieve a quenching axis of 45$^{o}$ (i.e. an equal importance of $M_{BH}$ and $M_*$ or $M_{\rm Halo}$ to quenching centrals) we would need to raise the differential noise level on $M_{BH}$ to $\sim$1 dex. Thus, in order for our primary conclusion (that black hole mass is more important for quenching than either stellar or halo mass) to be invalid, black hole masses would need to be measured a factor of {\it ten times} more accurately than stellar or halo masses!

More realistically, it is likely that the average total error on the stellar masses is $\sim$0.3 dex (see Mendel et al. 2014); and the total error on the halo masses is $\sim$0.5 dex (taking into account uncertainties in the abundance matching and assignment of galaxies to groups; see Yang et al. 2009). The error on the central velocity dispersions are relatively low, $\sim$0.1-0.2 dex, but the total error on the estimated black hole masses is likely much higher at $\sim$0.5-0.6 dex (taking into account the scatter on the $M_{BH} - \sigma_c$ relationship, see Saglia et al. 2016). Consequently, the level of differential error considered in Fig. A5 is most probably (far) greater than that actually present in the data. Thus, the fact that we find the quenched fraction to still depend much more tightly on black hole mass than on stellar or halo mass leads us to conclude that measurement errors cannot be dominating this main result of the paper. It is important to highlight that this need not have been the case. In the event that adding random noise to $M_{BH}$ resulted in an inversion of the trend with quenching, one ought then to conclude that the results are potentially determined by the accuracy of the measurements. Fortunately, with the data from the SDSS and MaNGA we are far removed from this uncertain regime.

Finally, we emphasise that the most important test to the rankings and conclusions of this paper will come when a large enough sample of dynamically measured halo and black hole masses may be compared in terms of their galaxy quenching (realistically, this is most probably at least several decades away for the sample sizes studied in this paper). However, preliminary steps in this direction, comparing stellar mass with dynamically measured black hole mass for $\sim$90 objects, are indeed completely consistent with our analysis (see Terrazas et al. 2016, 2017).

%
%   APPENDIX - THEORY
%

\section{Towards a Theoretical Explanation of Central Galaxy Quenching -- An Analytic Approach}

To understand more concretely the connection between stellar, halo and black hole mass and the quenching of central galaxies we construct a meta-model\footnote{A theoretical framework for comparing different models, which share common features.} of quenching in this appendix. In general, the probability of a galaxy being quenched may be parameterized as follows (e.g. Bluck et al. 2016):

\begin{equation}
P_{Q} = \sum\limits_{j} \big( f_{j} (W_{QM,j} - \phi_{{\rm act}, j}) \big)
\end{equation} 

\noindent where $W_{QM}$ is the work done to the system by the quenching mechanism, and $\phi_{{\rm act}}$ is the activation threshold for quenching to take effect. This type of model can be naturally applied to heating vs. cooling, outflows vs. inflows, and ejecta kinetic energy vs. the binding energy of the ejected material (amongst many other possible scenarios). However, in this appendix we will focus on possible quenching mechanisms via heating. 

To ensure that the probability of quenching is sufficiently flexible in its response to the energetics of the system, we express the quenching probability as an arbitrary function ($f$) of the net energy ($W_{QM} - \phi_{{\rm act}}$). Additionally, we take a sum over all $j$-quenching processes, i.e. supernova heating, AGN heating, and virial shock heating. For a single quenching process, the probability of a galaxy being quenched is given simply by

\begin{equation}
P_{Q} = f(W_{QM} - \phi_{\rm act}) = f(\epsilon E_{QM} -  \phi_{\rm act})
\end{equation} 

\noindent where in the second expression $\epsilon$ indicates the coupling efficiency of the quenching mechanism, i.e. the fraction of energy released by the physical process which couples to the halo to engender quenching. For concreteness, we may consider the sigmoid activation function (S-curve) as a likely functional form, which has the following properties:

\begin{equation}
f(x) \approx S(x) \equiv \frac{1}{1 + e^{-\kappa x}}
\end{equation}

\noindent where we take $x = (W_{QM} - \phi_{\rm act})$ and $\kappa$ is an arbitrary scaling parameter controlling the rate of transition (e.g., accounting for the possibility of fast or slow quenching). The sigmoid function has several desirable properties: at energies well below the activation threshold the probability of quenching is zero; at values well above the activation threshold the probability of quenching is 1; and at values around the activation threshold there is a monotonic transition from star forming to quenched, with a variable rate of transition. However, it is important to note that the following arguments are not function independent, and thus the sigmoid function is purely illustrative of how one might construct a toy model of this type. We next consider AGN feedback, supernova feedback and virial shocks in turn as sources of heating in haloes. \\

\subsection{AGN Feedback}

\noindent For AGN feedback, the total energy released in forming a black hole (of mass $M_{BH}$) is given by (e.g. Soltan et al. 1982, Silk \& Rees 1998):

\begin{equation}
E_{AGN} = \int\limits_{t_{f}}\limits^{t_{0}} L_{AGN}(t) \hspace{0.1cm} dt   =  \int\limits_{t_{f}}\limits^{t_{0}} c^2 \eta_{AGN} \frac{dM_{BH}}{dt} \hspace{0.1cm} dt 
\end{equation}

\noindent where the luminosity of an AGN is given by $L_{AGN} = c^2\eta_{AGN}\Gamma_{BH}$, and  $\Gamma_{BH}$ is the accretion rate of the black hole ($\equiv dM_{BH}/dt$). $\eta_{AGN}$ is an efficiency parameter which quantifies the fraction of mass converted to energy (electromagnetic radiation) in the accretion process. Observational and theoretical constraints set the value of $\eta_{AGN}$ = $\sim$0.1 (e.g. Thorne et al. 1974, Soltan et al. 1982, Elvis et al. 1994). 

The resulting functional form of the above equation motivates a change of variables from time to mass of black hole, and thus we find:

\begin{equation}
E_{AGN} = c^2 \int\limits_{0}\limits^{M_{BH}} \eta_{AGN} \hspace{0.1cm} dM'_{BH} 
\end{equation}

\noindent To proceed further, it would be advantageous to remove the efficiency parameter ($\eta_{AGN}$) from the integral. Even if $\eta_{AGN}$ is variable in $M_{BH}$, by assuming the existence of a mass (or equivalently time) average of the variable $\eta_{AGN}$, we may write

\begin{equation}
E_{AGN} = \langle \eta_{AGN} \rangle c^2 \int\limits_{0}\limits^{M_{BH}} dM'_{BH} =  \langle \eta_{AGN} \rangle c^2 M_{BH}
\end{equation}

\noindent and thus

\begin{equation}
E_{AGN} \propto M_{BH}  \hspace{0.2cm}  \Rightarrow  \hspace{0.2cm}  W_{AGN} \propto M_{BH}
 \end{equation}

\noindent where the constant of proportionality is given formally by $c^2$ multiplied by

\begin{equation}
\langle \eta_{AGN} \rangle \equiv \frac{\int\limits_{0}\limits^{M_{BH}} \eta_{AGN} \hspace{0.1cm} dM_{BH}}{\int\limits_{0}\limits^{M_{BH}} dM_{BH} }
\end{equation}

\noindent If the above integral exists, then the logic of the preceding steps are sound. Note that, in general, $\langle \eta_{AGN} \rangle$ may depend on the final mass of the black hole ($M_{BH}$) but not the time dependent mass ($M'_{BH}(t)$). Thus, we arrive at a very simple expression for the total energy released in forming a black hole, which is independent of both the accretion history and the detailed physics of the accretion process (as in Silk \& Rees 1998). Consequently, we may interpret our estimated black hole masses throughout the Discussion section as being directly proportional to the AGN feedback energy (modulo the efficiency), and hence also to the work done to the system via AGN feedback (see eq. B2).\\

\subsection{Supernova Feedback}

\noindent Analogous arguments to those presented above for AGN may be applied to the energy released by supernovae (SN). The total energy released by supernovae over the history of a galaxy is given by:

\begin{equation}
E_{SN} =  \int\limits_{t_{f}}\limits^{t_{0}} L_{SN}(t) \hspace{0.1cm} dt  =  \int\limits_{t_{f}}\limits^{t_{0}} c^2 \eta_{SN} f_{SN} \frac{dM_{*}}{dt} \hspace{0.1cm} dt
\end{equation}

\noindent where $f_{SN}$ is the fraction of new stellar material which will undergo a supernova (Type I or II) in each time step, and $\eta_{SN}$ quantifies the efficiency of mass-to-energy conversion in the nuclear processes of a supernova. It is important to emphasize that supernovae are {\it far} less efficient than AGN at converting mass to energy, with $\eta_{SN}$ found to be in the range 0.0001 - 0.001 (e.g. Burrows et al. 2006, Dalla Vecchia et al. 2012). By changing variables, combining $f_{SN}$ and  $\eta_{SN}$ into a single parameter ($\tilde{\eta}_{SN}$), and using the assumption of the existence of a time (or, equivalently, mass) average of $\tilde{\eta}_{SN}$, we find:

\begin{equation}
E_{SN} =  c^2 \langle \tilde{\eta}_{SN} \rangle \int\limits_{0}\limits^{M_{*}} dM'_{*} = c^2 \langle \tilde{\eta}_{SN} \rangle M_{*} 
\end{equation}

\noindent and thus

\begin{equation}
E_{SN} \propto M_{*}  \hspace{0.2cm}  \Rightarrow  \hspace{0.2cm}  W_{SN} \propto M_{*}
 \end{equation}

\noindent where the average modified efficiency parameter is given formally by:

\begin{equation}
\langle \tilde{\eta}_{SN} \rangle \equiv  \frac{\int\limits_{0}\limits^{M_{*}} \eta_{SN} f_{SN} \hspace{0.1cm} dM'_{*}}{\int\limits_{0}\limits^{M_{*}} dM'_*}
\end{equation}

\noindent Providing the above integral exists, and may be solved at least numerically, the above line of reasoning leads inevitably to the conclusion that the total energy released from supernova explosions scales with the total stellar mass of the galaxy, with the constant of proportionality set by $\langle \tilde{\eta}_{SN} \rangle$. This result is in exact analogy to the result for AGN, discussed above. Once again, we arrive at a simple expression which is independent of both the star formation history of the galaxy, and the detailed nuclear physics of supernovae. Consequently, to leading order, we may interpret the stellar masses used throughout the Discussion section as a proxy for the total integrated energy from supernovae in each system. 

However, note that in general $\langle \tilde{\eta}_{SN} \rangle$ may vary as a function of the final stellar mass of the galaxy ($M_{*}$). Although the efficiency of mass-to-energy transfer in SN ($\eta_{SN}$) is unlikely to vary significantly, the fraction of new stellar material added to the galaxy which will undergo a supernovae ($f_{SN}$) will depend on both the initial mass function (IMF) and the fraction of the established stellar population added via mergers. Hence, a variable IMF with stellar mass (e.g., Conroy et al. 2009, 2012, 2013) would lead to a second order modification to the above simple direct proportionality between mass and energy. Furthermore, since mergers contribute a greater fraction of stars to higher mass systems (e.g., Bluck et al. 2009, 2012; Ownsworth et al. 2014, Duncan et al. 2014), the direct proportionality between mass and energy may be reduced (and hence become sub-linear) at higher masses. Thus, in full generality, the total energy released from supernovae may be given as a function of the total stellar mass of the galaxy, which simplifies further to direct proportionality in the case of constant efficiency, IMF and merger fraction.\\

\subsection{Virial Shock Heating}

\noindent For halo mass quenching, the above arguments are not directly applicable because the formation of virial shocks is not a result of mass-to-energy transfer. Instead, halo mass quenching from virial shocks arises from: first, the conversion of gravitational potential energy into ordered kinetic energy of an in-falling clump of matter; and second, the conversion of some fraction of the in-fall kinetic energy to a disordered form (i.e. heat).  Nevertheless, we may construct an analogous line of reasoning, in the hope of yielding a similar simple expression for the energy released via shock heating of a halo, which is independent of the detailed formation history of the halo and the complex physics of shocks. 

We begin with the virial relations for a dark matter halo (e.g. Mo, van den Bosch \& White 2010):

\begin{equation}
\Phi_{H} = -\frac{GM_{v}}{R_v} , \hspace{0.5cm} M_{v} = \frac{4\pi}{3}\Delta_{v} \rho_{\rm crit}(a) R_{v}^3
\end{equation}

\noindent where $M_v$ and $R_v$ are the virial mass and radius, respectively. $\rho_{\rm crit}$ is the critical density of the Universe (which varies as a function of the scale factor $a \equiv 1/(1+z)$). $\Delta_{v}$ is the virialised over-density, which is equal to 200 for a flat matter dominated Universe, and in general (for a more complex cosmological energy budget) varies weakly as a function of $a$. By substituting $R_v$ from the second expression above into the first we obtain:

\begin{equation}
\Phi_{H} = -G \big( \frac{4\pi \Delta_{v}}{3} \big)^{1/3} M_{v}^{2/3}  \rho_{\rm crit}(a)^{1/3}
\end{equation}

\noindent Thus, the gravitational potential is a function of the virial mass of the halo {\it and} the critical density of the Universe. 

In general, the dependence of the critical density on the scale factor is given by (e.g., Hogg 1999):

\begin{equation}
\rho_{\rm crit}(a) = \frac{3H(a)^2}{8\pi G} = \frac{3 H_{0}^2}{8\pi G} \bigg( \frac{\Omega_{r,0}}{a^4} + \frac{\Omega_{M,0}}{a^3} + \frac{\Omega_{K,0}}{a^2} + \Omega_{\Lambda} \bigg)
\end{equation}

\noindent where $\Omega_{X}$ for $X = (r, M, K, \Lambda)$ indicates the ratio of energy density to the critical energy density for radiation, matter, curvature and a cosmological constant, respectively. The subscripts of $0$ indicate the present value of those parameters, and $H$ indicates the Hubble parameter ($\equiv \dot{a}/a$). Proceeding with simple analytics utilising the above formula is not possible. However, the contribution to the energy density of the Universe of both curvature and radiation is negligible throughout the vast majority of the assembly history of dark matter haloes. Furthermore, the mass density of the Universe has dominated until relatively recently in the Universe's history (at z $\sim$ 0.3, e.g. Huterer \& Shafer 2018). As such, we may proceed with the simplifying assumption that the Universe is both flat and matter dominated and still yield useful results. In this simple case the over-density parameter yields $\Delta_v$=200, which sets $M_v = M_{200} \equiv M_{H}$ (as defined in this paper and in Yang et al. 2007, 2009). Using these assumptions, eq. B15 (above) simplifies considerably to:

\begin{eqnarray}
\rho_{\rm crit}(a) = \frac{3H_{0}^2\Omega_{M,0}}{8\pi G} \frac{1}{a^3}  \hspace{0.3cm} & \\
\Rightarrow \hspace{0.3cm} \Phi_{H} = -(GH_{0})^{2/3} (100 \Omega_{M,0} )^{1/3} \frac{M_{H}^{2/3}}{a} 
\end{eqnarray}

\noindent where the second expression above results from inserting the first expression above into eq. B14, and setting $M_v = M_H$. 

For a clump of baryonic material falling from rest at infinity on a radial plunge orbit into a virialised dark matter halo, the total kinetic energy of the clump entering the halo will be:

\begin{equation}
E_{\rm K, clump} = -\Phi_{H} \cdot M_{\rm clump} 
\end{equation}

\noindent and thus (inserting eq. B17 into the above),

\begin{equation}
E_{\rm K, clump} = (GH_{0})^{2/3} (100 \Omega_{M,0} )^{1/3} \frac{M_{H}^{2/3} M_{\rm clump}}{a} 
\end{equation}

\noindent Since this is the source of energy for virial shocks, we may define the incremental increase in the thermal energy of the halo from shock heating to be

\begin{equation}
\delta E_{\rm shock} = \eta_{\rm shock} \cdot E_{\rm K, clump}
\end{equation}

\noindent where $\eta_{\rm shock}$ is the fraction of kinetic energy of the in-falling clump of material which is converted into heat via the formation of virial shocks. By inserting eq. B19 into eq. B20, and furthermore noting that the baryonic mass of the clump may be expressed as $M_{\rm clump} \approx f_b \cdot \delta M_H$ (assuming the cosmological baryon fraction for the clump, $f_b$), we arrive at:

\begin{equation}
\delta E_{\rm shock} = \eta_{\rm shock} \cdot (GH_{0})^{2/3} (100 \Omega_{M,0} )^{1/3} \cdot \frac{M_{H}^{2/3}}{a} \cdot f_b \delta M_{H} 
\end{equation}

\noindent Integrating over time (or equivalently integrating over the incremental mass accreted), we find that the total energy released via shocks into the halo is:

\begin{equation}
E_{\rm shocks} = \langle \tilde{\eta}_{\rm shock} \rangle \int\limits_{0}\limits^{M_H} \frac{(M'_H)^{2/3}}{a} \hspace{0.1cm} dM'_{H}
\end{equation}

\noindent where we have simplified the above expression by collating the constants into the efficiency parameter by defining

\begin{equation}
\langle \tilde{\eta}_{\rm shock} \rangle = (GH_{0})^{2/3} (100 \Omega_{M,0} )^{1/3} f_b \langle \eta_{\rm shock} \rangle
\end{equation}

\noindent The formal solution for the time averaged conversion efficiency is given by

\begin{equation}
\langle \eta_{\rm shock} \rangle = \frac{ \int\limits_{0}\limits^{M_v} \eta_{\rm shock} ((M'_v)^{2/3} / a) \hspace{0.1cm} dM'_v}{ \int\limits_{0}\limits^{M_v} ((M'_v)^{2/3}) / a) \hspace{0.1cm} dM'_v}
\end{equation}

\noindent It is important to stress that the efficiency of shock heating ($\eta_{\rm shock}$) will in general depend of the mass of the halo, and hence the average shock heating efficiency ($\langle \eta_{\rm shock} \rangle$) will depend on the final mass of the halo. More specifically, Dekel \& Birnboim (2006) and Dekel et al. (2009) find through hydrodynamical simulations that the efficiency of shock formation is low (essentially zero) in low mass haloes, and rises to high values (of order unity) in high mass haloes. The critical transition in shock formation occurs at $M_{\rm crit} \sim 10^{12} M_{\odot}$ (Dekel \& Birnboim 2006). Hence, to leading order, we may model  $\eta_{\rm shock}$ as a step function centred on the critical halo mass. In such a simple picture, the above arguments motivate a change in the lower integration limit in eq. B22 from zero to $M_{\rm crit}$. However, a change to the lower integration limit will not alter the proportionality with final halo mass (only offset this by a constant value), and hence we leave the integral in eq. B22 unchanged. As noted before (e.g. Woo et al. 2013), this type of model predicts a sharp transition in quenching around the critical halo mass, which it is crucial to note is not observed at a fixed black hole mass (see Figs. 14 \& 15).

Unlike for AGN and SN feedback (discussed above), the resultant integrals (eqs. B22 \& B24) are not trivially a function of mass alone. Hence, we have demonstrated that it is not possible to remove the dependence of the energy from virial shocks on the accretion history of material into the halo. This is perhaps unsurprising in this case, given that the energy transfer process of shocks is necessarily dependent on the evolving gravitational potential. Ultimately, the reason for the dependence on the exact accretion history is that the radius of a dark matter halo (in a matter dominated flat Universe) is given by $R_H \propto M_H^{1/3} a$, and hence haloes of a given fixed mass were smaller at earlier cosmic times, resulting in a higher gravitational potential for their mass. The presence of the scale factor in the above integrals is a direct consequence of this well known effect.

To proceed further, we must assume something about the accretion of material into haloes as a function of the scale parameter (or equivalently time). It has been shown that the halo accretion history is well approximated by a simple power-law, at least up to z $\sim$ 2, where the vast majority of the mass in haloes is accreted (e.g. Dekel \& Birnboim 2006). Furthermore, quenching occurs almost exclusively at z $<$ 2 (e.g. Bauer et al. 2011, Lang et al. 2014, Barro et al. 2014), which supports the applicability of the power law approximation. As such, we posit $M_H \propto a^{\alpha}$ (and hence $a \propto M_H^{1/\alpha}$), where $\alpha$ is found to vary slowly between 4 - 5 from z = 2 to the present (e.g. Springel et al. 2005, Dekel \& Birnboim 2005). Consequently,

\begin{eqnarray}
E_{\rm shocks} = {\rm const} \times \langle \tilde{\eta}_{\rm shock} \rangle \int\limits_0\limits^{M_H} (M'_H)^{2/3 - 1/\alpha} dM'_H \\
=  \frac{ {\rm const} \times \langle \tilde{\eta}_{\rm shock} \rangle}{(5\alpha-3)/3\alpha} M_H^{(5\alpha-3)/3\alpha}
\end{eqnarray}

\noindent and hence,

\begin{eqnarray}
E_{\rm shocks} \propto M_H^{(5\alpha-3)/3\alpha} \approx M_H^{1.5} \\
\Rightarrow W_{\rm shocks} \propto  M_H^{1.5} 
\end{eqnarray}

\noindent where $\alpha$ is the exponent of a power-law parameterization of the mass growth of a dark matter halo as a function of the scale factor, and in the last step we assume $\alpha \approx$ 4.5 (e.g. Dekel \& Birnboim 2006). The above expression holds for $M_{H} > M_{\rm crit}$, and to leading order we may assume that the work done via shocks is zero below this threshold. Finally, we arrive at a simple expression for the total energy released in a halo via virial shocks, which is a function solely of the mass of the halo (under the assumptions of a geometrically flat, matter dominated Universe with a power law mass accretion history). Deviation from the above expression is expected due to the existence of a cosmological constant. However, given that the cosmological constant has only dominated the energy density of the Universe since $z \sim$ 0.3 (e.g. Huterer \& Shafer 2018), virial shocks throughout the vast majority of the formation history of dark matter haloes must follow the above simple relation.\\

\subsection{Summary of Theoretical Insights}

\noindent Bringing together all of the results from the preceding derivations (eqs. B7, B11 \& B28), we find that the work done to quench a galaxy via heating of the halo is given by:

\begin{eqnarray*}
\log(W_{QM}) \propto \log(M_{X}) \hspace{0.3cm} {\rm for} \hspace{0.1cm} M_{X} = & M_{BH} {\rm (AGN)}\\
&M_* {\rm (SN)}\\
& M_H {\rm (shocks)}\\
& \rm{[\gtrapprox 10^{12} M_{\odot}]}
\end{eqnarray*}

\noindent The direct proportionality between the logarithms of the work done on the halo to quench star formation and the mass of each component is quite remarkable. The arguments that lead to this point are highly general for AGN and SN, and are applicable for the majority of the accretion history of haloes for virial shocks. Even with variable efficiencies, the above results hold exactly provided the dependence of efficiency on mass can be modelled as a power law (which is almost invariably true, at least approximately). Thus, the accretion history of black holes and the detailed accretion physics are not relevant for determining the scaling of the total energy released from AGN feedback with $M_{BH}$ (as originally demonstrated in Soltan 1982, Silk \& Rees 1998). Similarly, the star formation history and the detailed nuclear physics of supernovae are not relevant for determining the scaling of the total energy released from SN feedback with $M_*$. For virial shocks, we were forced to parameterize the accretion history to make progress, but even here we arrive at a simple power-law dependence of energy on mass, and succeed in finding a result independent of the detailed shock physics. Ultimately, the power of the above approach is in its simplicity, i.e. by not being model dependent, the above results necessarily hold regardless of the myriad details involved in quenching galaxies. 

Our final step is to connect the above theoretical results to the observations of the rest of this paper. Evoking our meta-model for quenching (outlined in eqs. B1 \& B2 and associated text), we posit that the probability of a galaxy being quenched is equal to a function of the work done to the system above some activation threshold. Utilising the theoretical results of this section we may update this parameterization to:

\begin{equation}
P_{Q} = f'(\tilde{\epsilon} \cdot \log(M_X)  -  \phi_{\rm act})
\end{equation}

\noindent This is a significant leap forwards from a function given in terms of total energy, since current masses of galaxy haloes, galaxy stellar components, and central supermassive black holes are far easier to obtain stringent observational constrains on than the integrated energy released in forming these components over the entirety of cosmic time. Note that in the above expression the arbitrary function and the coupling efficiency parameter are both changed to new, arbitrary values to allow for the use of $\log(M_X)$ in the place of $W_{QM}$. 

Adopting the frequentist definition of probability, the fraction of objects with a given property will be equal to the probability that each object has the property in question, in the limit where the number of objects tends to infinity. (Note that all predictions from the frequentist and the Bayesian approach to statistics are identical in the case where the latter adopts a flat distribution of priors). Hence, in the case of galaxy and spaxel quenching this statement becomes:

\begin{eqnarray}
P_{QG} = \lim_{N_{\rm gal}\to\infty} \big( f_{QG} \big) \\
P_{QS} = \lim_{N_{\rm spax}\to\infty} \big( f_{QS} \big) 
\end{eqnarray}

\noindent where the subscripts $QG$ and $QS$ refer to quenched galaxies and quenched spaxels, respectively. For MaNGA we have a sample of $\sim$ 5 million spaxels, and for the SDSS parent sample we have a sample of over 500 000 galaxies. In both cases the data size is very large and hence we may assert that it is, in a meaningful sense, {\it tending} to infinity. Thus,

\begin{equation}
f_{QG} \approx P_{QG} \hspace{0.3cm} \& \hspace{0.3cm} f_{QS} \approx P_{QS}
\end{equation}

\noindent where the approximately equal to signs allow for the possibility that the fraction of quenched objects as a function of other parameters may vary with different (larger, more complete) datasets. At this point we have achieved our goal to link theory with observations. Summarising the entirety of this appendix, we may now write

\begin{equation} 
f_{Q} = f'(\tilde{\epsilon} \cdot \log(M_X)  -  \phi_{\rm act}) \approx \tilde{f}(M_{X}) + \mathcal{O}(\tilde{\epsilon}, \phi_{\rm act})
\end{equation}

\noindent where in the above expression, $f_{Q}$ can be read as the quenched galaxy or spaxel fraction (as desired) with the caveat that the variables and function may vary as a result. In the second step we emphasize that this is a function of $M_X$ primarily, with a corrective term based on the coupling efficiency and activation threshold (as could be derived from grouping terms in a Taylor expansion). 

We have arrived at a general prediction from our (highly flexible) meta-model of quenching: {\it the fraction of quenched galaxies (or spaxels) must scale primarily with the mass of the source of the quenching energy.} For AGN feedback (in any form) this mass is $M_{BH}$; for supernova (and all other forms of stellar feedback) this mass is $M_*$; and for virial shocks (and all other gravitational-potential-driven quenching mechanisms) this mass is $M_{H}$. Note that the above prediction still holds exactly, even if the efficiencies of each process vary with mass (due to the unspecified form of the function). Of course, all three of these masses are highly correlated with each other for central galaxies. High mass haloes host high stellar mass galaxies which host high mass central black holes. As such, it is important to test the impact on quenching of varying each of these masses at fixed values of the other masses.

In Figs. 14 \& 15 (presented in the Discussion) we directly test how the fraction of quenched spaxels (in MaNGA) and galaxies (in the SDSS) vary with $M_{BH}$ at fixed $M_{*}$ and $M_H$, and vice versa. For both datasets, and scales of quenching, we find that the fraction of quenched objects varies much more significantly with  $M_{BH}$ than $M_{*}$ or $M_H$, at fixed values of the other parameters. Therefore, given the energetics arguments of this appendix, our observational results lend strong support to the AGN heating paradigm of quenching, and pose serious problems to models which seek to explain quenching via either supernovae or virial shock heating.

Finally, it is of course still possible that the activation energy ($\phi_{\rm act}$) and the coupling efficiency ($\tilde{\epsilon}$) may be variables, dependent on other parameters than the mass of the quenching source. For example, the coupling efficiency of AGN feedback energy will likely increase with increasing temperature of the halo, and hence as a function of $M_H$. Additionally, the cooling rate of the hot gas halo (set to a large extent by the density of gas in the halo) is likely also a function of halo mass. Note that these two effects are opposite in impact - increasing the halo mass increases the coupling efficiency but it also increases the energy needed to reach the activation threshold (i.e. to offset cooling). However, the extremely tight coupling between quenched fraction and $M_{BH}$ (and its insensitivity to $M_{H}$) found observationally in this work suggest that these secondary dependencies on other parameters are subdominant to the heating source, as we have implicitly assumed throughout this section. To explore further the dependence of coupling efficiency and activation energy on various observables, different theoretical approaches and different observational data will be required. Hence, we postpone this interesting avenue of further research to future publications.

\end{document}